\documentclass[twocolumn]{aastex61}
\usepackage{graphicx, graphics, amsmath,amssymb, amsfonts}%Include figure fiels
\usepackage{dcolumn}% Align table columns on decimal point
\usepackage{bm}% bold math (modern LaTeX standard that works for all symbols)
\usepackage{natbib}
\usepackage{enumerate}
\usepackage[colorlinks=true]{hyperref}
\usepackage{cleveref}
\usepackage[utf8]{inputenc}

\begin{document}

\title{Zero net flux MRI-turbulence in disks -- sustenance scheme and magnetic Prandtl number dependence}

\correspondingauthor{G. Mamatsashvili}
\email{g.mamatsashvili@hzdr.de}

\author{George Mamatsashvili}

\affiliation{Helmholtz-Zentrum Dresden-Rossendorf, Bautzner Landstraße 400, D-01328 Dresden, Germany}
\affiliation{Niels Bohr International Academy, Niels Bohr Institute,
Blegdamsvej 17, 2100 Copenhagen, Denmark}
\affiliation{E. Kharadze Georgian National Astrophysical Observatory, Abastumani 0301, Georgia}
\affiliation{Institute of Geophysics, Tbilisi State University, Tbilisi 0193, Georgia}

\author{George Chagelishvili}
\affiliation{E. Kharadze Georgian National Astrophysical Observatory, Abastumani 0301, Georgia}
\affiliation{Institute of Geophysics, Tbilisi State University, Tbilisi 0193, Georgia}

\author{Martin E. Pessah}
\affiliation{Niels Bohr International Academy, Niels Bohr Institute,
Blegdamsvej 17, 2100 Copenhagen, Denmark}

\author{Frank Stefani}
\affiliation{Helmholtz-Zentrum Dresden-Rossendorf, Bautzner Landstraße 400, D-01328 Dresden, Germany}

\author{Gianluigi Bodo}
\affiliation{INAF Osservatorio Astrofisico di Torino, Strada
Osservatorio 20, I-10025 Pino Torinese, Italy}

\begin{abstract}
We investigate sustenance and dependence on magnetic Prandtl number (${\rm Pm}$) for magnetorotational instability (MRI)-driven turbulence in Keplerian disks with zero net magnetic flux using standard shearing box simulations. We focus on the turbulence dynamics in Fourier space, capturing specific/noncanonical anisotropy of nonlinear processes due to disk flow shear. This is a new type of nonlinear redistribution of modes over wavevector orientations in Fourier space  --  the nonlinear transverse cascade -- which is generic to shear flows and fundamentally different from usual direct/inverse cascade. The zero flux MRI has no exponentially growing modes, so its growth is transient, or nonmodal. Turbulence self-sustenance is governed by constructive cooperation of the transient growth of MRI and the nonlinear transverse cascade. This cooperation takes place at small wavenumbers (on the flow size scales) referred to as the vital area in Fourier space. The direct cascade transfers mode energy from the vital area to larger wavenumbers. At large ${\rm Pm}$, the transverse cascade prevails over the direct one, keeping most of modes' energy contained in small wavenumbers. With decreasing  ${\rm Pm}$, however, the action of the transverse cascade weakens and can no longer oppose the action of direct cascade which more efficiently transfers energy to higher wavenumbers, leading to increased resistive dissipation. This undermines the sustenance scheme, resulting in the turbulence decay. Thus, the decay of zero net flux MRI-turbulence with decreasing ${\rm Pm}$ is attributed to topological rearrangement of the nonlinear processes when the direct cascade begins to prevail over the transverse cascade.
\end{abstract}

\section{Introduction}
\label{sec:Introduction}

The challenge of understanding accretion processes and associated angular momentum transport in astrophysical disks requires a comprehensive study of the nonlinear dynamics of perturbations in this kind of flows. This has prompted, over the past decades, a vast number of theoretical and numerical analyses. Keplerian differential rotation plays a significant role in the linear stability and nonlinear dynamics of the disk flow. In fact, in the purely hydrodynamic case, Keplerian flows are both linearly and nonlinearly stable, ruling out self-sustained turbulence in them \citep[e.g.,][]{Hawley_etal99, Lesur_Longaretti05,Shen_Stone06,Rincon_etal07}. Recent
developments revealed, however, other physical processes, such as
vertical shear instability, convection and vortices capable of driving weak
turbulence and transport in non-magnetic regions of disks
\citep[e.g.,][]{Pfeil_Klahr19,Lyra_Umurhan19}.

The situation is fundamentally different in magnetized regions of
disks -- magnetic field imposed on Keplerian rotation of a
conducting disk matter gives rise to dynamic activity, which is
primarily manifested in the linear magnetorotational instability
(MRI) and its nonlinear outcome -- magnetohydrodynamic (MHD)
turbulence \citep{Balbus_Hawley91,Hawley_etal95,Balbus_Hawley98}.
The self-sustaining mechanism, statistical characteristics and
transport properties of this MRI-driven turbulence strongly
depend on the configuration of background magnetic field itself.

Disks threaded by a weak net vertical magnetic field are
unstable to MRI -- axisymmetric perturbations can grow exponentially
on a fast orbital time at a linear stage, that greatly facilitates
the onset of turbulence
\citep[e.g.,][]{Hawley_etal95,Sano_Inutsuka01,Lesur_Longaretti07,
Pessah_Goodman09,Pessah10,Longaretti_Lesur10}. In addition to these exponentially growing axisymmetric
(channel) modes, the system contains transiently, or
\emph{nonmodally} growing non-axisymmetric MRI modes due to shear
flow nonnormality
\citep{Balbus_Hawley92,Pessah_Chan12,Mamatsashvili_etal13,
Squire_Bhattacharjee14,Squire_Bhattacharjee14a,Zhuravlev_Razdoburdin14,Razdoburdin_Zhuravlev17},
which, despite being transient, strongly affect
the statistical properties of MRI-turbulence
\citep[][hereafter Paper II]{Longaretti_Lesur10, Bodo_etal08,
Murphy_Pessah15, Gogichaishvili_etal18}.

Disks threaded by a net azimuthal magnetic field lack linear
exponentially growing axisymmetric modes and hence the turbulence
can be energetically fueled only by the linear process of
transient/nonmodal growth of non-axisymmetric MRI \citep[e.g.,
][]{Hawley_etal95,Brandenburg_Dintrans06,Simon_Hawley09,Pessah_Chan12,Squire_Bhattacharjee14,Mamatsashvili_Stefani17}.
In this case, the role of a specific, or noncanonical nonlinearity
-- the so-called nonlinear transverse cascade (see below) -- is
crucial, since it ensures positive feedback for nonmodally growing MRI
modes \citep[][hereafter Paper I]{Gogichaishvili_etal17}.

\subsection{An overview of zero net flux MRI-turbulence}

Disks with zero net magnetic flux occupy a special place in
MRI-turbulence research and have been very actively studied since
the seminal paper by \citet{Hawley_etal96}, which had demonstrated
for the first time a link between MRI and self-sustaining magnetic
dynamo action in disks. The zero net flux MRI has also spurred much
debate in the last decade because of the nonconvergence
issue in numerical simulations \citep{Pessah_etal07,
Fromang_Papaloizou07,
Guan_etal09,Davis_etal10,Bodo_etal11,Bodo_etal14,Ryan_etal17}. The
chief appeal of the zero net flux case is that it offers the
prospect of universal state of MRI-turbulence in which the disk
generates and maintains its own magnetic field via joint action of
dynamo mechanism and MRI, not requiring any imposed net field. The
resulting transport of angular momentum would then depend on the
disk properties, but not on the magnitude and direction of magnetic
field in the disk.

Unlike the net vertical field case, but similar to the net azimuthal
field case, there are no purely exponentially growing large-scale
linear MRI modes for the zero net flux configuration. The growth of
MRI is then of transient type and hence the resulting turbulence and
associated dynamo action are subcritical by nature
\citep{Rincon_etal08}, being energetically powered by this linear
nonmodal MRI growth mechanism. But, the self-sustaining schemes and
statistical characteristics of the subcritical MRI-turbulence generally differ for these two configurations of magnetic field. Analyzing the self-sustaining schemes of such subcritical MHD turbulence in disks, one is tempted to appeal to mathematical concepts and tools that lie at the basis of self-sustenance of subcritical turbulence in hydrodynamic (Couette, Poiseuille, pipe, etc.) shear flows. Yet, such an approach can be
misleading because these flows are bounded and the corresponding self-sustaining schemes \citep[elaborated by the hydrodynamic community over the past three decades, see
e.g.,][]{Hamilton_etal95,Waleffe95, Schmid_Henningson01}
assign a key role to the effect of boundaries. However, in astrophysical disks, especially in local shearing box considerations, there are effectively no rigid boundaries. Therefore, the self-sustaining scheme of the subcritical turbulence in the disks should be built on the interplay between linear nonmodal MRI and nonlinear processes.

Over the past ten years, a large number of numerical studies have
extensively explored statistical properties of unstratified zero net
flux MRI-turbulence in disks. Using the spherical shell-averaging
technique in Fourier space
\citep[e.g.,][]{Verma04,Alexakis_etal07,Alexakis_Biferale18}, the
spectra of magnetic and kinetic energies as well as Maxwell stress
were analyzed first without explicit dissipation, where the infamous
issue of numerical non-convergence arises \citep[e.g.,][]{Pessah_etal07,Fromang_Papaloizou07,Simon_etal09,Guan_etal09,Bodo_etal11,Shi_etal16}. This prompted further studies, where different configurations and physical ingredients were also considered, for
example, explicit/physical viscous and resistive dissipation,
different boundary conditions and box aspect ratios
\citep[e.g.,][]{Fromang_etal07,Liljestrom_etal09,Fromang10,Kapyla_Korpi11,Herault_etal11,Riols_etal13,Riols_etal15,Bhat_Ebrahimi16,Shi_etal16,Walker_etal16,Riols_etal17,Potter_Balbus17,Walker_Boldyrev17,Nauman_Pessah16,Nauman_Pessah18}. These studies revealed a quite complex behavior, where convergence may depend on several factors. It was also demonstrated that at sufficiently high Reynolds and magnetic Reynolds numbers, the saturated value of the total turbulent stress appears to depend only on the magnetic Prandtl number, ${\rm Pm}$, scaling as its
power-law at ${\rm Pm}\gtrsim 1$, whereas the turbulence is generally not sustained at ${\rm Pm}\lesssim 1$ in standard boxes (i.e., with radial/azimuthal sizes larger than vertical one). Recently, \citet{Guseva_etal17} have reported sustained zero net flux MRI-turbulence and dynamo at large ${\rm Pm}$ in a magnetized cylindrical Taylor-Couette flow, going beyond the Cartesian shearing box.

Despite the above numerous theoretical efforts that have gone into
the analysis of the zero net flux MRI-turbulence phenomenon,
relatively less attention has been devoted to properly understanding
an underlying self-sustaining mechanism and consequently elucidating
physics behind the ${\rm Pm}$-dependence of the turbulence. Clearly,
one could not do the latter without first doing the former. Perhaps
this is because most of these studies have concentrated in physical
space, hence offering a limited insight into mode dynamics, while a
much richer dynamical picture unfolds in Fourier space, which is at
the basis of the turbulence self-sustenance. Only a few papers
looked into the dynamics of zero net flux MRI-turbulence in Fourier
(${\bf k}$-) space
\citep[e.g.,][]{Fromang_Papaloizou07,Fromang_etal07,Simon_etal09,Davis_etal10}.
However, they used various types of averaging techniques (over
spherical shells of constant wavevector magnitude $k=|{\bf k}|$ or
over slices with different directions of ${\bf k}$, etc.), which in
the end have led to overlooking a key ingredient of the dynamics --
specific spectral anisotropy of nonlinear transfers due to flow
shear, i.e., the transverse cascade -- and thus somewhat hindered
uncovering the full dynamical story. With a similar approach
\cite{Lesur_Longaretti11} considered the spectral dynamics of
MRI-turbulence in Fourier space but with nonzero net vertical field.
They emphasized the nonlocality of nonlinear transfers, and  focused
only on explaining the dependence on ${\rm Pm}$, rather than clarifying the self-sustaining mechanism. In fact, in this case, there are exponentially growing MRI channel modes in the flow
and hence no shortage of energy supply to the
turbulence. So, the role of nonlinearity in the self-sustenance of
perturbations is not as crucial as in the net azimuthal or zero net
flux cases.

In this quest for the self-sustenance in the zero net flux case,
\cite{Herault_etal11} and \cite{Riols_etal13,Riols_etal15,
Riols_etal17} carried out a detailed study of the spectral dynamics
of nonlinear MRI-dynamo states in Fourier space in the shearing box. They also interpreted the dependence of the nonlinear state and associated transport on ${\rm Pm}$ as the effect of turbulent, or nonlinear magnetic diffusion, which increases with decreasing ${\rm Pm}$. However, these studies adopted relatively small Reynolds and magnetic Reynolds numbers, and specially designed azimuthally or vertically elongated boxes. Due to this, their models comprise only a relatively low number of active modes (degrees of freedom) with small wavenumbers and therefore are non-turbulent, low-order models of zero net flux MRI. So, the self-sustenance scheme of the MRI-dynamo states elaborated by these authors is based on the interplay of the dominant large-scale axisymmetric mode, responsible for the dynamo azimuthal field, and the next large-scale non-axisymmetric modes. On the other hand, the number of active modes generally involved in the fully developed MRI-turbulence is orders of magnitude higher, spanning a broader range of wavenumbers (see also Papers I and II). Consequently, the overall self-sustenance scheme of the MRI-turbulence is more complex, being determined by the dynamical interplay of much larger number of these active modes. 

\subsection{Goals of this study}

For a deeper understanding of the self-sustaining process of the zero net flux MRI-turbulence a detailed analysis of scale-to-scale nonlinear interactions (transfers), linear processes and their interplay in Fourier ${\bf k}$-space is necessary without resorting to low-order models. Based on this, one can then explain the dependence of the turbulence level on ${\rm Pm}$. This is exactly the program we set as the main goal in this paper. We reveal and analyze the structural type of nonlinear action in disks with zero net magnetic flux. It ensures the replenishment of modes capable of drawing shear flow energy through the linear nonmodal MRI mechanism and, in this way, a continual energy supply to the turbulence. Performed here spectral analysis indicates that the main nonlinear process in this case is an angular, that is, over wavevector orientations redistribution of modes in Fourier space. As mentioned above, we call this anisotropic type of nonlinear action in shear flows, the {\it nonlinear transverse cascade}. Such an angular redistribution in Fourier space due to nonlinearity was first found for a simplified two-dimensional (2D) model problem in constant shear flows by \citet{Chagelishvili_etal02}. Later, analyzing the interplay of linear transient growth and nonlinear processes, \citet{Horton_etal10} and  \citet{Mamatsashvili_etal14} revealed and distinctly described the action of the nonlinear transverse cascade in spectrally stable 2D hydrodynamic and MHD plane shear flows, which is mainly responsible for the turbulence sustenance therein. The similar analysis, showing the importance of the transverse cascade for the nonzero net flux MRI-turbulence in Keplerian shear flows, is carried out in Papers I and II. It should be emphasized that the nonlinear transverse cascade, being induced by flow shear, essentially differs from the canonical direct/inverse cascades in classical (i.e., without mean shear flow), Kolmogorov, Iroshnikov–Kraichnan or Goldreich-Sridhar theories of turbulence \citep[e.g.,][]{Biskamp03,Alexakis_Biferale18}.

In this paper, we first demonstrate the significance of the
nonlinear transverse cascade for the sustenance of zero net flux
MRI-turbulence at ${\rm Pm}> 1$ (specifically, we consider the ${\rm
Pm} = 4$ case). Then, we show that with decreasing ${\rm Pm}$, the
action of the transverse cascade weakens, so that it can no longer interfere with the action of the direct cascade, which transfers energy from small wavenumber modes (from the so-called vital area, see e.g., Papers I and II) towards higher wavenumber ones. Such a course of events ultimately leads to the increased resistive dissipation of the mode energy and, consequently, to the decay of the turbulence. In other words, we attribute the decay of the zero net flux MRI-turbulence with decreasing ${\rm Pm}$ to structural/topological changes in the nonlinear processes when the direct cascade begins to dominate the nonlinear transverse cascade, which is crucial for the turbulence sustenance.

The paper is organized as follows. The physical model and main
equations in Fourier space are given in Section
\ref{sec:Basicequations}. Simulations of zero net flux
MRI-turbulence are described in Section \ref{sec:DNS}. Spectral dynamics of the turbulence in Fourier space and the
proposed sustenance scheme are presented in
Section \ref{sec:Spectraldynamics}, the dependence of the turbulence
dynamics on magnetic Prandtl number is analyzed in Section
\ref{sec:Pmdependence}. Summary and relation to other
works are given in Section \ref{sec:Conclusion}.

\section{Physical model and basic equations}
\label{sec:Basicequations}

We adopt the shearing box model of an accretion disk \citep{Goldreich_LyndenBell65}, where a local Cartesian coordinate system $(x,y,z)$ centered at a radius $r_0$ orbits with angular velocity $\Omega$ of the disk at this radius. This reference frame has the unit vectors $({\bf e}_x, {\bf e}_y, {\bf e}_z)$, respectively, in the radial ($x$), azimuthal ($y$) and vertical ($z$) directions. The main equations of incompressible non-ideal MHD in the shearing box are
\begin{multline}\label{eq:mom}
\frac{\partial {\bf U}}{\partial t}+({\bf U}\cdot \nabla) {\bf
U}=-\frac{1}{\rho}\nabla P +\frac{\left({\bf B}\cdot\nabla\right){\bf B}}{4\pi \rho} -\\ 2\Omega{\bf e}_z\times{\bf U}+2q\Omega^2 x {\bf e}_x +\nu\nabla^2 {\bf U},
\end{multline}
\begin{equation}\label{eq:ind}
\frac{\partial {\bf B}}{\partial t}=\nabla\times \left( {\bf U}\times {\bf B}\right)+\eta\nabla^2{\bf B},
\end{equation}
\begin{equation}\label{eq:div}
\nabla\cdot {\bf U}=0,~~~~\nabla\cdot {\bf B}=0,
\end{equation}
where $\rho$ is the fluid density, ${\bf U}$ is the velocity in the
rotating frame, ${\bf B}$ is the magnetic field and $P$ is the sum of the thermal and magnetic pressures. The explicit
kinematic viscosity, $\nu$, and Ohmic resistivity, $\eta$, of the fluid
are constant. In the shearing box, Keplerian rotation of the disk is represented as an azimuthal flow ${\bf U}_0=-q\Omega x{\bf e}_y$ with constant radial shear parameter $q=3/2$, spatially uniform pressure $P_0$ and density $\rho_0$. The flow domain is a rectangle with sizes $(L_x,L_y,L_z)$.

We normalize time by $\Omega^{-1}$, length by vertical box size $L_z$, velocities by $\Omega L_z$, magnetic field by $\Omega L_z\sqrt{4\pi\rho_0}$ and the pressure by $\rho_0\Omega^2L_z^2$. The relevant parameters of the problem are the Reynolds number, ${\rm Re}=\Omega L_z^2/\nu$, the magnetic Reynolds number, ${\rm Rm}=\Omega L_z^2/\eta$, and the magnetic Prandtl number, ${\rm Pm}=\nu/\eta={\rm Rm}/{\rm Re}$, which plays a very important role in MRI-turbulence. 

An initial field configuration represents a purely vertical magnetic field, which varies only along $x$ as
\[
{\bf B}=B_0\sin\left(\frac{2\pi x}{L_x}\right){\bf e}_z,
\]
and hence has zero net flux within the domain. The field amplitude,
$B_0$, is specified by the parameter $\beta=2\Omega^2L_z^2/v_A^2=100$, where $v_A=B_{0}/(4\pi \rho_0)^{1/2}$ is the amplitude of the associated Alfv\'en speed. 

In the following, we divide the total velocity ${\bf U}$ into the stationary background Keplerian flow ${\bf U}_0$ and perturbation ${\bf u}$ on top of it, ${\bf U}={\bf U}_0+{\bf u}$, and work with ${\bf u}$ and ${\bf B}$ (Appendix A gives the equations for these variables as derived from main Equations \ref{eq:mom}-\ref{eq:div}). Our primary goal is to deepen our understanding of the self-sustaining dynamics of zero net flux MRI-turbulence and its dependence on viscous and resistive dissipation. To this end, we focus on the turbulence dynamics in Fourier space, applying the similar tools as used in Paper I.

\subsection{Equations in Fourier space}
\label{sec:Fourierspace}
The second part of our study starts with the derivation of \emph{spectral dynamical equations} for the velocity and magnetic field components in Fourier space. All variables are decomposed into spatial Fourier modes
\begin{equation}\label{eq:Fourier}
f({\bf r},t)=\int \bar{f}({\bf k},t)\exp\left({\rm i}{\bf
    k}\cdot{\bf r} \right)d^3{\bf k}
\end{equation}
where $f$ stands for $({\bf u}, P, {\bf B})$ and $\bar{f}$ for the corresponding Fourier transforms $(\bar{\bf u}, \bar{P}, \bar{\bf B})$. Substituting (\ref{eq:Fourier}) into Equations (\ref{eq:App-ux})-(\ref{eq:App-divB}), we obtain Equations (\ref{eq:App-uxk1})-(\ref{eq:App-uzk1}) for the spectral velocity and  Equations (\ref{eq:App-bxk})-(\ref{eq:App-bzk}) for the magnetic field  components. Below we give the final equations for the quadratic forms of these quantities, while their derivation is given in Appendix A.

Multiplying Equations (\ref{eq:App-uxk1})-(\ref{eq:App-uzk1}),
respectively, by $\bar{u}_x^{\ast}$, $\bar{u}_y^{\ast}$,
$\bar{u}_z^{\ast}$, and summing with their conjugates, we
have
\begin{equation}\label{eq:uxk2}
\frac{\partial}{\partial t}\frac{|\bar{u}_x|^2}{2} = -
qk_y\frac{\partial}{\partial k_x}\frac{|\bar{u}_x|^2}{2} + {\cal
    H}_x  +{\cal
    D}_x^{(u)}+{\cal N}_x^{(u)},
\end{equation}
\begin{equation}\label{eq:uyk2}
\frac{\partial}{\partial t} \frac{|\bar{u}_y|^2}{2} = -
qk_y\frac{\partial }{\partial k_x}\frac{|\bar{u}_y|^2}{2} + {\cal
    H}_y +{\cal
    D}_y^{(u)}+{\cal N}_y^{(u)},
\end{equation}
\begin{equation}\label{eq:uzk2}
\frac{\partial}{\partial t} \frac{|\bar{u}_z|^2}{2} = -
qk_y\frac{\partial}{\partial k_x}\frac{|\bar{u}_z|^2}{2} + {\cal
    H}_z +{\cal
    D}_z^{(u)}+{\cal N}_z^{(u)},
\end{equation}
where the linear terms are\footnote{Although these terms are quadratic, we still refer to them as ``linear'', because they have a linear origin, being obtained via multiplication of the corresponding linear terms of spectral Equations (\ref{eq:App-uxk})-(\ref{eq:App-bzk}) by the complex conjugates of the Fourier amplitudes of the velocity and magnetic field.}
\begin{equation*}\label{eq:Hx}
{\cal H}_x=\left(1-\frac{k_x^2}{k^2}\right)(\bar{u}_x\bar{u}_y^{\ast}+\bar{u}_x^{\ast}\bar{u}_y)+2(1-q)\frac{k_xk_y}{k^2}|\bar{u}_x|^2,
\end{equation*}
\begin{equation*}\label{eq:Hy}
{\cal
    H}_y=\frac{1}{2}\left[q-2-2(q-1)\frac{k_y^2}{k^2}\right](\bar{u}_x\bar{u}_y^{\ast}+\bar{u}_x^{\ast}\bar{u}_y)
- 2\frac{k_xk_y}{k^2}|\bar{u}_y|^2
\end{equation*}
\begin{equation*}\label{eq:Hz}
{\cal H}_z =
(1-q)\frac{k_yk_z}{k^2}(\bar{u}_x\bar{u}_z^{\ast}+\bar{u}_x^{\ast}\bar{u}_z)
-\frac{k_xk_z}{k^2}(\bar{u}_y\bar{u}_z^{\ast}+\bar{u}_y^{\ast}\bar{u}_z),
\end{equation*}
and the negative terms of viscous dissipation
\begin{equation*}\label{eq:Dui}
{\cal D}_i^{(u)}=-\frac{k^2}{\rm Re}|\bar{u}_i|^2,
\end{equation*}
while the modified nonlinear transfer functions in these spectral equations are
\begin{equation*}\label{eq:Nui}
{\cal
    N}^{(u)}_i=\frac{1}{2}(\bar{u}_iQ^{\ast}_i+\bar{u}_i^{\ast}Q_i),
\end{equation*}
with  the index $i=x,y,z$ here and everywhere below. The quantity $Q_i$ (given by Equation \ref{eq:App-Qi}) describes the nonlinear redistributions via triad interactions for the spectral velocities $\bar{u}_i$ in Equations (\ref{eq:App-uxk1})-(\ref{eq:App-uzk1}). The sum of ${\cal H}_i$ is equal to the Reynolds stress spectrum multiplied by $q$, ${\cal H}={\cal H}_x+{\cal H}_y+{\cal H}_z=q(\bar{u}_x\bar{u}_y^{\ast}+\bar{u}_x^{\ast}\bar{u}_y)/2$.

Next, multiplying Equations (\ref{eq:App-bxk})-(\ref{eq:App-bzk}),
respectively, by $\bar{B}_x^{\ast}$, $\bar{B}_y^{\ast}$,
$\bar{B}_z^{\ast}$, and summing with their conjugates, we
obtain
\begin{equation}\label{eq:bxk2}
\frac{\partial}{\partial
    t}\frac{|\bar{B}_x|^2}{2}=-qk_y\frac{\partial}{\partial k_x}
\frac{|\bar{B}_x|^2}{2} +{\cal D}_x^{(b)}+{\cal
    N}^{(b)}_x
\end{equation}
\begin{equation}\label{eq:byk2}
\frac{\partial}{\partial
    t}\frac{|\bar{B}_y|^2}{2}=-qk_y\frac{\partial}{\partial k_x}
\frac{|\bar{B}_y|^2}{2}+{\cal M}+{\cal
    D}_y^{(b)}+{\cal N}^{(b)}_y
\end{equation}
\begin{equation}\label{eq:bzk2}
\frac{\partial}{\partial
    t}\frac{|\bar{B}_z|^2}{2}=-qk_y\frac{\partial}{\partial k_x}
\frac{|\bar{B}_z|^2}{2}+{\cal D}_z^{(b)}+{\cal
    N}^{(b)}_z,
\end{equation}
where the linear terms are the Maxwell stress spectrum ${\cal M}$ multiplied by $q$,
\begin{equation*}\label{eq:M}
{\cal
    M}=-\frac{q}{2}(\bar{B}_x\bar{B}_y^{\ast}+\bar{B}_x^{\ast}\bar{B}_y),
\end{equation*}
and the negative terms of resistive dissipation
\begin{equation*}\label{eq:Dbi}
{\cal D}_i^{(b)}=-\frac{k^2}{\rm Rm}|\bar{b}_i|^2,
\end{equation*}
while the modified nonlinear terms in these equations are,
\begin{eqnarray*}\label{eq:Nbx}
{\cal N}^{(b)}_x=\frac{\rm
    i}{2}\bar{B}_x^{\ast}[k_y\bar{F}_z-k_z\bar{F}_y]+c.c., \\
{\cal N}^{(b)}_y=\frac{\rm
    i}{2}\bar{B}_y^{\ast}[k_z\bar{F}_x-k_x\bar{F}_z]+c.c., \\
{\cal N}^{(b)}_z=\frac{\rm
    i}{2}\bar{B}_z^{\ast}[k_x\bar{F}_y-k_y\bar{F}_x]+c.c. .
\end{eqnarray*}
Here $\bar{F}_x, \bar{F}_y, \bar{F}_z$ (given by Equations
\ref{eq:App-Fxk}-\ref{eq:App-Fzk}) are the Fourier transforms of the
respective components of the perturbed electromotive force, which
describe nonlinear transfers for the spectral magnetic field
components in Equations (\ref{eq:App-bxk})-(\ref{eq:App-bzk}) via triad interactions.

Equations (\ref{eq:uxk2})-(\ref{eq:uzk2}) and (\ref{eq:bxk2})-(\ref{eq:bzk2}) are central for our analysis. The similar spectral equations in Fourier space, together with the description of each dynamical term, were already given in Paper I but for the case of nonzero net azimuthal field. The energy exchange between perturbation modes and the background shear flow is described by the linear terms ${\cal H}_i({\bf k},t)$ for the velocity and by the Maxwell stress ${\cal M}$ for the (azimuthal) magnetic field. These linear exchange processes represent the central energy suppliers for the perturbations. Since the linear terms depend on the shear, $q$, these processes originate from the nonnormality of the Keplerian shear and hence constitute the linear nonmodal dynamics of MRI in the zero net flux case \citep[see also][Paper I]{Squire_Bhattacharjee14,Squire_Bhattacharjee14a,Zhuravlev_Razdoburdin14}. On the other hand, the kinetic-magnetic energy exchange, in contrast to MRI in the nonzero net flux case, is part of the nonlinear dynamics described by ${\cal N}_i^{(u)}({\bf k},t)$ and ${\cal N}_i^{(b)}({\bf k},t)$, in the above spectral equations. These nonlinear terms cannot be a source of new energy for the turbulence, since their net contribution, integrated over Fourier space, is zero
\begin{equation}\label{eq:sum}
\sum_{i=x,y,z}\int[{\cal N}_i^{(u)}({\bf k},t)+{\cal N}_i^{(b)}({\bf k},t)]d^3{\bf k}=0,
\end{equation}
meaning that the kinetic and magnetic nonlinear terms only
redistribute power among wavenumbers in ${\bf k}$-space
as well as among velocity and magnetic field components, while leaving the total (magnetic+kinetic) spectral energy integrated over all wavenumbers unchanged. Still, these nonlinear transfers, especially magnetic ones, are of great importance in the self-sustenance process.

The other linear process is the shear-induced drift of modes parallel to the $k_x$-axis, described by the first terms $-qk_y\partial /\partial k_x$
which are proportional to the product of the shear parameter $q$ and azimuthal wavenumber on the rhs of Equations (\ref{eq:uxk2})-(\ref{eq:bzk2}). These terms merely advect the amplitudes of velocity and magnetic field components of non-axisymmetric ($k_y\neq 0$) modes along (opposite) the $k_x$-axis for $k_y>0$ ($k_y<0$) at a speed $q|k_y|$, without producing new energy for them, because the integral of the drift terms over $k_x$-axis is zero, $-\int qk_y\partial (...)/\partial k_x dk_x=0$. As a result, Eulerian radial wavenumber of each non-axisymmetric mode (shearing wave) varies linearly with time, $k_x(t)=k_x(0)+qk_yt$, in the shearing box. So, the effect of this drift is straightforward -- it only makes the growth of individual modes transient as it sweeps them in ${\bf k}$-space. For this reason, we do not show this term in the spectral plots below.

For the finite simulation box, the grid in Fourier space is determined by
the box sizes $L_i$ and numerical resolution $N_i$, $i=x,y,z$, so
that the cell sizes are given by $\Delta k_i=2\pi/L_i$, and hence the
wavenumbers run through values $k_i=2\pi n_i/L_i$ where $n_i=0,\pm
1, \pm 2..., \pm N_i/2$.  So, everywhere below we use these wavenumbers normalized by the corresponding grid cell sizes after which they become all integers, $k_i/\Delta k_i \rightarrow k_i=n_i$. However, the radial wavenumber of non-axisymmetric modes, varying with time as $k_x(t)=2\pi n_x/L_x+qt(2\pi n_y/L_y)$, becomes integer at discrete time moments $t_n=nL_y/(q|n_y|L_x)$, where $n$ is an integer multiple of $n_y$ \citep{Hawley_etal95}.

\begin{figure}
\includegraphics[width=\columnwidth]{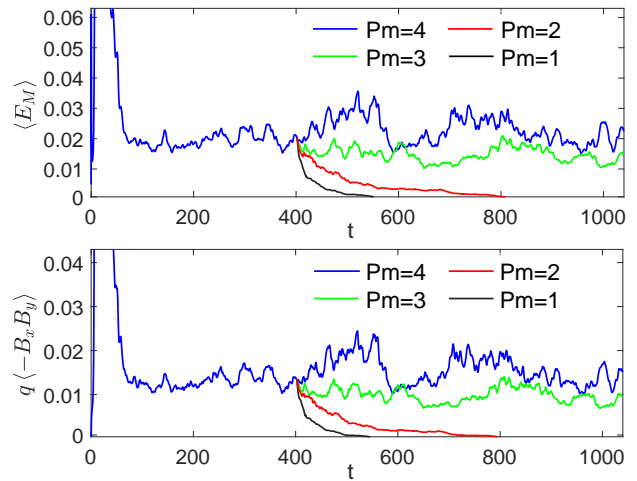}
\caption{Evolution of the volume-averaged magnetic energy (top) and
Maxwell stress (bottom) at different ${\rm Pm}=1~(black), 2~(red), 3~
(green)$ and $4~(blue)$. The runs at ${\rm Pm}=1,2$ and $3$ are started
from th flow snapshot of the ${\rm Pm}=4$ run at $t=400$. The runs
for ${\rm Pm}=3$ and $4$ are self-sustained, with the turbulence level decreasing with ${\rm Pm}$, whereas the runs at ${\rm Pm}=1$ and $2$ decay slowly.}\label{fig:evol_diff_Pm}
\end{figure}

\begin{table*}[t]
\caption{Simulation characteristics: box size, number
of grid points, the Reynolds number, the magnetic Reynolds number,
the magnetic Prandtl number, volume- and time- averaged kinetic energy, magnetic energy, Reynolds stress and Maxwell stress.}
\begin{ruledtabular}
    \begin{tabular}{ccccccccccccc}
$(L_x,L_y,L_z)$ & $(N_x, N_y, N_z)$ &  ${\rm Re}$ & ${\rm Rm}$ & ${\rm Pm}$ &  $\left\langle \left\langle E_K\right\rangle \right\rangle$ & $\left\langle \left\langle E_M\right\rangle \right\rangle$ & $\left\langle \left\langle u_xu_y\right\rangle \right\rangle$ & $\left\langle \left\langle -B_xB_y\right\rangle \right\rangle$ & \\\hline
$(3,3,1)$ & $(512,512,128)$ &  $1.2\times10^4$ & $1.2\times10^4$ & 1 & - &  - & - & - &\\
$(3,3,1)$ & $(512,512,128)$ &   $0.6\times10^4$  &  $1.2\times10^4$ & 2 & - &  - & - & - &\\
$(3,3,1)$ & $(512,512,128)$ &  $0.4\times10^4$  &  $1.2\times10^4$ & 3  & $4.2\times10^{-3}$ &  $1.31\times10^{-2}$ & $8.62\times10^{-4}$ & $5.8\times10^{-3}$ &&\\
$(3,3,1)$ & $(512,512,128)$ &  $0.3\times10^4$  &  $1.2\times10^4$ & 4 & $7\times10^{-3}$ &  $2.3\times10^{-2}$ & $1.5\times10^{-3}$ & $1.04\times10^{-2}$ &&\\
    \end{tabular}
\end{ruledtabular}
\end{table*}

\section{General properties of the simulations}
\label{sec:DNS}

We solve the main Equations (\ref{eq:mom})-(\ref{eq:div}) using the pseudo-spectral code SNOOPY
\citep{Lesur_Longaretti07}. The computational domain has sizes $(L_x,L_y,L_z)=(3,3,1)$ and numerical resolution $(N_x,N_y,N_z)=(512,512,128)$. Note that in contrast to previous studies on zero net flux MRI-turbulence \citep[e.g.,][]{Fromang_Papaloizou07, Lesur_Ogilvie08,Shi_etal16, Herault_etal11, Riols_etal17}, we adopt a box with equal radial and azimuthal sizes, $L_x=L_y$, since it is itself isotropic in $(x,y)$-slice and does not cause ``numerical deformation'' of the inherent anisotropic dynamics of the MRI-turbulence (a more detailed analysis of the effect of the aspect ratio $L_x/L_y$ on the numerical
deformation due to box asymmetry is presented in subsection 5.3 of Paper I). The standard shearing box boundary conditions \citep{Hawley_etal95} are used in the code that keep the net magnetic flux in the box to zero at all times.

We initialize our simulations by imposing random velocity perturbations on the Keplerian flow and trace their subsequent evolution. First we ran the case with ${\rm Rm}=1.2\times10^4$, ${\rm Re}=3\times 10^3$, having ${\rm Pm}=4$. After it has settled down into a quasi-steady turbulent state, from the flow snapshot of this run at $t=400$, we started three additional runs with lower $Pm=1,2,3$ and the same ${\rm Rm}$. These new runs have thus higher Reynolds numbers ${\rm Re}=[1.2, 0.6, 0.4]\times10^4$, respectively. Table I summarizes the volume- and time-averaged values of the kinetic, $E_K=\rho_0{\bf u}^2/2$, and magnetic, $E_M={\bf B}^2/8\pi$, energy densities as well as Reynolds, $u_xu_y$, and Maxwell, $-B_xB_y$, stresses in all the four runs. 

Figure  \ref{fig:evol_diff_Pm} shows the evolution of the volume-averaged magnetic energy and Maxwell stress for various ${\rm Pm}=1,2,3$ and $4$, which recovers classical results of other simulations of zero net flux MRI-turbulence in the literature \citep[e.g.,][]{Fromang_etal07,Nauman_Pessah16, Walker_etal16,Shi_etal16,Walker_Boldyrev17, Nauman_Pessah18}. Turbulence is sustained at ${\rm Pm}=3$ and $4$, but its level decreases with ${\rm Pm}$. It no longer persists at lower ${\rm Pm}=1$ and $2$, gradually decaying during a time which is the shorter, the smaller is ${\rm Pm}$. Thus, for our box, the critical value separating sustained and decaying cases is between ${\rm Pm}=2$ and $3$ consistent with \citet{Riols_etal17}. But \citet{Nauman_Pessah16}  do find sustained turbulence for ${\rm Pm}<1$ provided the vertical size of a box, $L_z$, is sufficiently large.

The structure of the MRI-turbulence is different in the sustained and decaying cases. As ${\rm Pm}$ decreases, the characteristic length-scale (correlation length) of the magnetic field structures decreases as well \citep{Riols_etal17,Walker_Boldyrev17}.
Performing the spectral analysis of the dynamical processes at different ${\rm Pm}$ below (Section \ref{sec:Pmdependence}), we show that this behavior in physical space is a consequence of the fact that the energy of large-wavenumber modes relative to that of small-wavenumber ones increases with decreasing ${\rm Pm}$. 

We also carried out simulations including thermal stratification along the vertical direction in the Boussinesq approximation. Its role, however, turned out to be negligible in the sustaining dynamics of MRI-turbulence. Thus, here we do not take into account the stratification, although it plays an important role in disk dynamo \citep[e.g.,][]{Davis_etal10, Bodo_etal12, Gressel13}.

\begin{figure*}
\centering
\includegraphics[width=\columnwidth]{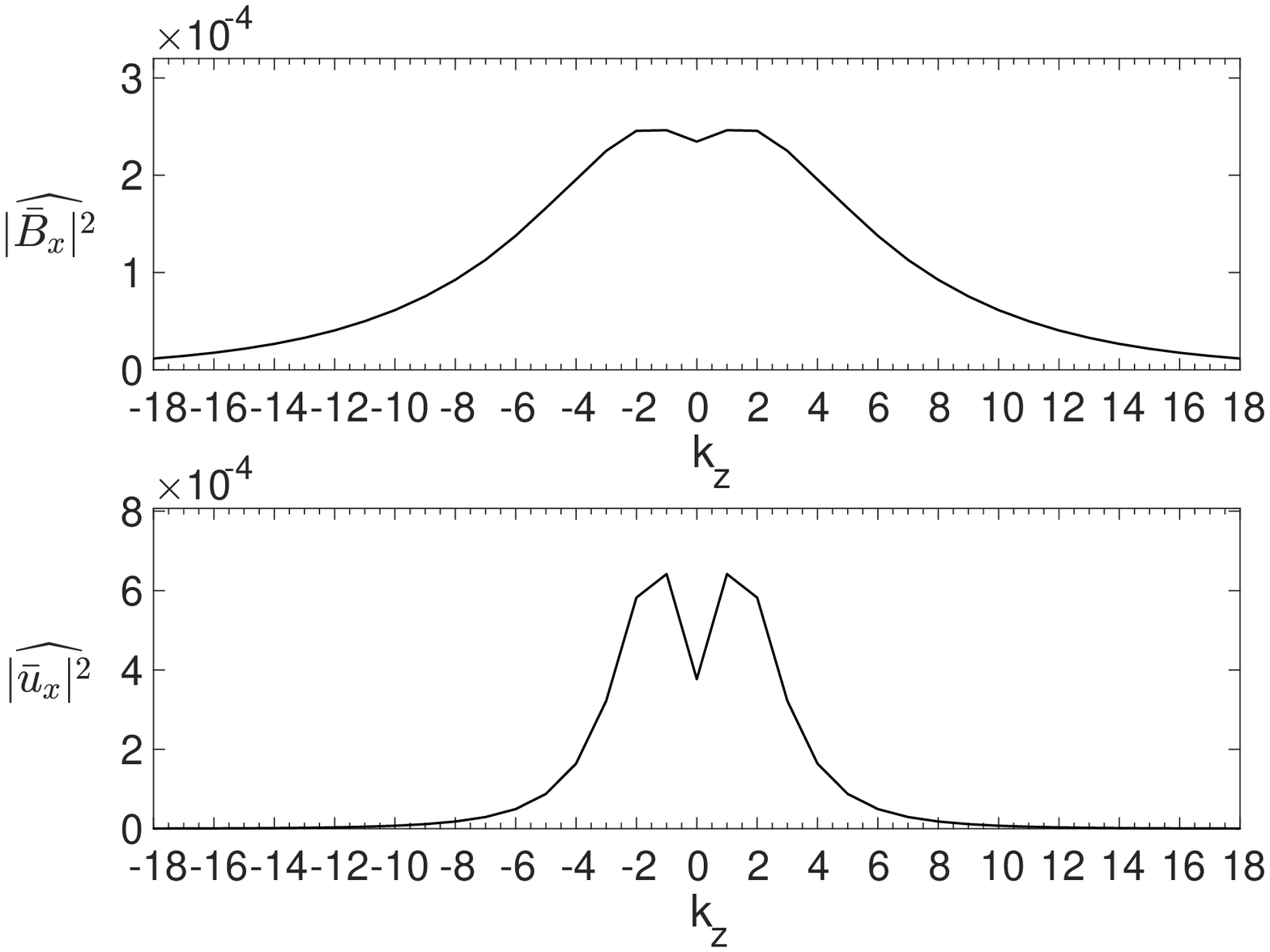}
\includegraphics[width=\columnwidth]{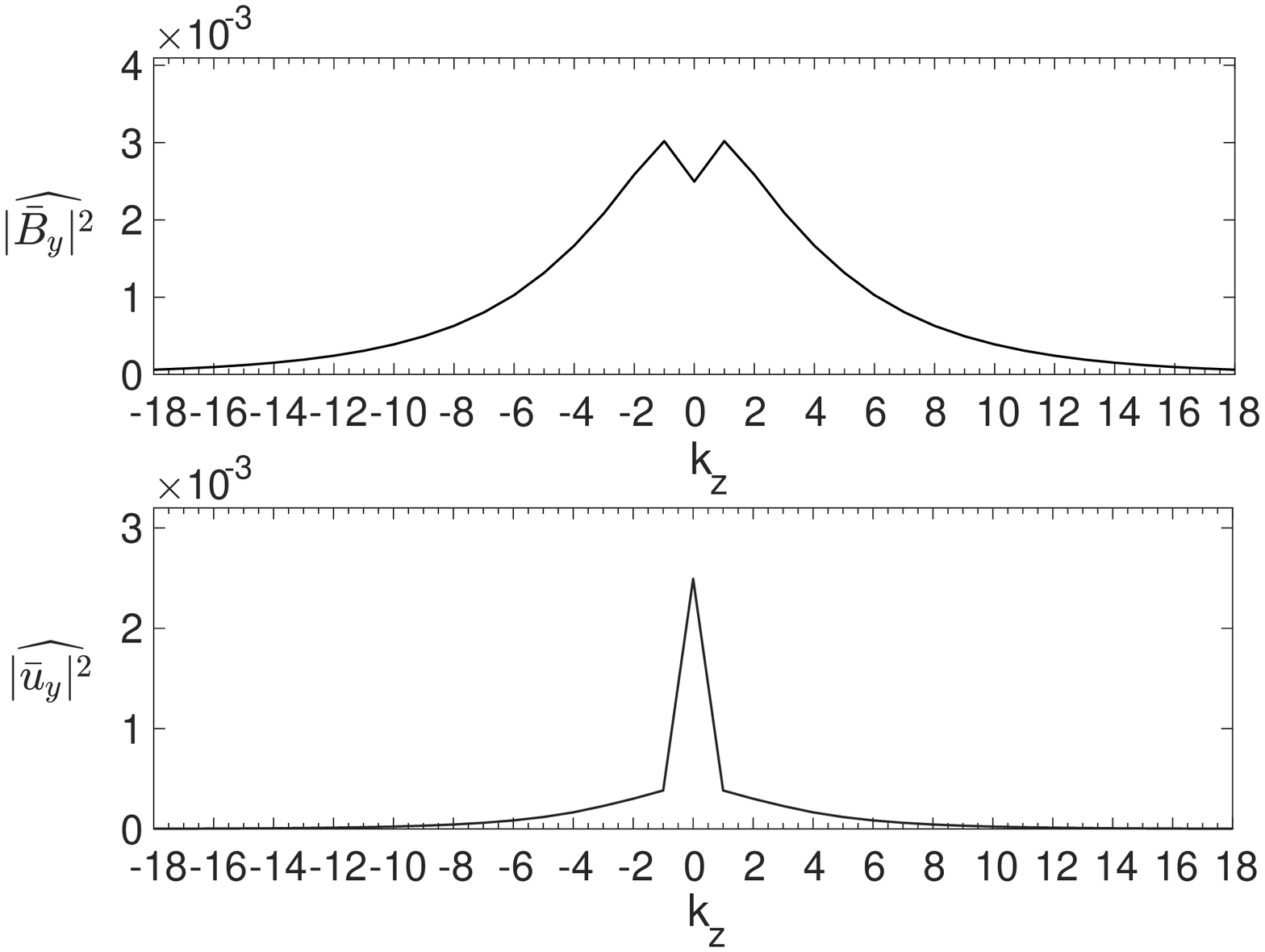}
\caption{Time-averaged quadratic forms of the spectral radial and azimuthal velocity and magnetic field components all integrated in $(k_x,k_y)$-slices and represented as a function of $k_z$ for ${\rm Pm}=4$.}\label{fig:integrated in plane-ub}
\end{figure*}

\begin{figure*}
\centering
\includegraphics[width=\columnwidth]{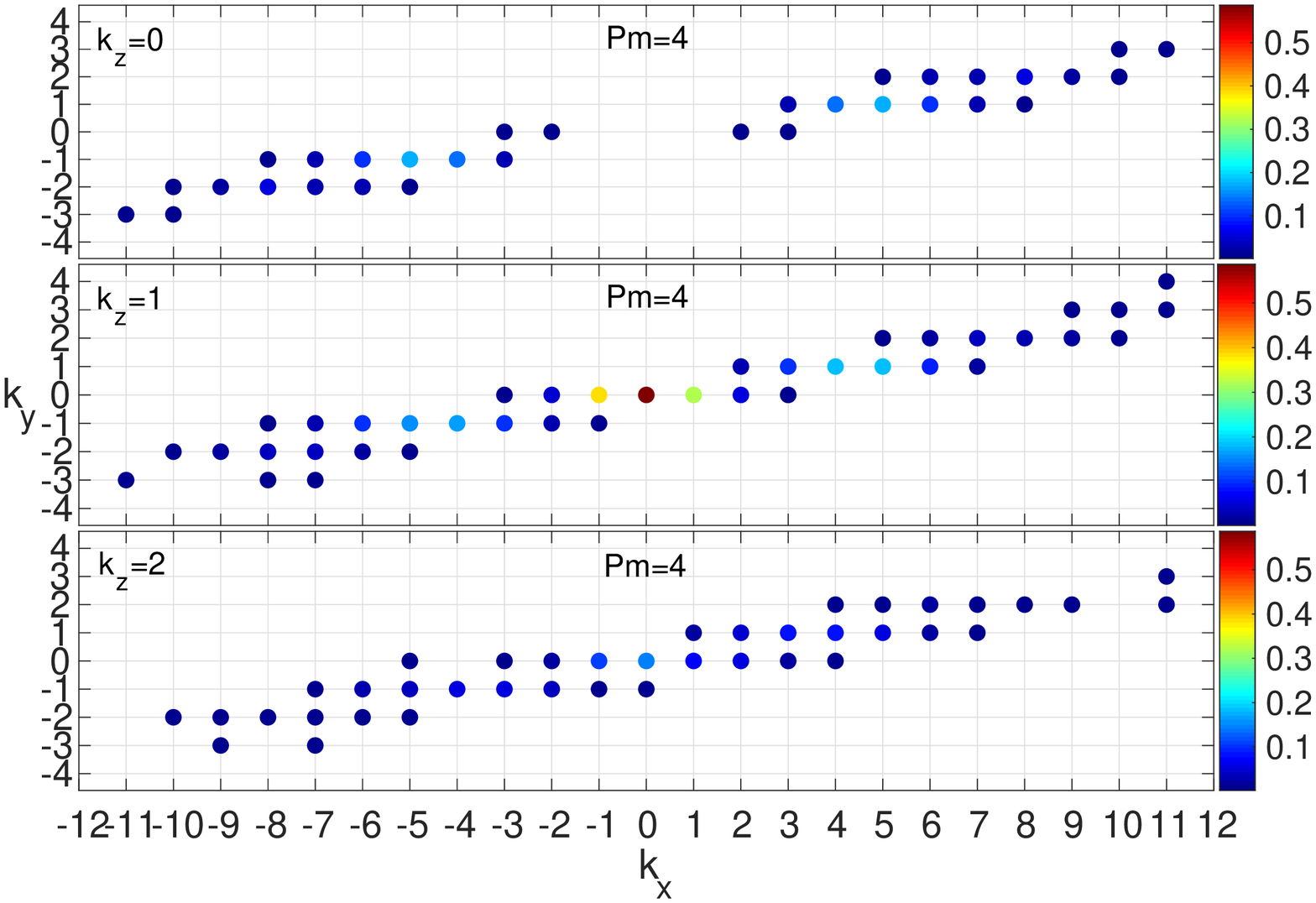}
\includegraphics[width=\columnwidth]{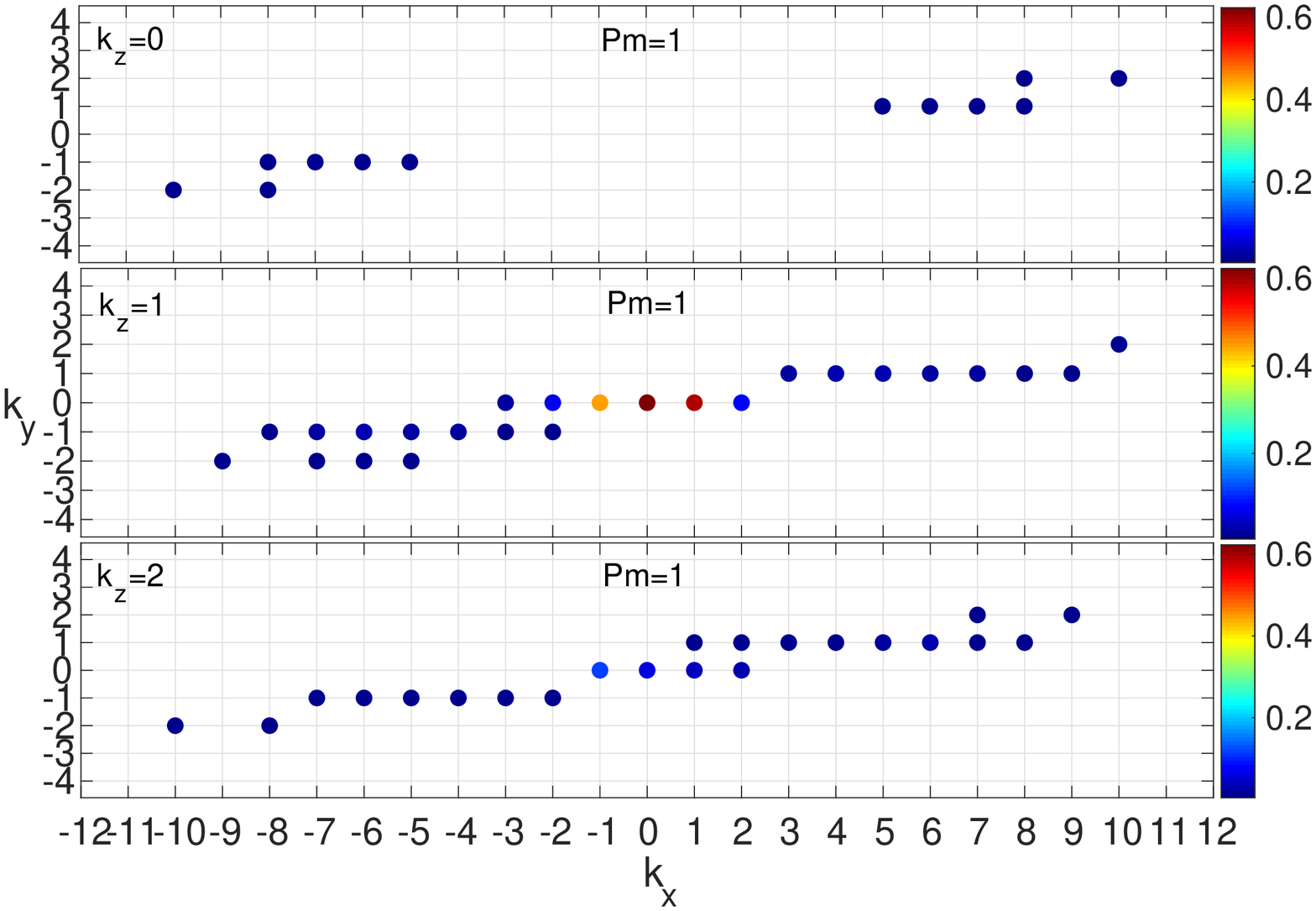}
\caption{Distribution of the active modes (color dots) in Fourier space at first three dominant $k_z=0,1,2$ for the sustained case ${\rm Pm}=4$ (left column) and for the decaying case ${\rm Pm}=1$ (right column). These active modes represent the modes whose magnetic energy ${\cal E}_M$ at least once attains values greater than 50\% of the maximum spectral magnetic energy at the same time. The colorbar shows the ratio of a time interval during which the mode carry this high a magnetic energy to the total run time.}\label{fig:modes}
\end{figure*}

\section{Turbulence dynamics in Fourier space}
\label{sec:Spectraldynamics}

In this section, adopting the approach of our previous studies
\citep[][Papers I and II]{Mamatsashvili_etal14}, we investigate the
spectral dynamics and self-sustenance of the zero flux MRI-turbulence by computing and visualizing the individual linear and nonlinear terms in spectral Equations (\ref{eq:uxk2})–(\ref{eq:bzk2}) from the simulation data. Below we focus on the dynamics of the radial and azimuthal components of the spectral velocity, $\bar{u}_x,\bar{u}_y$, and magnetic field, $\bar{B}_x,\bar{B}_y$, because these are the most important ones for the sustenance process, making up the Reynolds and Maxwell stresses that extract energy from the flow. Here we consider the ${\rm Pm}=4$ case, while the dependence on ${\rm Pm}$ is explored in the next section. Since we deal with the quasi-steady turbulence in this section, we time-average the spectra of the velocity, magnetic field and the dynamical terms from $t=80$ till $t=1000$.

\subsection{$k_z$-spectra of the magnetic field and velocity}

We begin the analysis in Fourier space by first looking at the distribution of the spectral quantities as a function of the vertical wavenumber. Figure \ref{fig:integrated in plane-ub} shows the integrated in $(k_x,k_y)$-slice and time-averaged spectra of the radial and azimuthal velocity and magnetic field components, $\widehat{|\bar{u}_{x,y}|^2}=\int |\bar{u}_{x,y}|^2dk_xdk_y$, $\widehat{|\bar{B}_{x,y}|^2}=\int |\bar{B}_{x,y}|^2dk_xdk_y$, versus $k_z$ for ${\rm Pm}=4$.
Both the magnetic field and velocities' spectra reach their highest
values at small $|k_z|$ and decrease as $|k_z|$
increases. The maximum of $\widehat{|\bar{B}_x|^2}$ and
$\widehat{|\bar{B}_y|^2}$ come, respectively, at $|k_z|=2$  and
$|k_z|=1$, while that of the velocity spectra,
$\widehat{|\bar{u}_x|^2}$ and $\widehat{|\bar{u}_y|^2}$,
respectively, at $|k_z|=1$ and $k_z=0$. Out of these four spectra,
the spectrum of the azimuthal velocity $\widehat{|\bar{u}_y|^2}$ noticeably differs from the remaining three in that the peak at $k_z=0$ is
much more pronounced. We show below that at $k_z=0$ most of the contribution in the azimuthal velocity spectrum comes from the largest scale axisymmetric $k_y=0$ mode with the smallest nonzero radial wavenumber in the box, $k_x=\pm 1$ ($\pm 2\pi/L_x$ in dimensional units). So, this peak, belonging to the mode ${\bf k}_{zf}=(\pm 1,0,0)$, corresponds to the zonal flow accompanying MRI-turbulence \citep[][Paper I and II]{Johansen_etal09,Walker_Boldyrev17}. Its formation is a consequence of the action of the nonlinearity on the perturbations. Thus, most of the power for velocities and magnetic field is contained in small $k_z$ which thus play a main, dynamically important role in the sustenance process of the turbulence.

\subsection{Active modes and the vital area}

Having analyzed the spectra of the velocity and magnetic field along
vertical wavenumbers, we move now to the analysis of the
spectral dynamics in the horizontal $(k_x,k_y)$-slices at a given
$k_z$. We choose the first few small vertical wavenumbers $|k_z|=0,1,2$, which carry most of the kinetic and magnetic energy and hence play a central role in the turbulence dynamics. Below we first identify those active modes which participate in and shape the self-sustaining dynamics of the turbulence.

Figure \ref{fig:modes} shows the energy-carrying, or active modes (color dots) in $(k_x,k_y)$-slices at $k_z=0,1,2$ for ${\rm Pm}=4$ and ${\rm Pm}=1$ placed next to each other for the sake of comparison. Here the active modes are defined as those modes whose spectral magnetic energy, ${\cal E}_M=(|\bar{B}_x|^2+|\bar{B}_y|^2+|\bar{B}_z|^2)/2$, becomes larger than 50\% of the maximum spectral magnetic energy, ${\cal E}_{M, max}$, at least once during the evolution. Although this definition is somewhat arbitrary, it provides information on the location and number of the dynamically important modes in Fourier space. Following Paper I, these modes are identified by tracing the temporal evolution of all the modes in the box during the entire simulation and recording at each instant only those ones whose magnetic energy is higher than the above given fraction of the maximum spectral magnetic energy at that moment. At the end of the simulation, the fraction of time during which a given mode from this set retains this high a magnetic energy relative to the total simulation time is calculated. The resulting ratio is shown on the colorbars.

\begin{figure*}[t!]
\centering
\includegraphics[width=0.32\textwidth, height=0.2\textwidth]{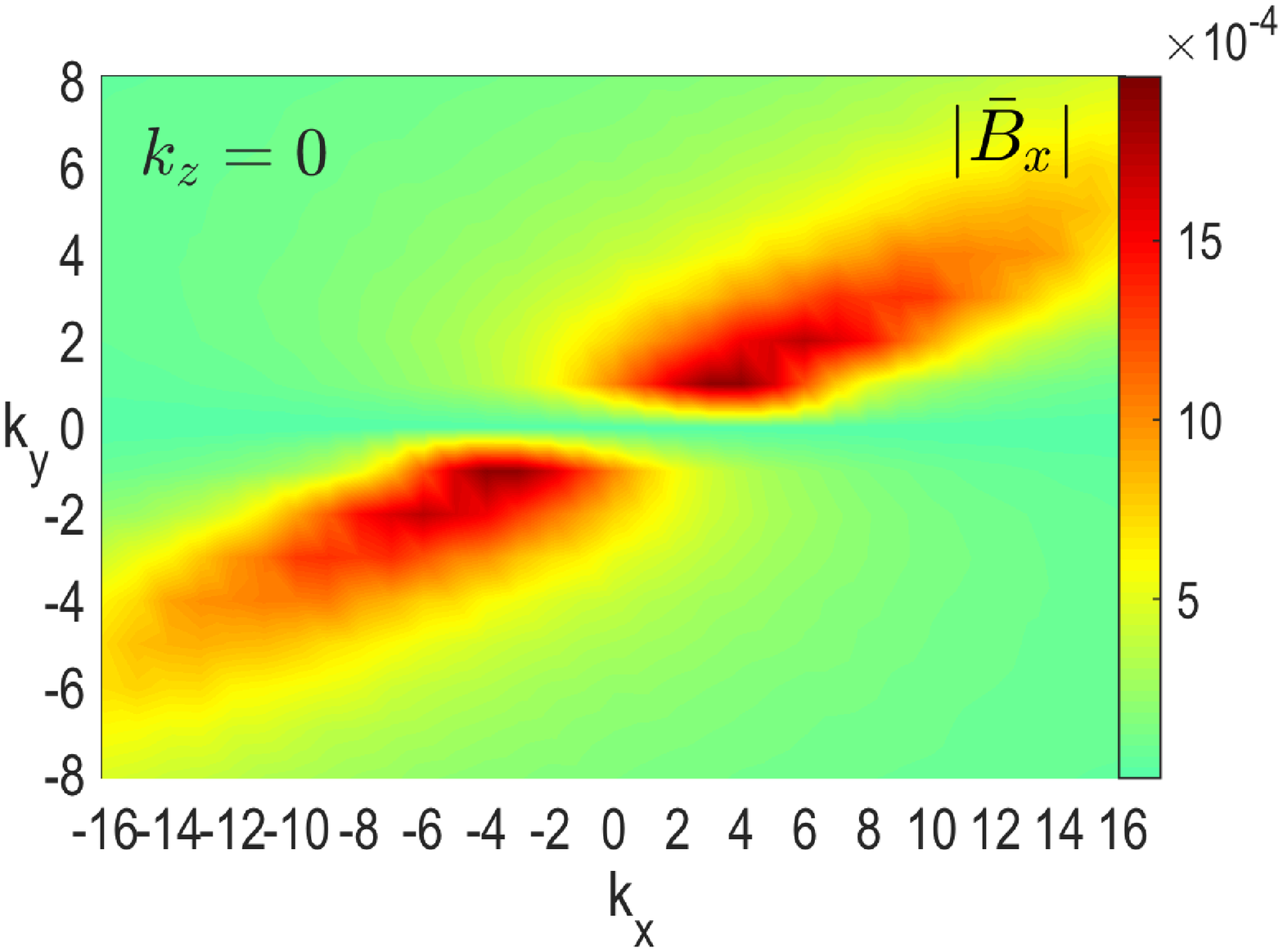}
\includegraphics[width=0.32\textwidth, height=0.2\textwidth]{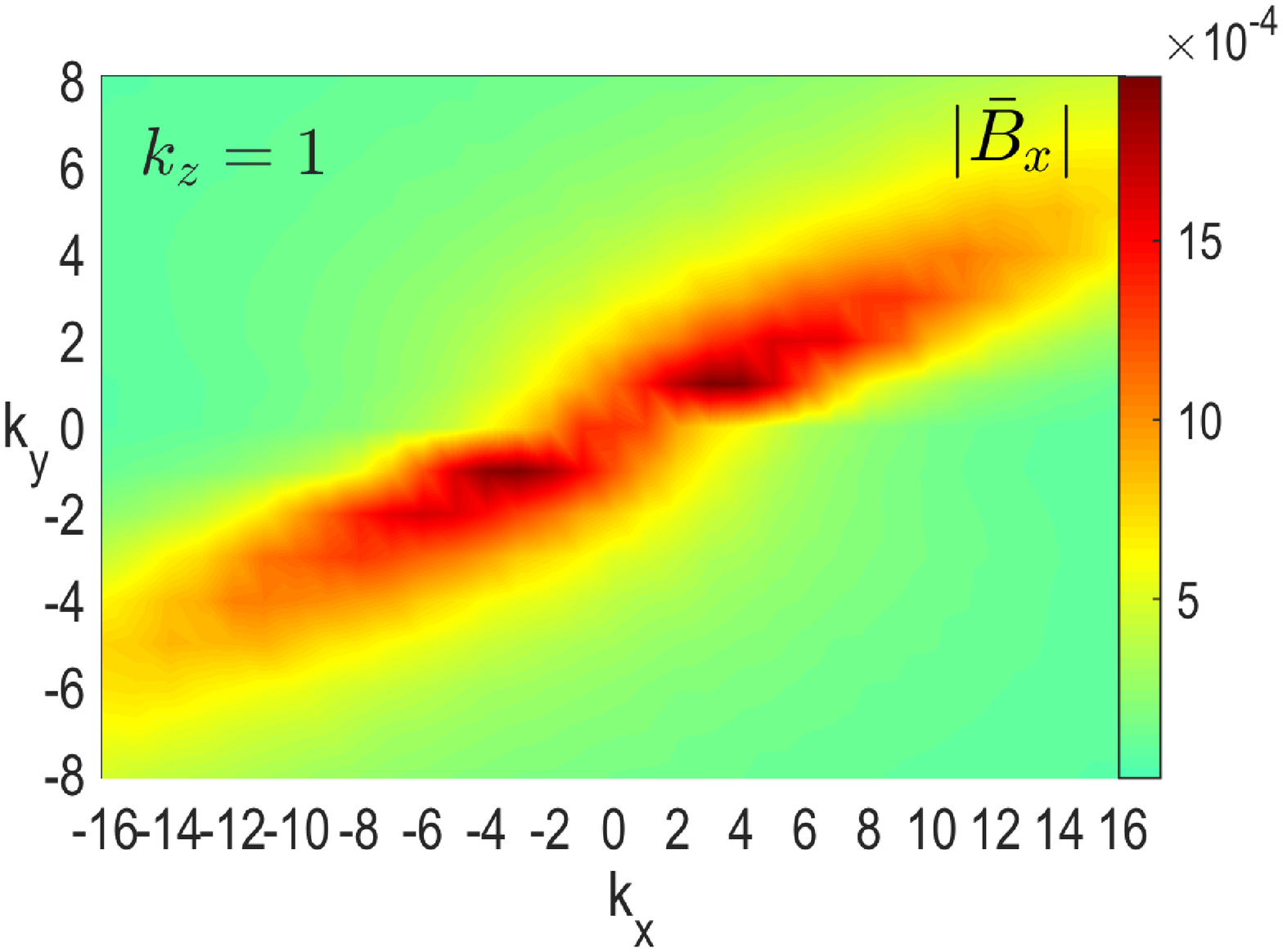}
\includegraphics[width=0.32\textwidth, height=0.2\textwidth]{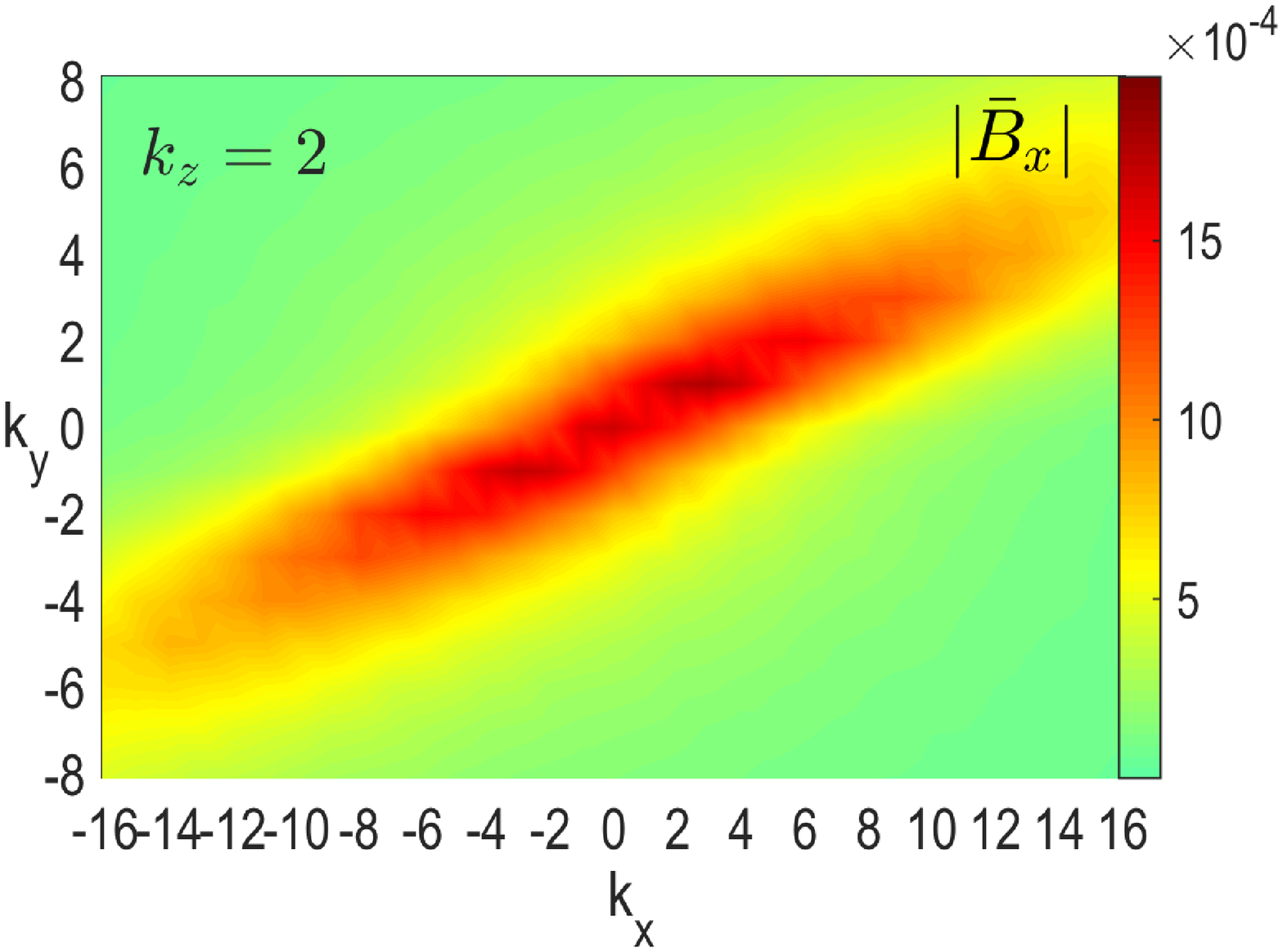}
\includegraphics[width=0.32\textwidth, height=0.2\textwidth]{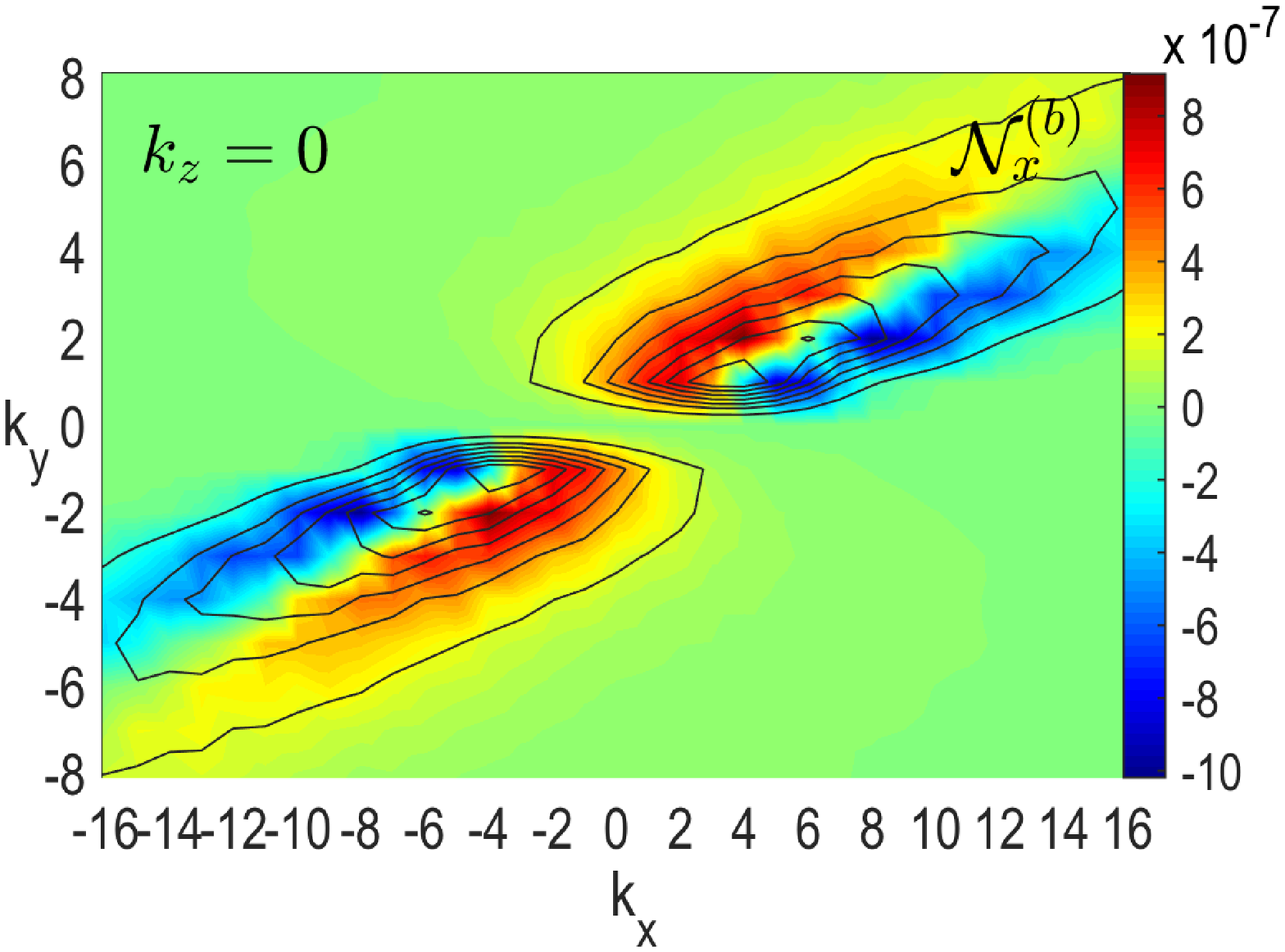}
\includegraphics[width=0.32\textwidth, height=0.2\textwidth]{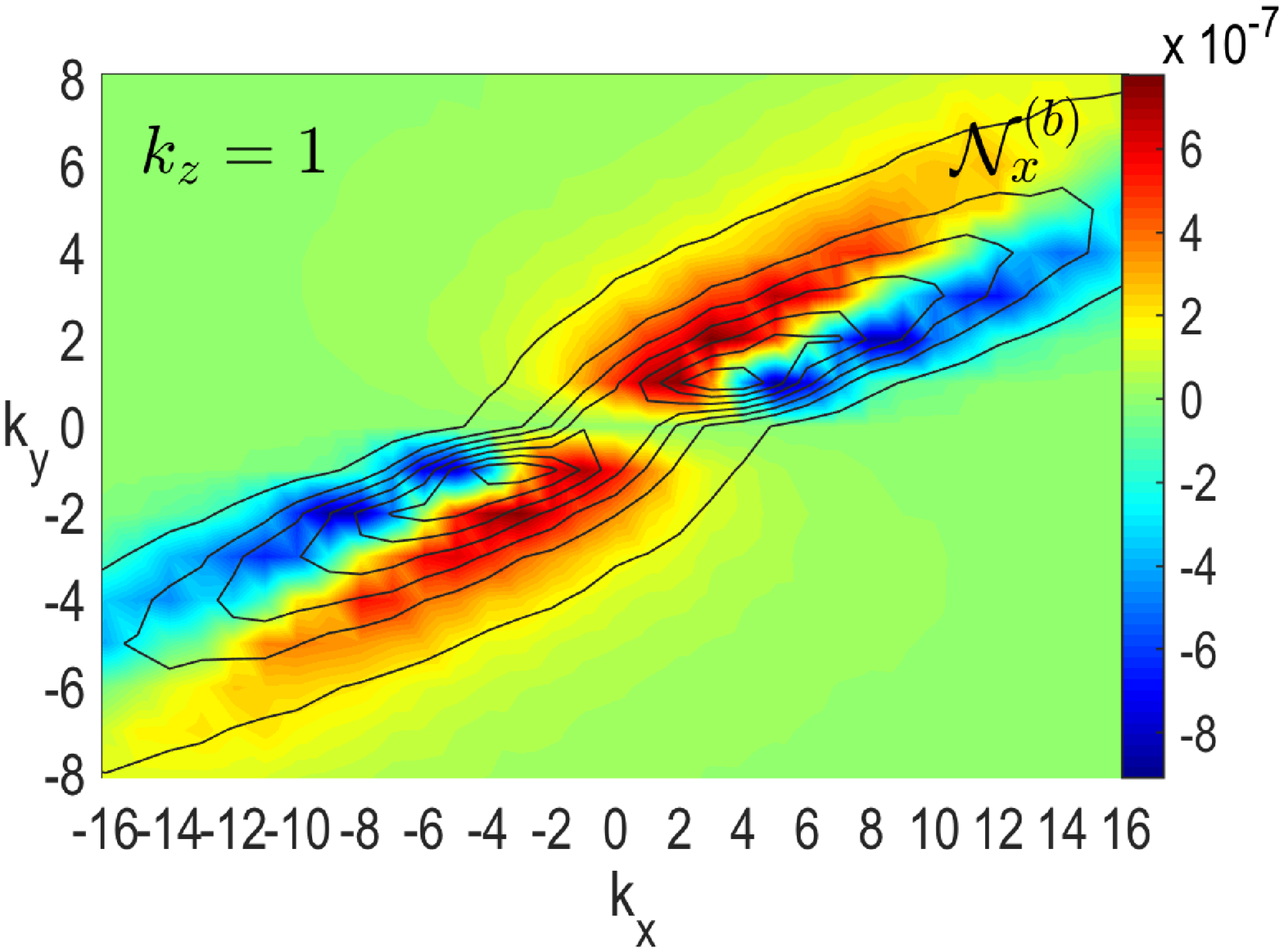}
\includegraphics[width=0.32\textwidth, height=0.2\textwidth]{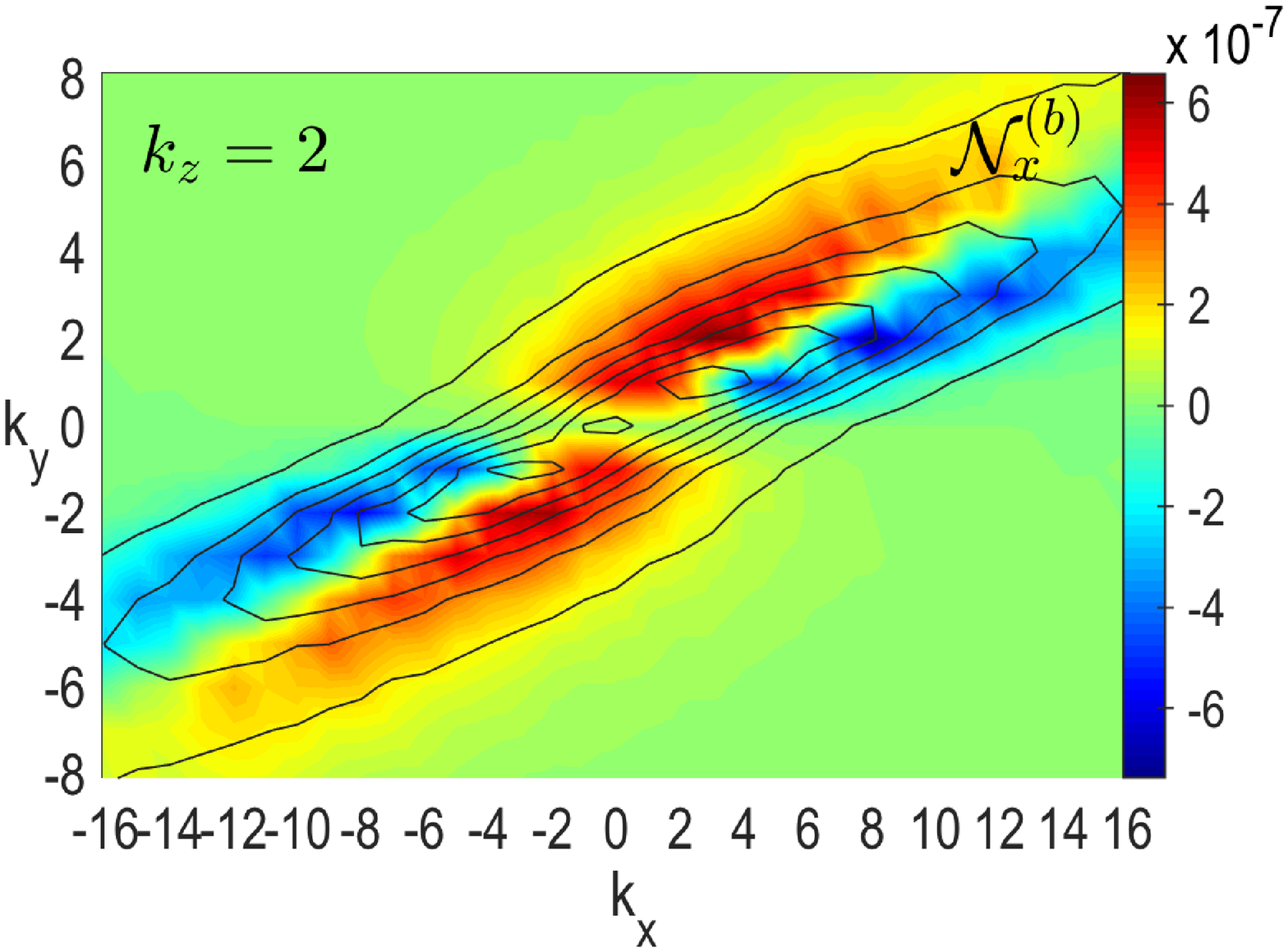}
\caption{Spectra of the radial field, $|\bar{B}_x|$, and the
governing nonlinear term ${\cal N}^{(b)}_x$ in $(k_x,k_y)$-slices at
$k_z=0 (left),~1 (middle),~2 (right)$. Both these spectra exhibit a characteristic similar anisotropy (i.e., variation over the polar angle of wavevector) due to the flow shear, which is also clearly illustrated by the contours of $|\bar{B}_x|$ overplotted on the spectral map of the nonlinear term. ${\cal N}^{(b)}_x$ is larger in the vital area $|k_x|\lesssim 11, |k_y|\lesssim 3$, and manifests the transverse cascade therein. That is, it transversely redistributes/transfers $|\bar{B}_x|^2/2$ from ``giver'' wavenumbers where ${\cal N}_x^{(b)}<0$ (blue) to ``receiver'' wavenumbers where ${\cal N}_x^{(b)}>0$ (red and yellow), draining and supplying the radial field spectral energy in these areas, respectively. This nonlinear transverse (over wavevector orientations/polar angles) redistribution in $(k_x,k_y)$-slices plays a key role, since it continually replenishes (seeds) radial field at the wavenumbers within the red and yellow areas, which we accordingly call the growth area. The linear drift process (the first rhs term in Equation \ref{eq:bxk2}) sweeps $|\bar{B}_x|$ of a given non-axisymmetric mode from the growth area back into the blue area, where ${\cal N}_x^{(b)}<0$ and, therefore, $|\bar{B}_x|$ decreases. This interplay between the nonlinear transfer and linear drift establishes the spectrum of $|\bar{B}_x|$ whose contours overlap with both -- red/yellow and blue -- areas. Exactly due to this transient nature the existence of permanent regeneration due to the transverse cascade is essential for self-sustaining perturbations (turbulence).}\label{fig:spectral_bx}
\end{figure*}
\begin{figure*}[t!]
\centering
\includegraphics[width=0.32\textwidth, height=0.2\textwidth]{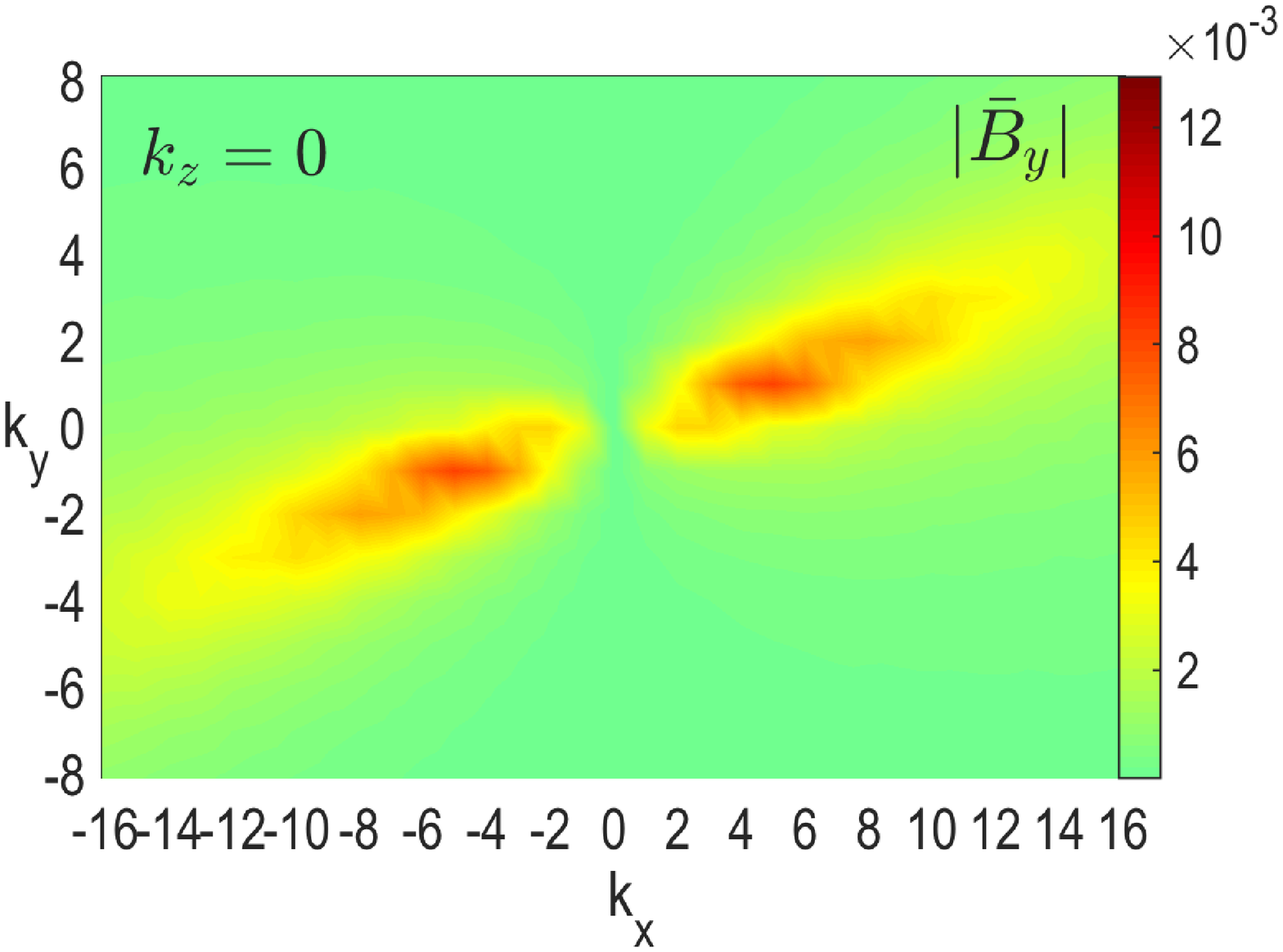}
\includegraphics[width=0.32\textwidth, height=0.2\textwidth]{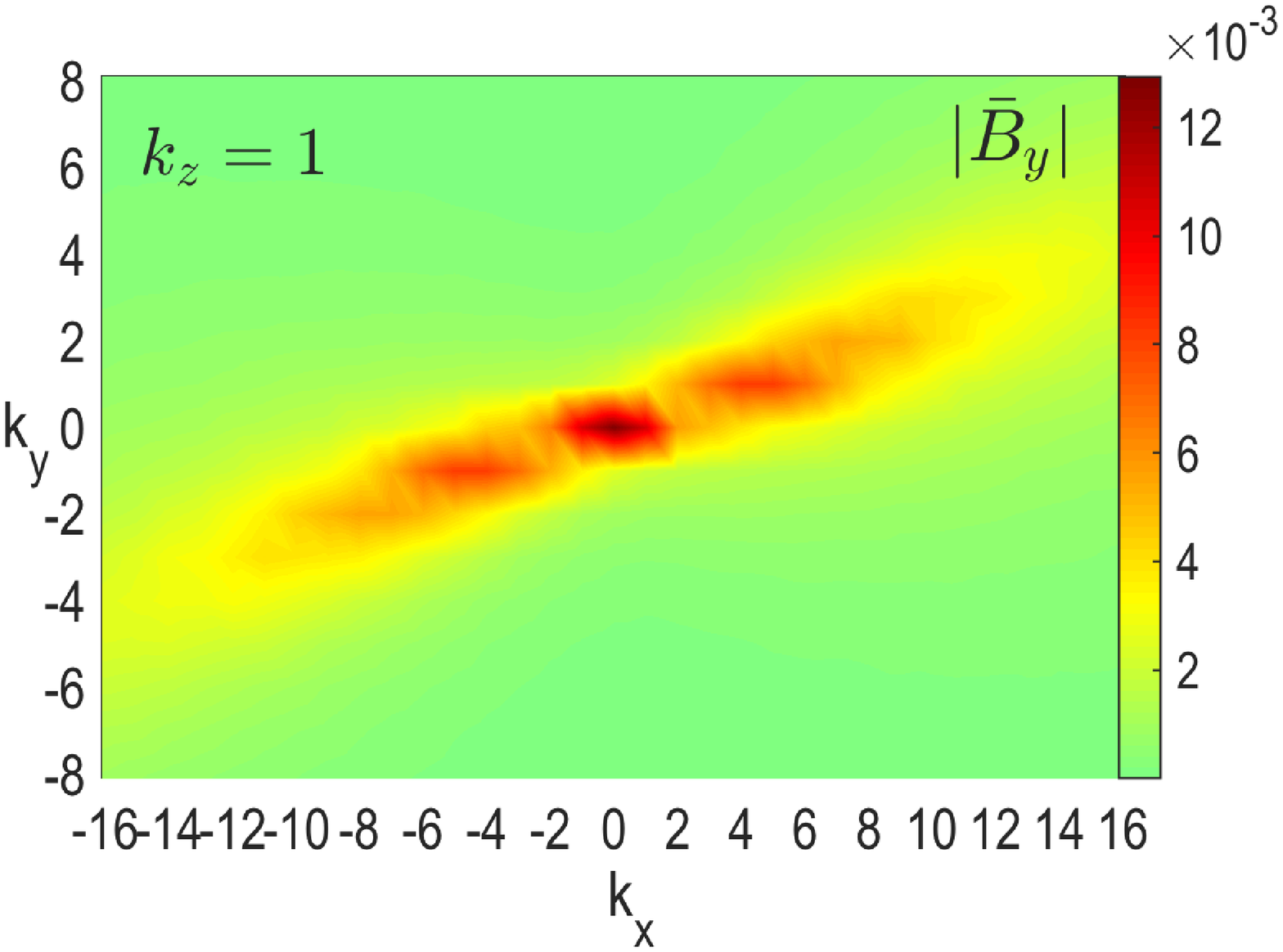}
\includegraphics[width=0.32\textwidth, height=0.2\textwidth]{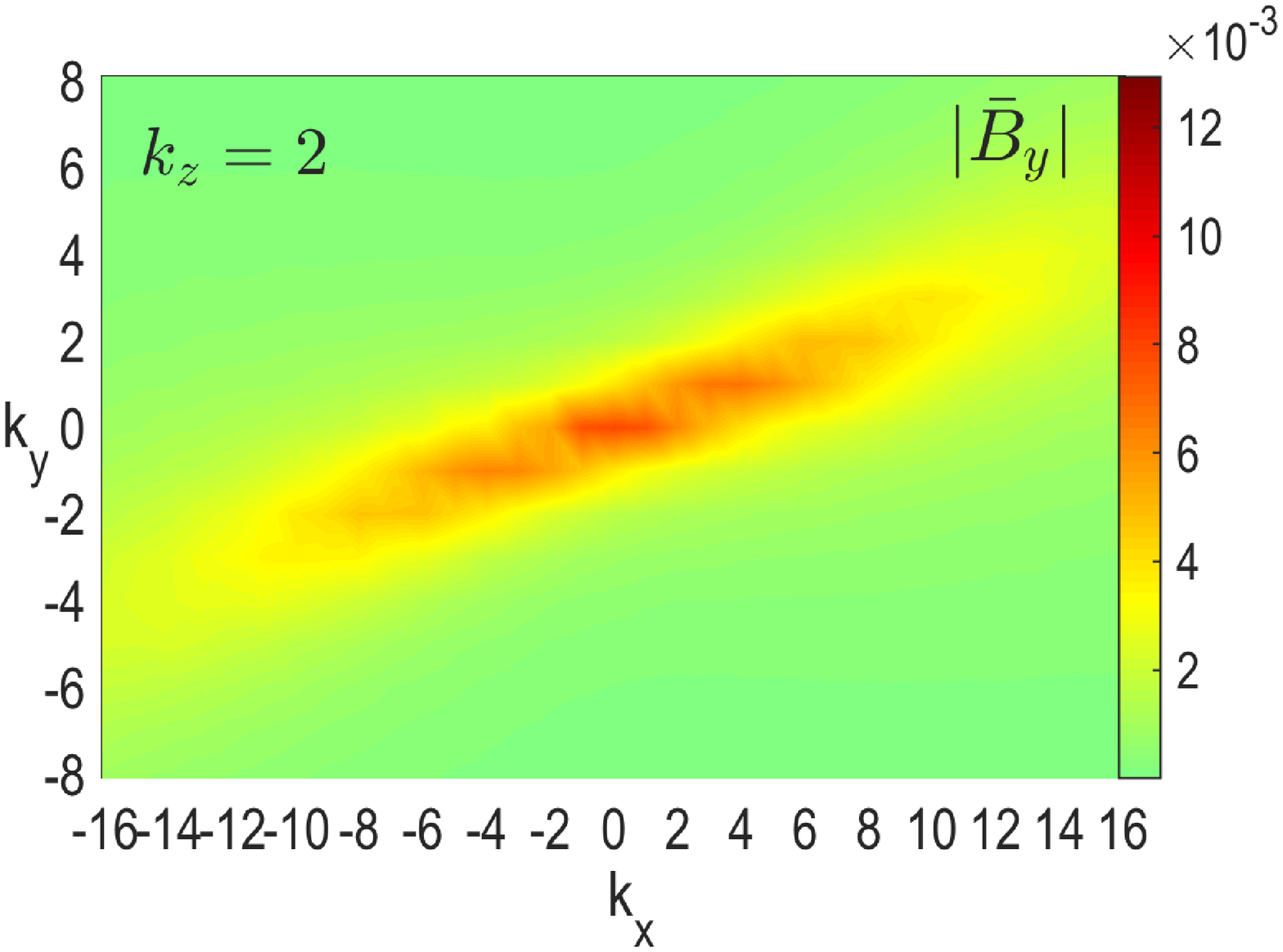}
\includegraphics[width=0.32\textwidth,height=0.2\textwidth]{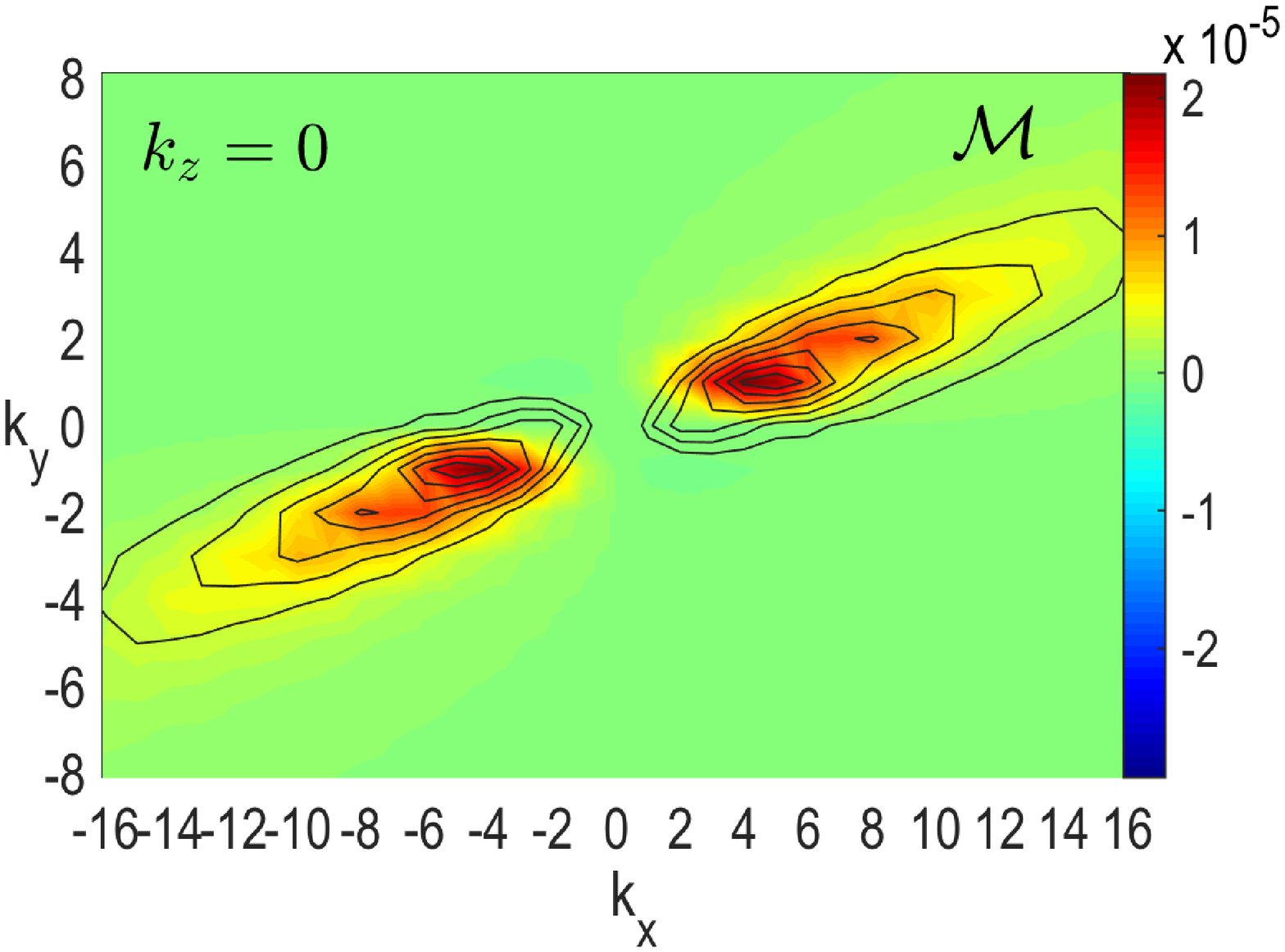}   
\includegraphics[width=0.32\textwidth, height=0.2\textwidth]{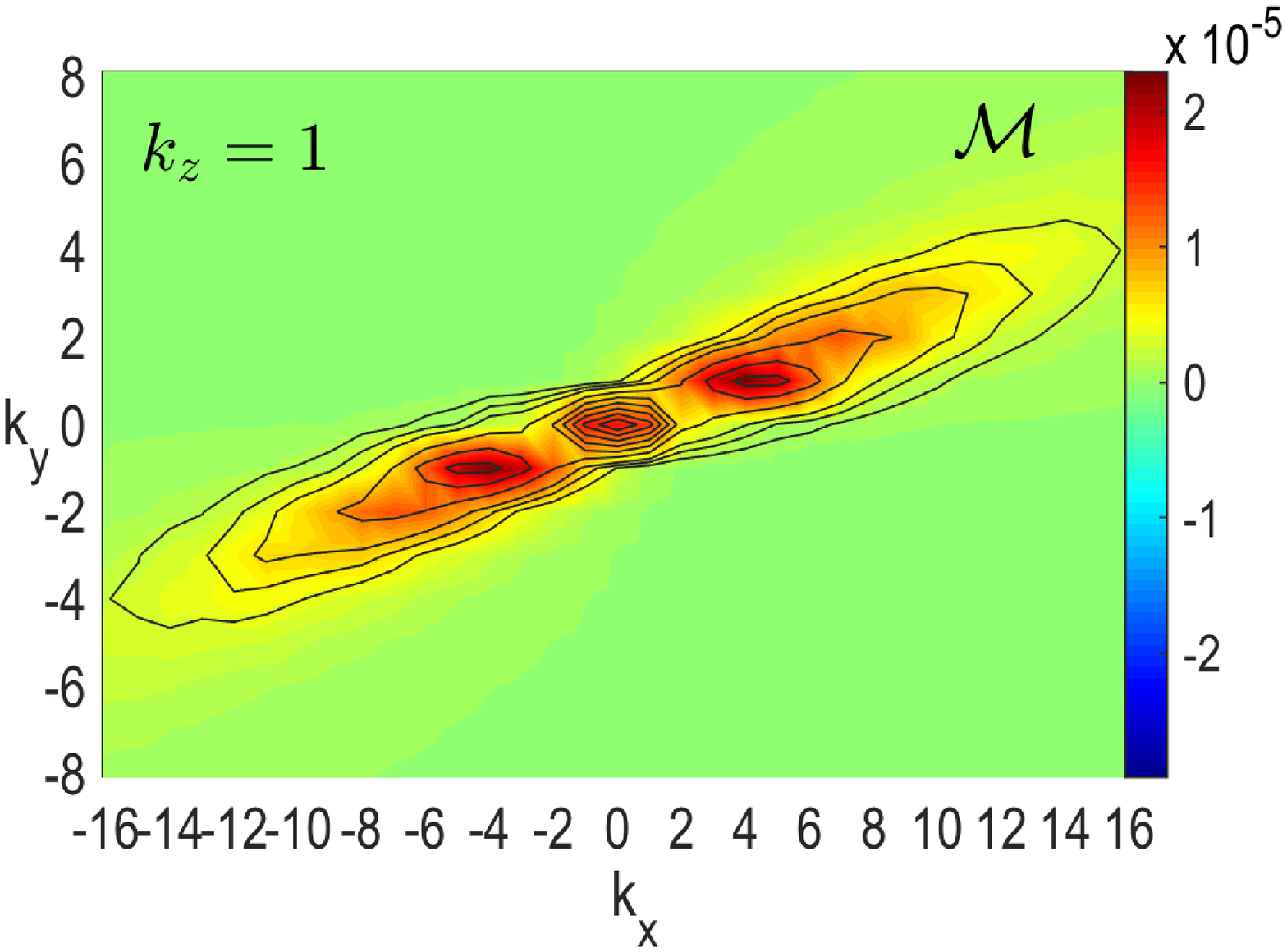}
\includegraphics[width=0.32\textwidth, height=0.2\textwidth]{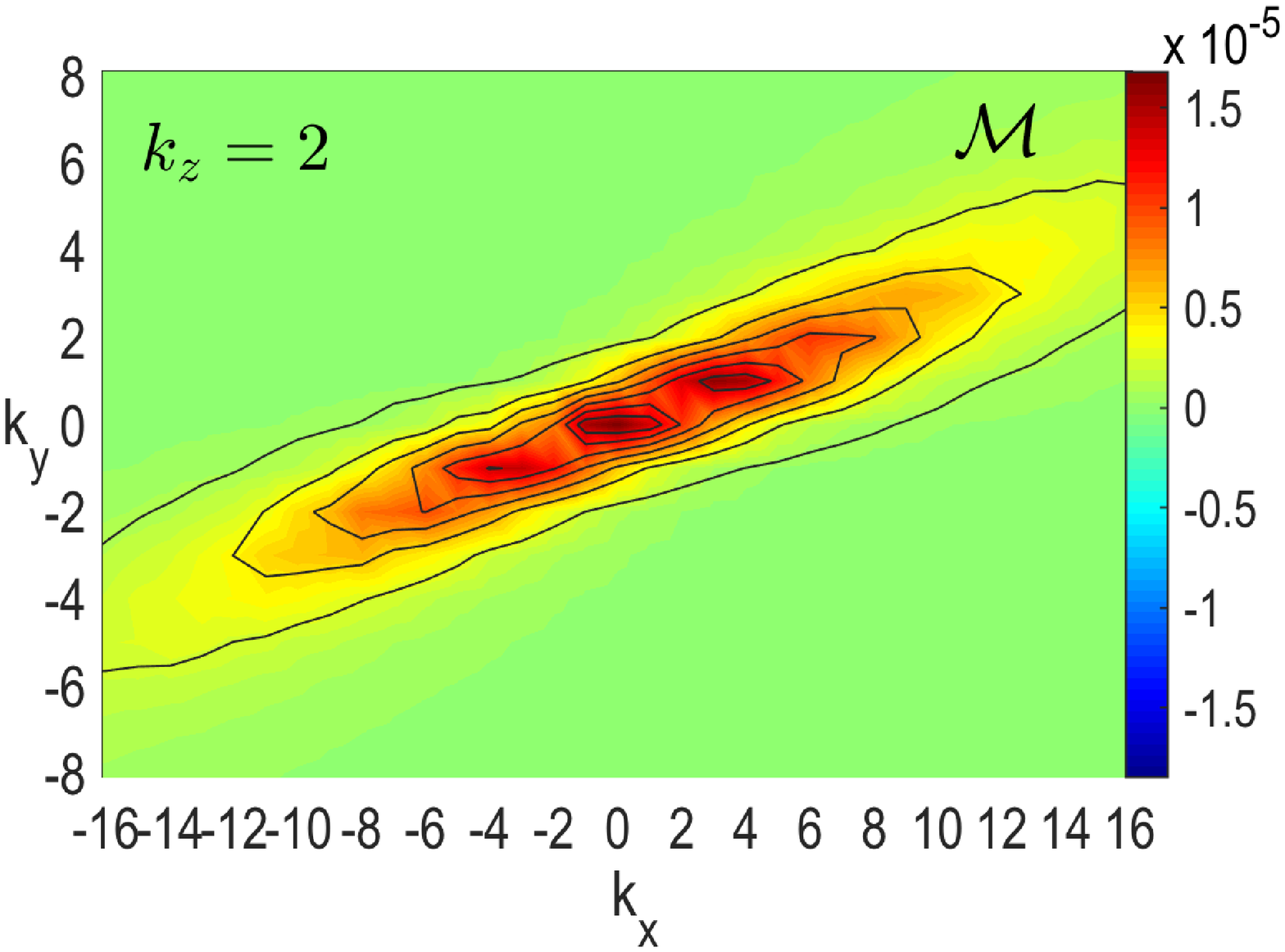}
\includegraphics[width=0.32\textwidth, height=0.2\textwidth]{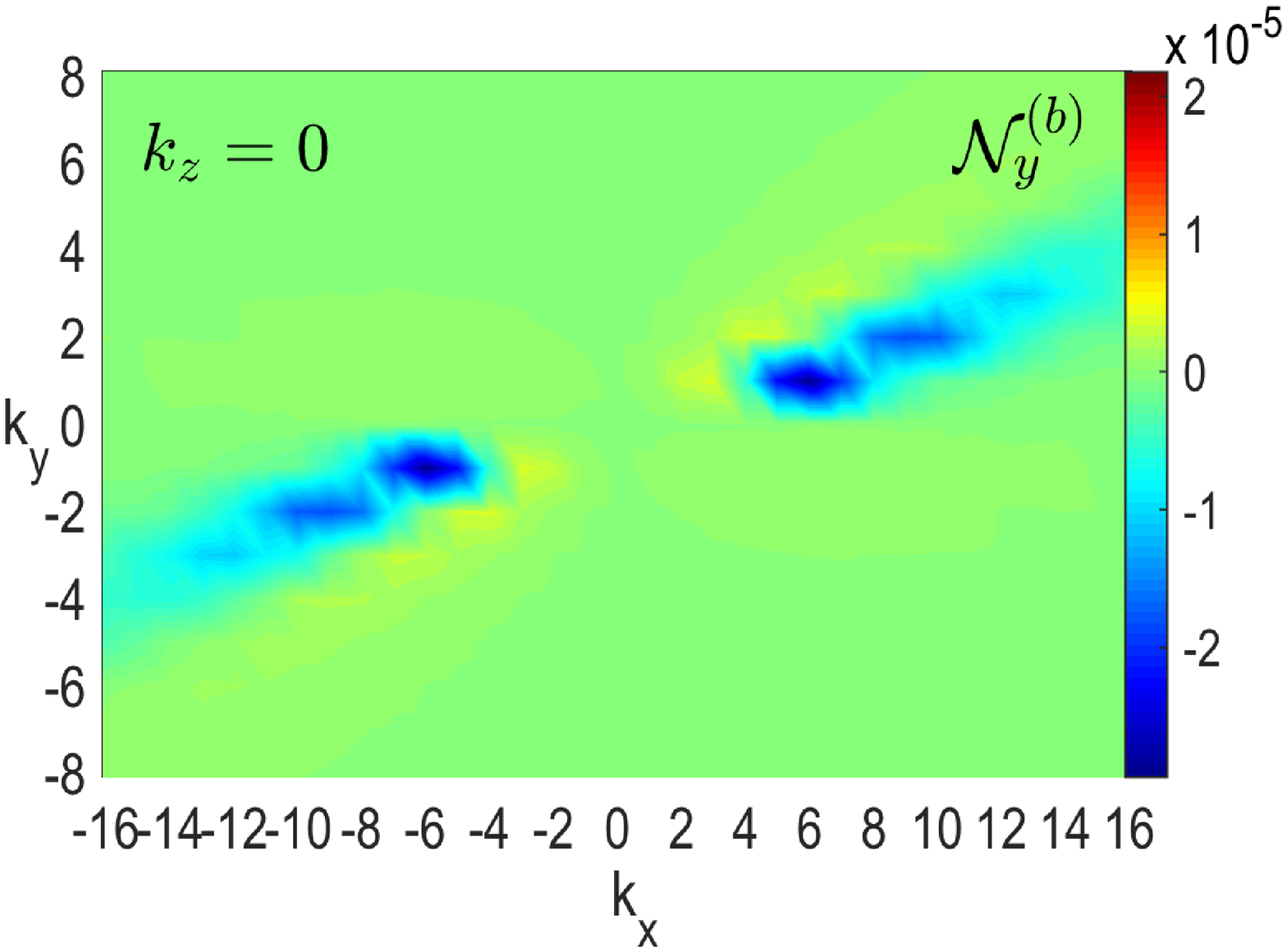}
\includegraphics[width=0.32\textwidth, height=0.2\textwidth]{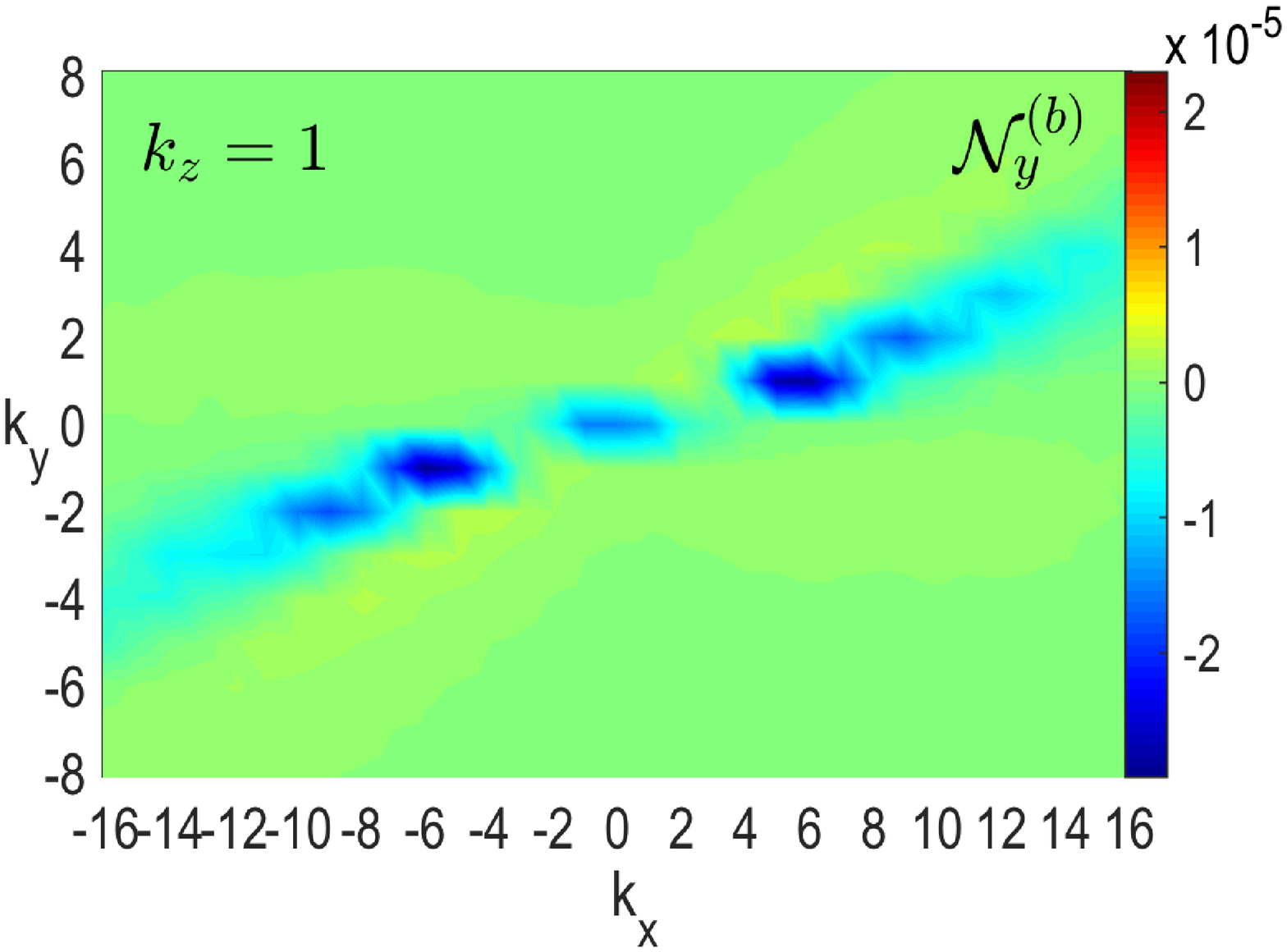}
\includegraphics[width=0.32\textwidth, height=0.2\textwidth]{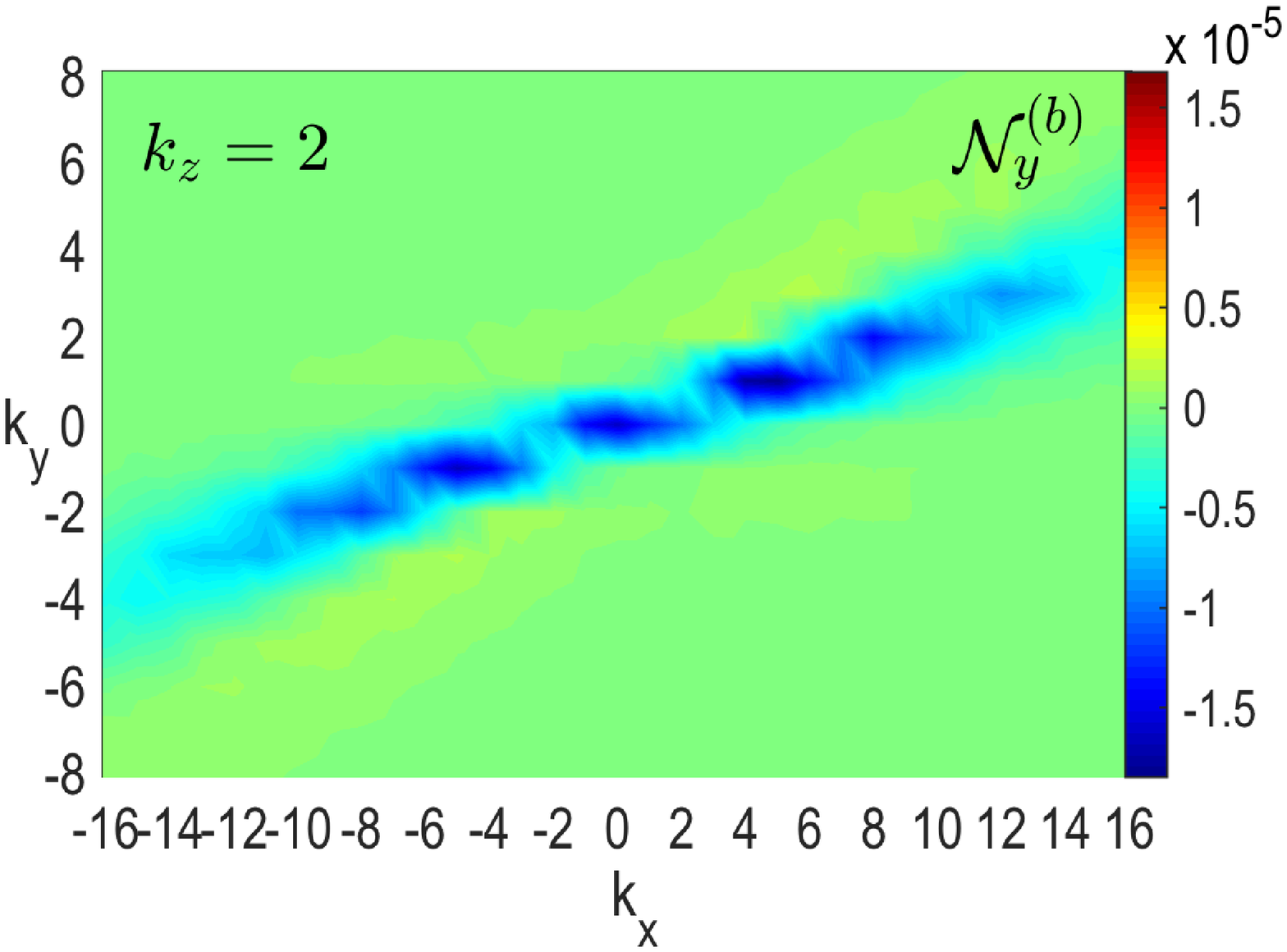}
\caption{Spectra of the azimuthal field $|\bar{B}_y|$ and the governing dynamical terms -- the Maxwell stress ${\cal M}$ and the nonlinear transfer term ${\cal N}_y^{(b)}$. The maximum azimuthal field amplitude comes at the large-scale ${\bf k}_d$ dynamo mode. In the vital area,
the positive Maxwell stress ${\cal M}$ (red and yellow) ensures energy injection and amplification of $|\bar{B}_y|^2/2$ for those modes, which have been regenerated by ${\cal N}_x^{(b)}$ (Figure \ref{fig:spectral_bx}). By contrast, ${\cal N}_y^{(b)}$ is mainly negative (blue) in the vital area, acting as a sink. These modes undergo drift through the vital area where ${\cal M}>0$ and, hence, the amplification of the azimuthal field for them is transient, induced by the nonmodal MRI process, while negative ${\cal N}_y^{(b)}<0$ there transfers the azimuthal field spectral energy from the vital area to larger wavenumbers. The spectra of $|\bar{B}_y|$, ${\cal M}$ and ${\cal N}_y^{(b)}$ exhibit a characteristic similar anisotropy (i.e., variation over the polar angle of wavevector). The contours of $|\bar{B}_y|$ overplotted on the map of ${\cal M}$ show that these two spectral quantities are well aligned with each other, though $|\bar{B}_y|$ is a bit more inclined towards the $k_x$-axis due to the action of the linear drift.}\label{fig:spectral_by}
\end{figure*}
\begin{figure*}[t!]
\centering
\includegraphics[width=0.32\textwidth, height=0.2\textwidth]{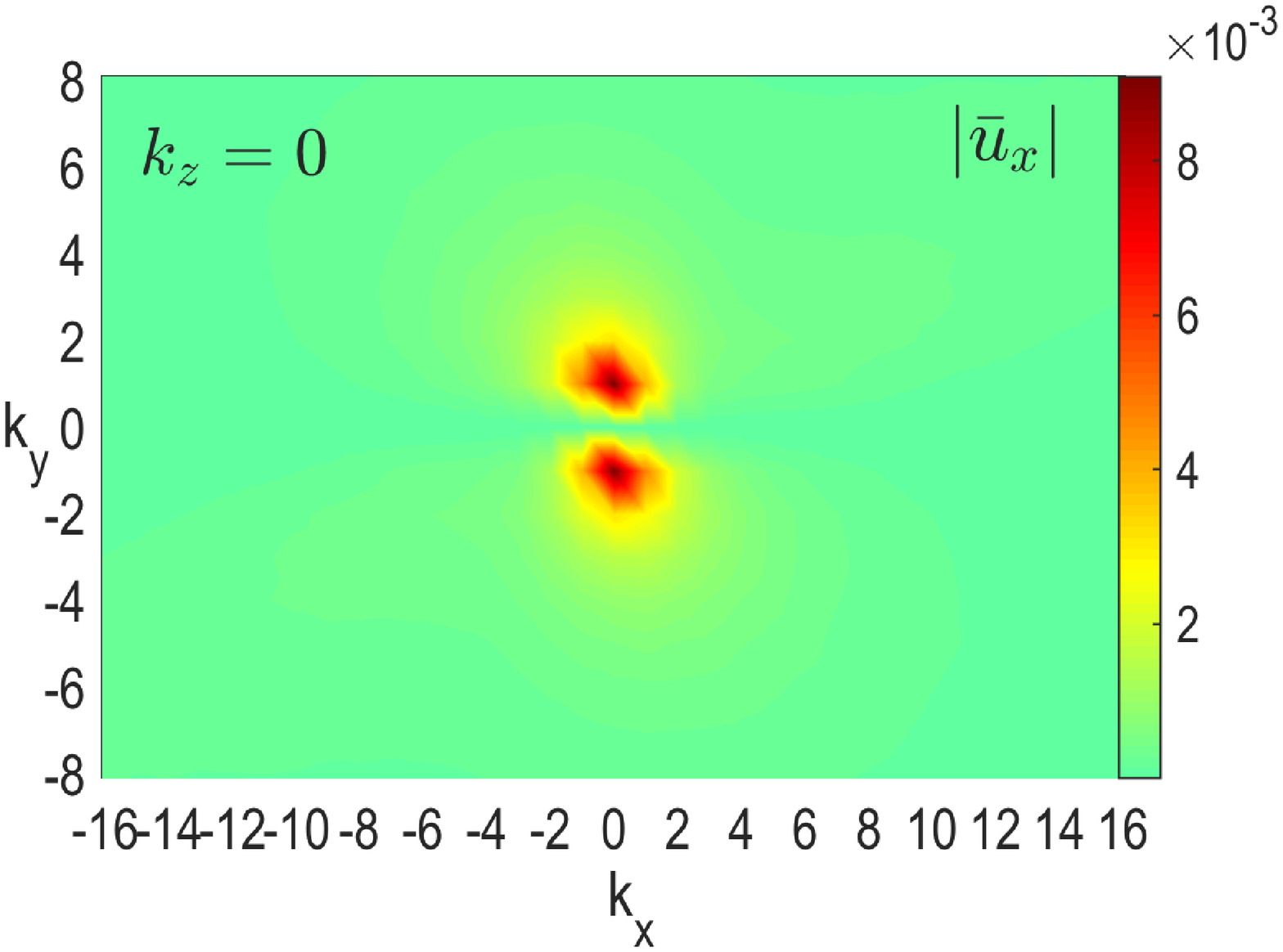}
\includegraphics[width=0.32\textwidth, height=0.2\textwidth]{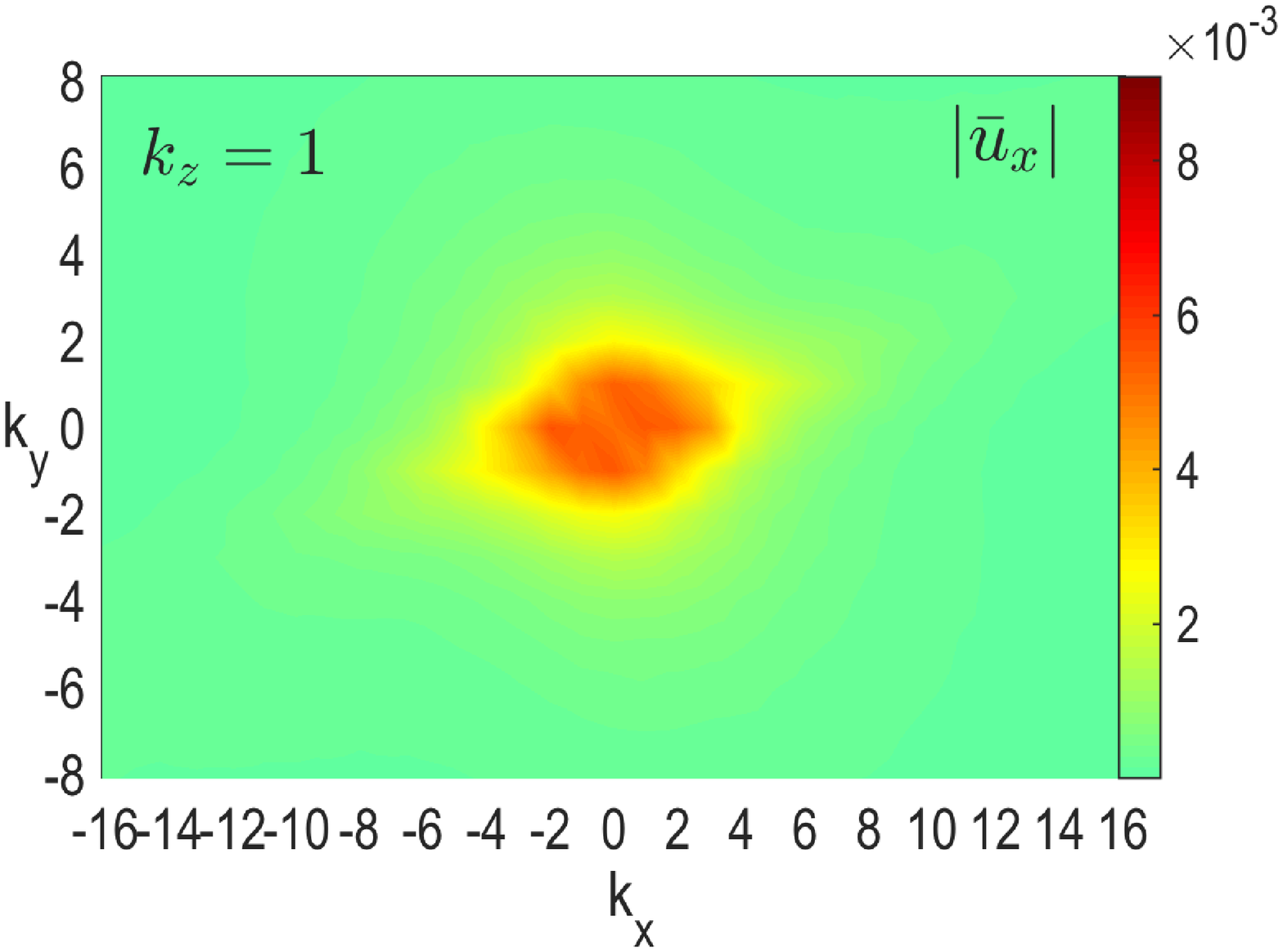}
\includegraphics[width=0.32\textwidth, height=0.2\textwidth]{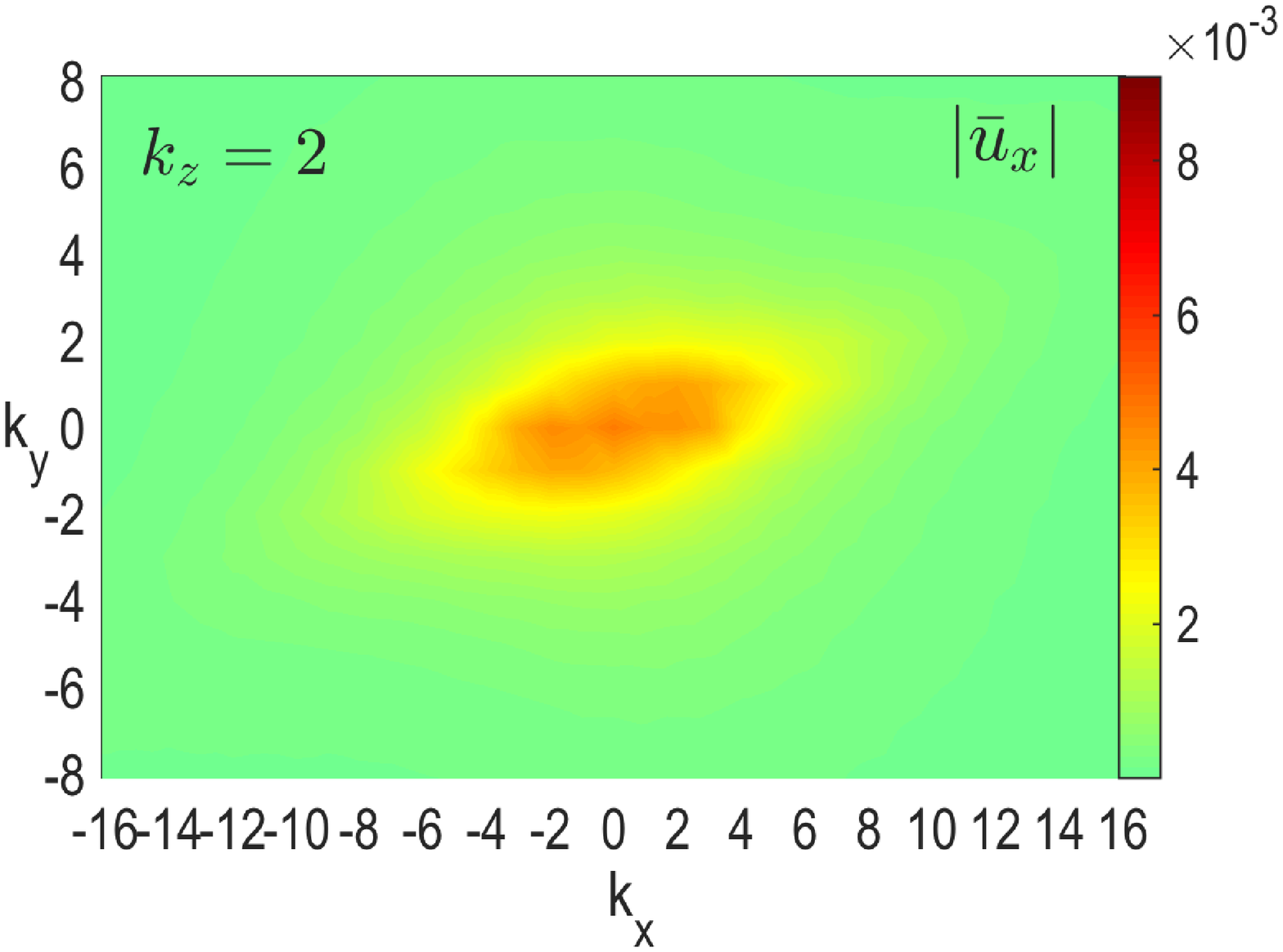}
\includegraphics[width=0.32\textwidth, height=0.2\textwidth]{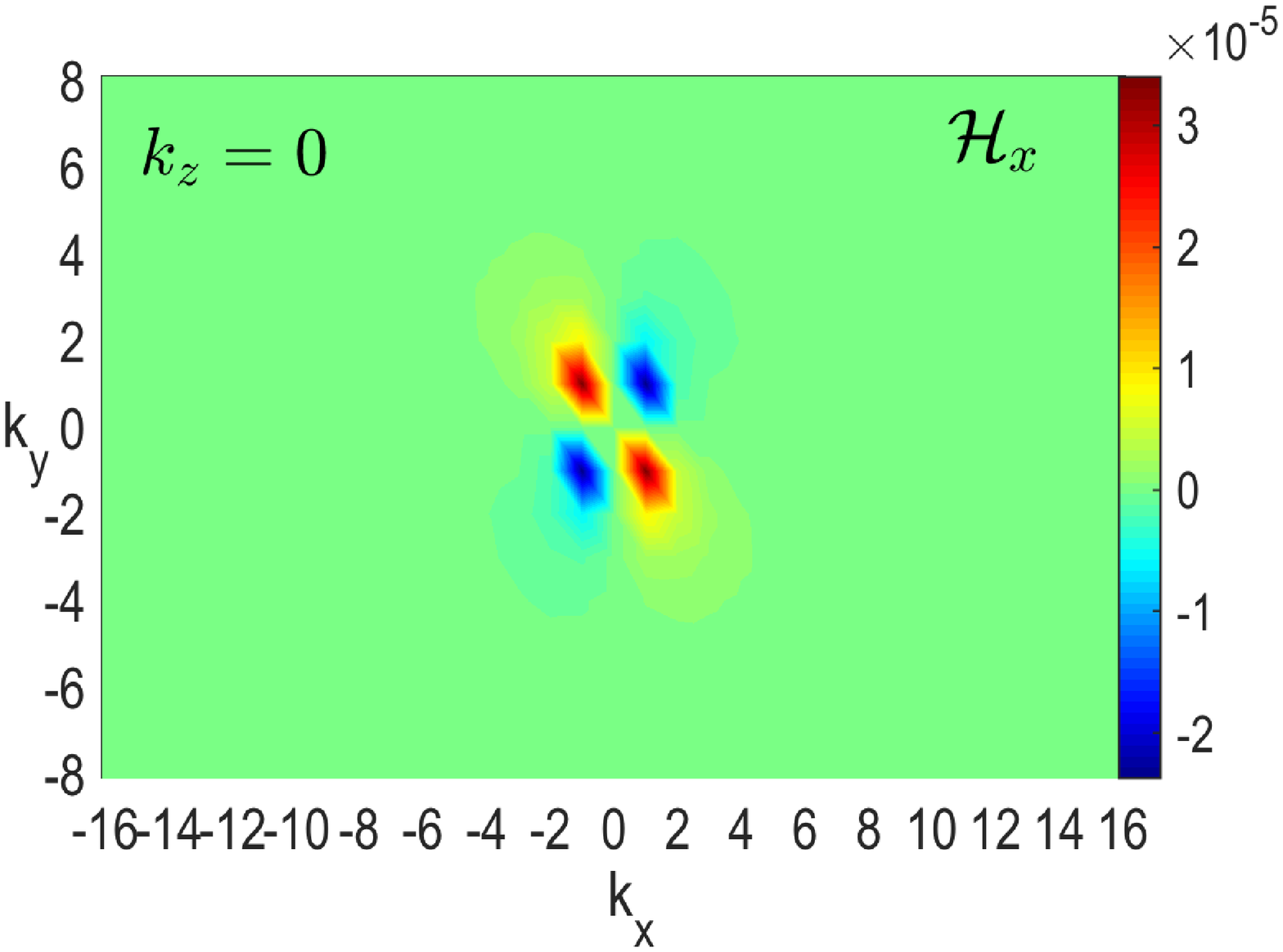}
\includegraphics[width=0.32\textwidth, height=0.2\textwidth]{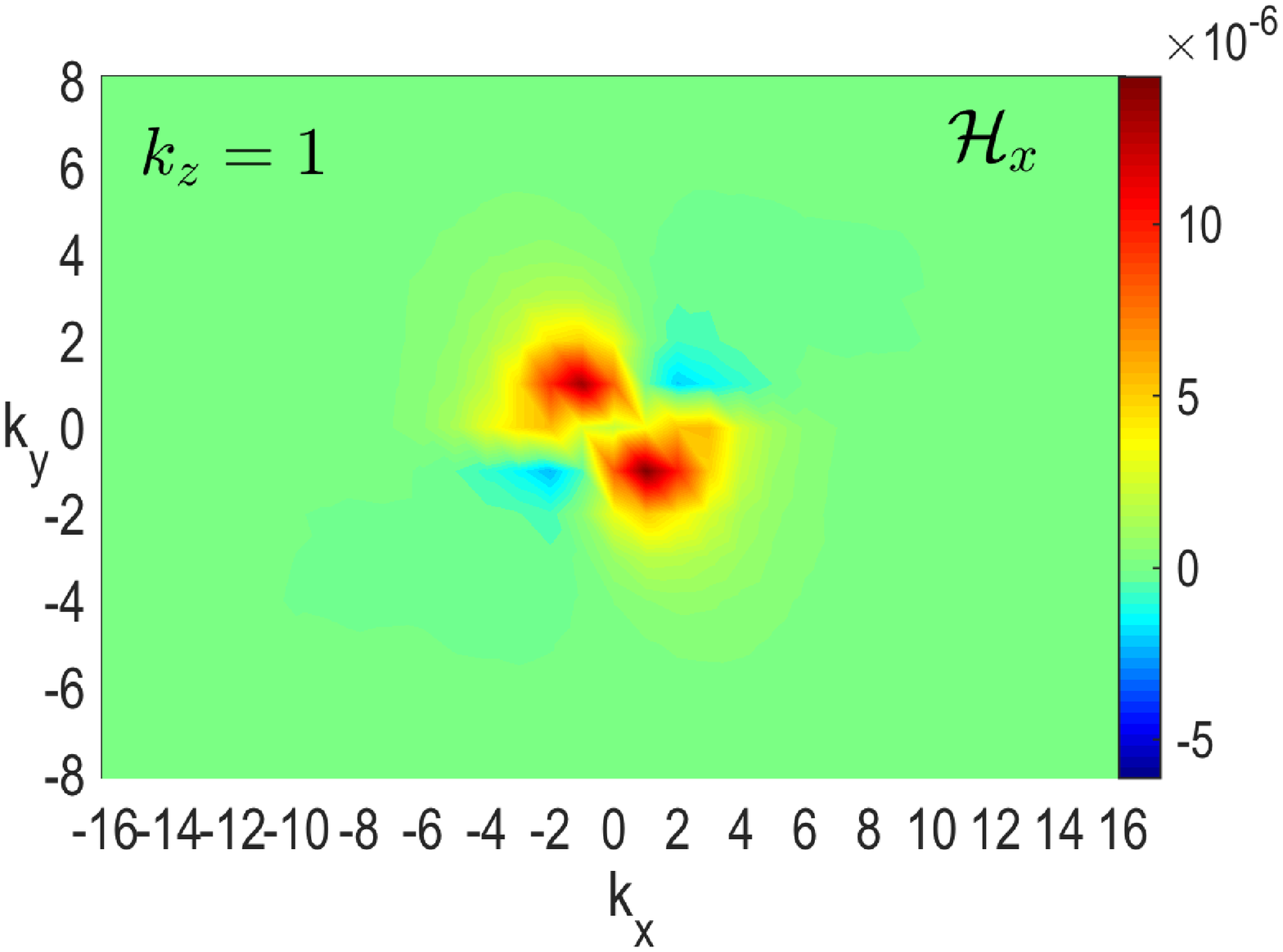}
\includegraphics[width=0.32\textwidth, height=0.2\textwidth]{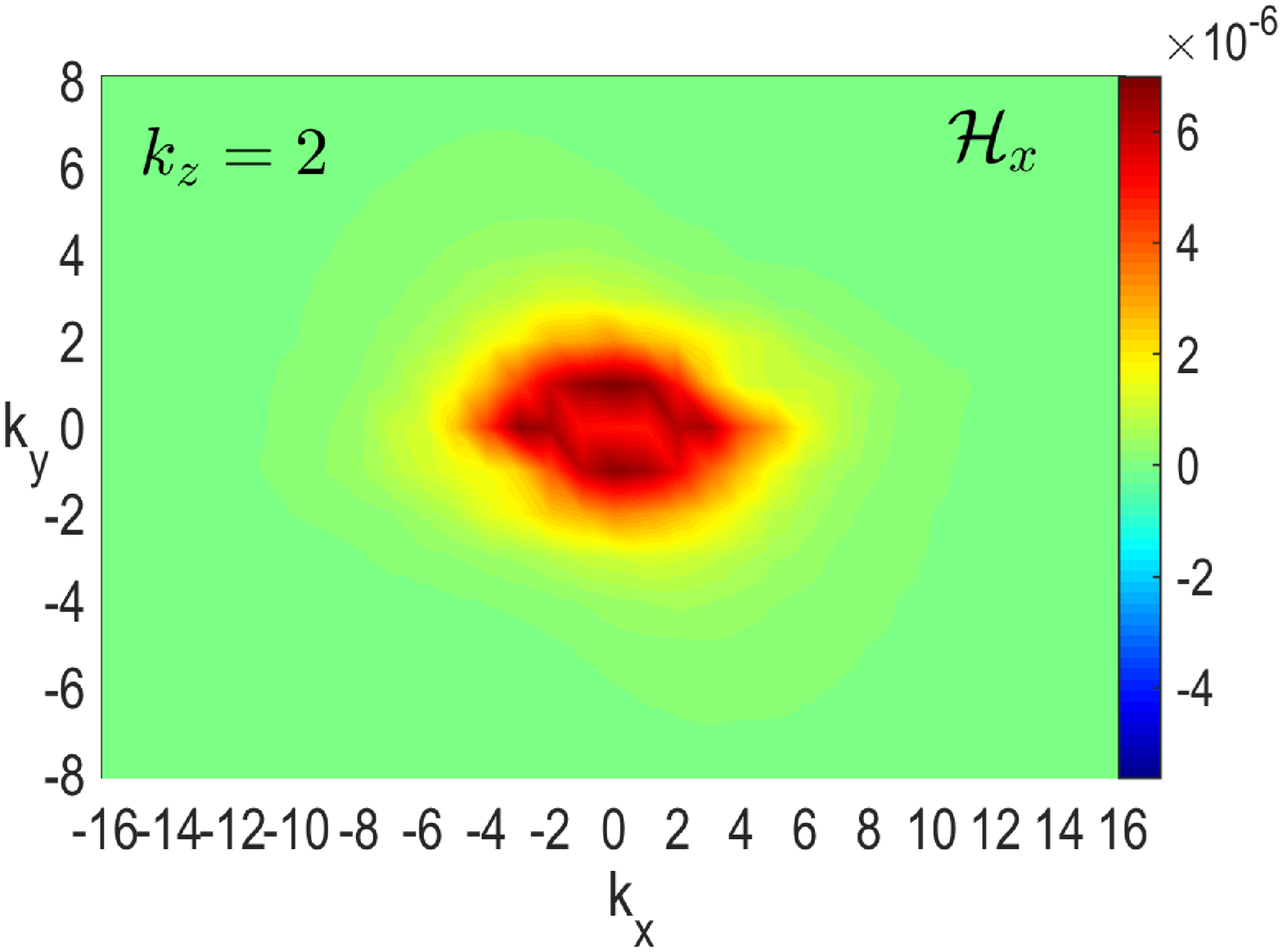}
\includegraphics[width=0.32\textwidth, height=0.2\textwidth]{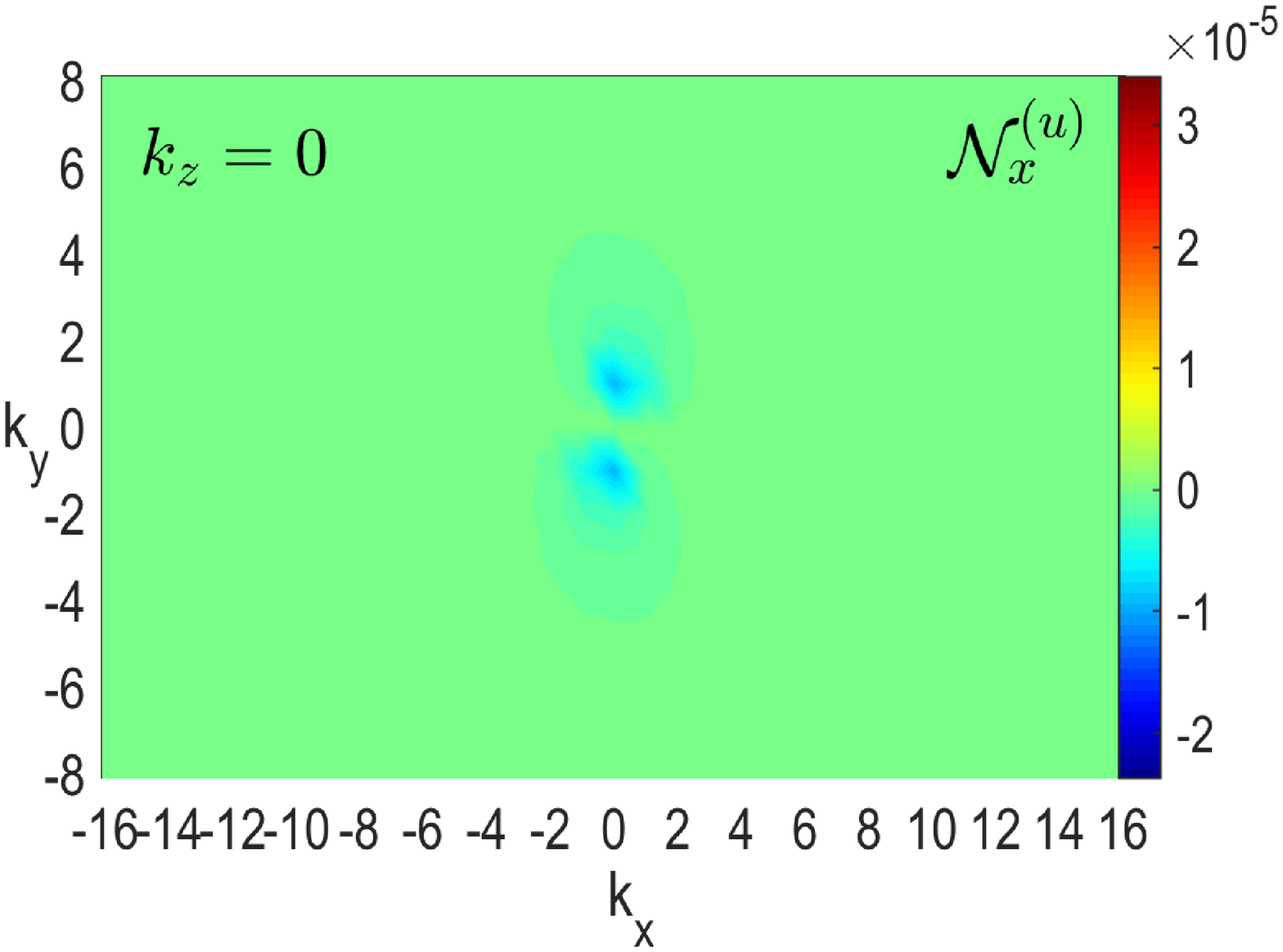}
\includegraphics[width=0.32\textwidth, height=0.2\textwidth]{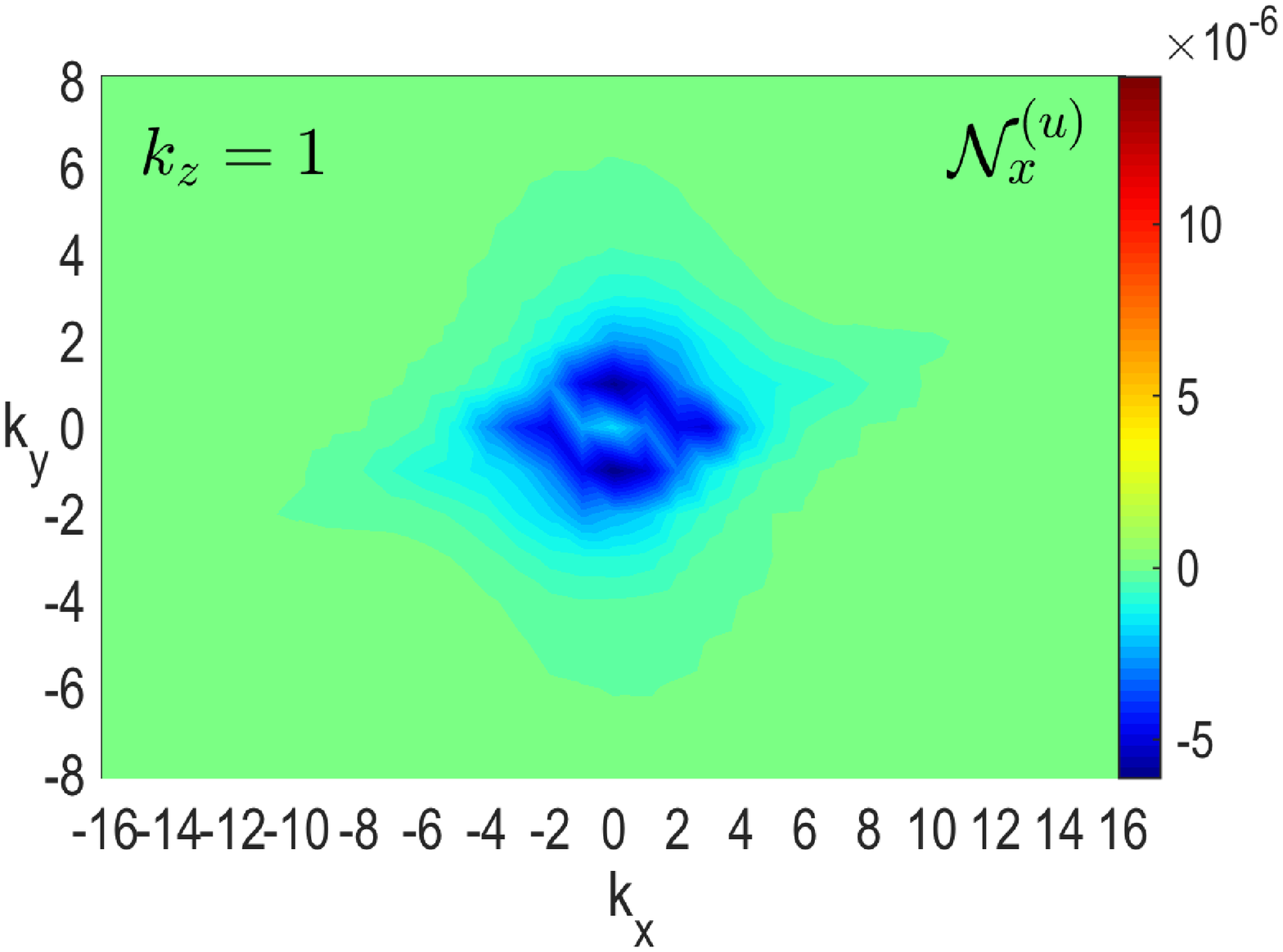}
\includegraphics[width=0.32\textwidth, height=0.2\textwidth]{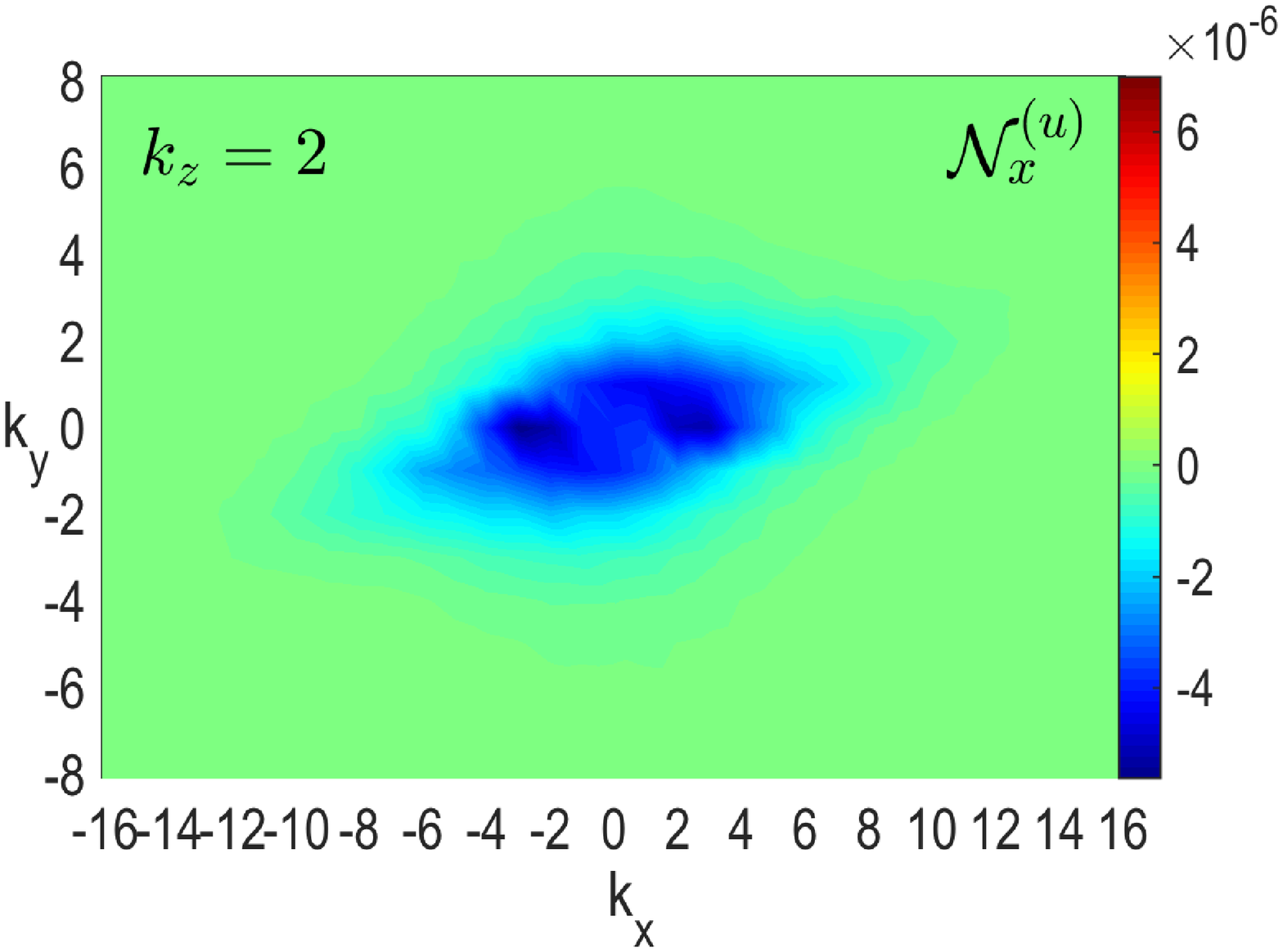}
\caption{Spectra of the radial velocity $|\bar{u}_x|$ and the
governing linear ${\cal H}_x$ and nonlinear ${\cal N}^{(u)}_x$ terms
in $(k_x, k_y)$-slice at $k_z = 0~(left), 1~(middle), 2~(right)$. In the vital area, where these terms reach higher values, the linear term acts as a source, extracting energy for this velocity component from the background flow, whereas the negative nonlinear transfer term as a sink.}\label{fig:spectral_ux}
\end{figure*}

\begin{figure*}[t!]
\centering
\includegraphics[width=0.32\textwidth, height=0.2\textwidth]{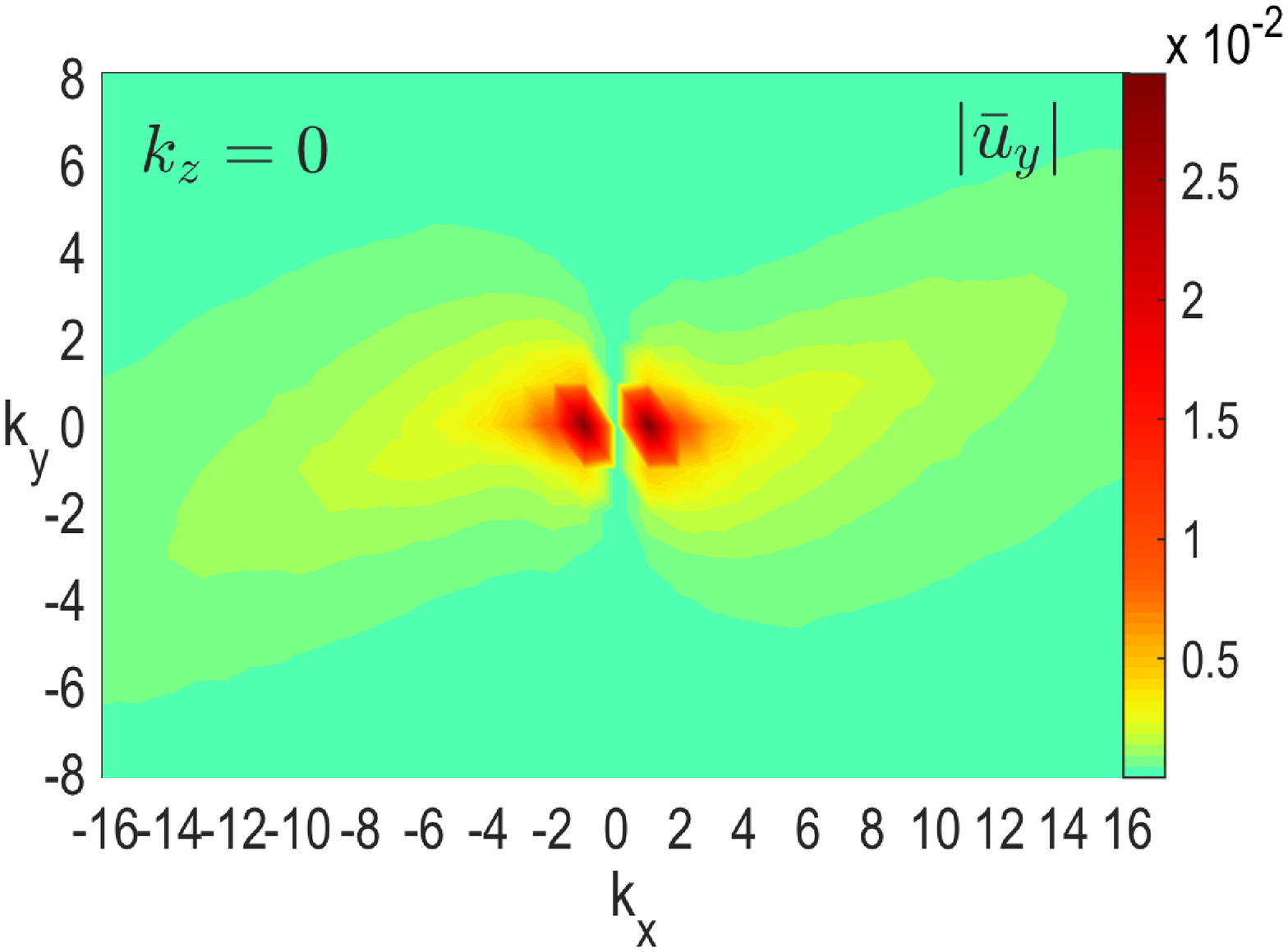}
\includegraphics[width=0.32\textwidth, height=0.2\textwidth]{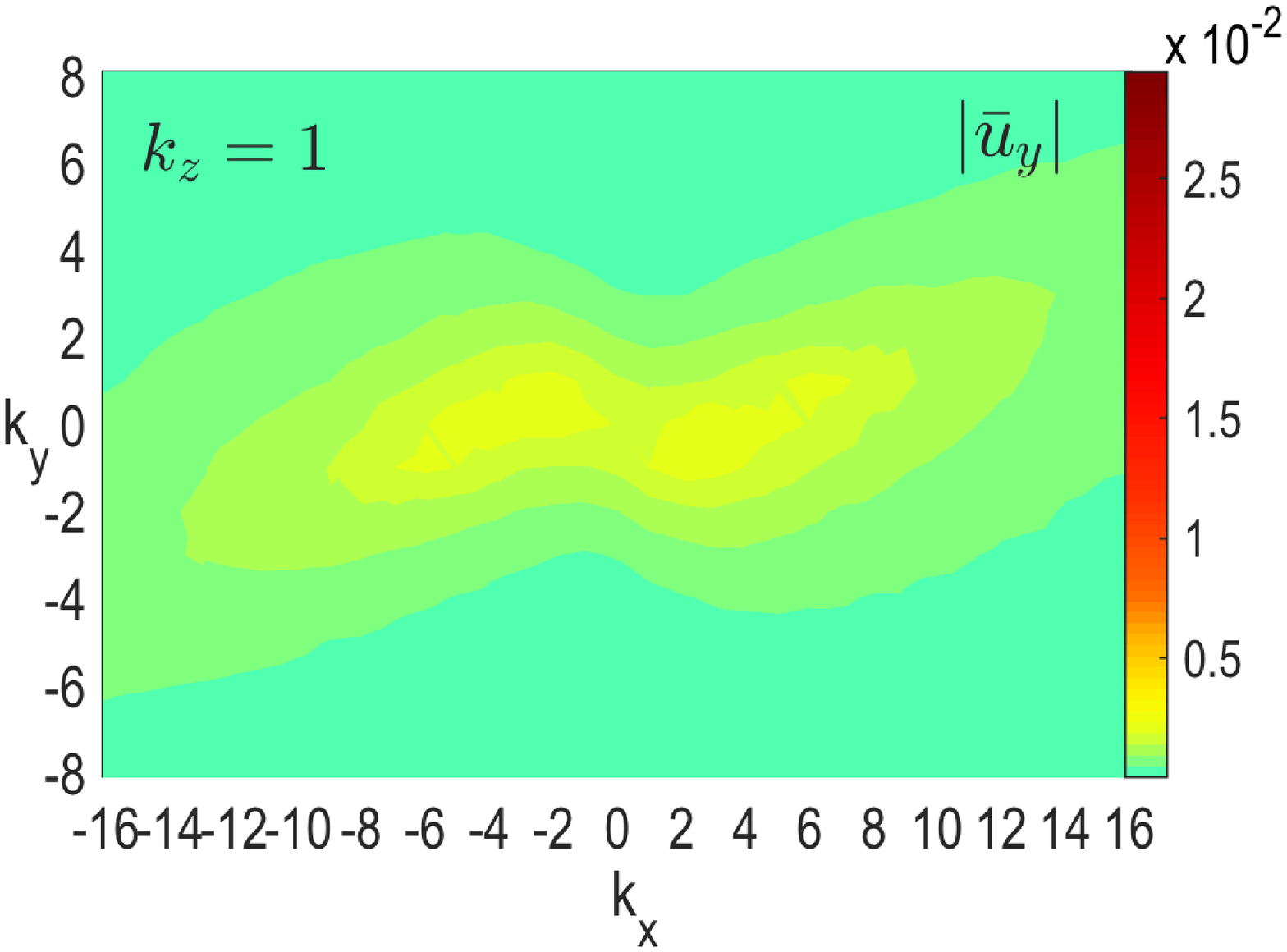}
\includegraphics[width=0.32\textwidth, height=0.2\textwidth]{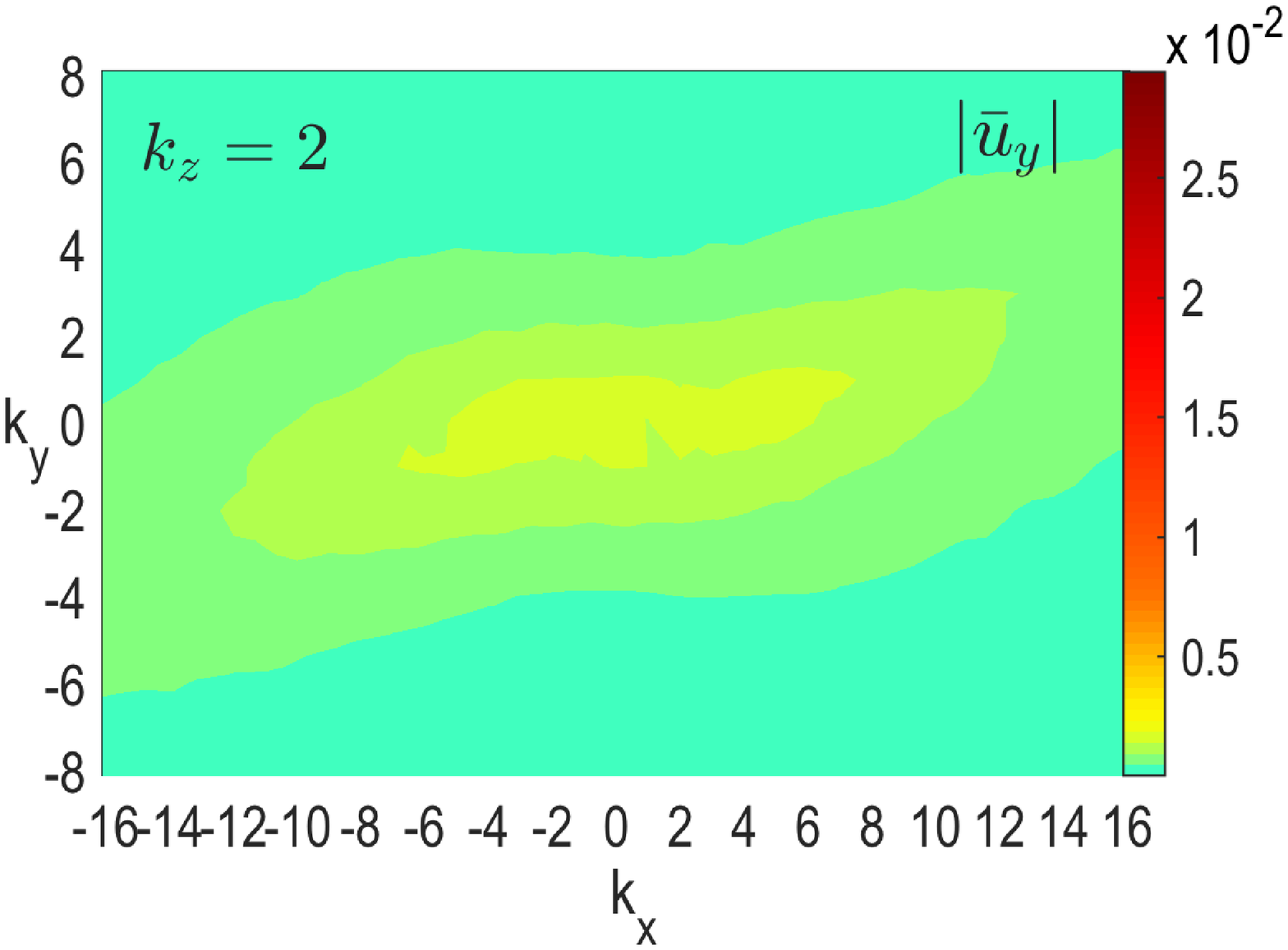}
\includegraphics[width=0.32\textwidth,height=0.2\textwidth]{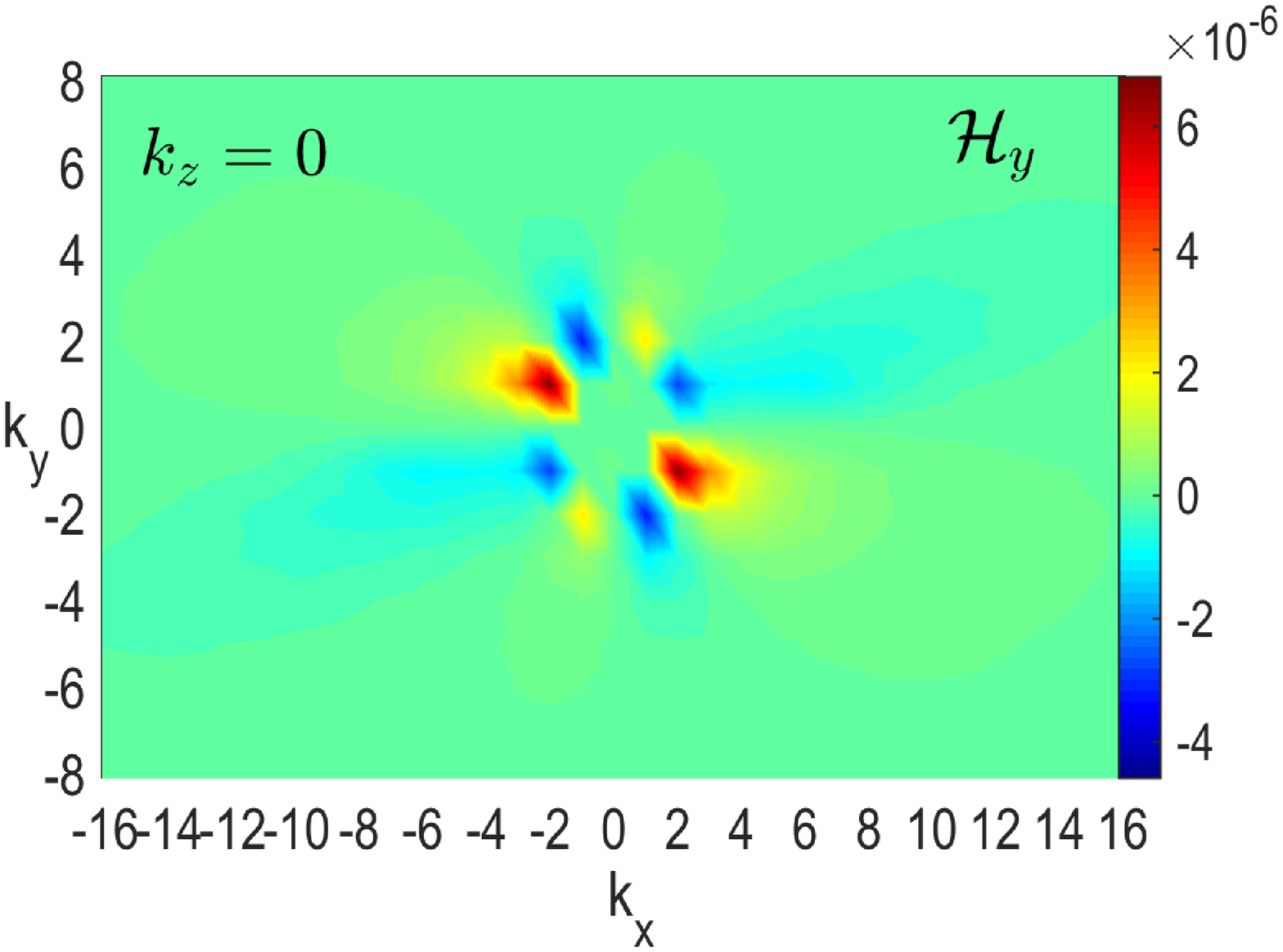}   
\includegraphics[width=0.32\textwidth, height=0.2\textwidth]{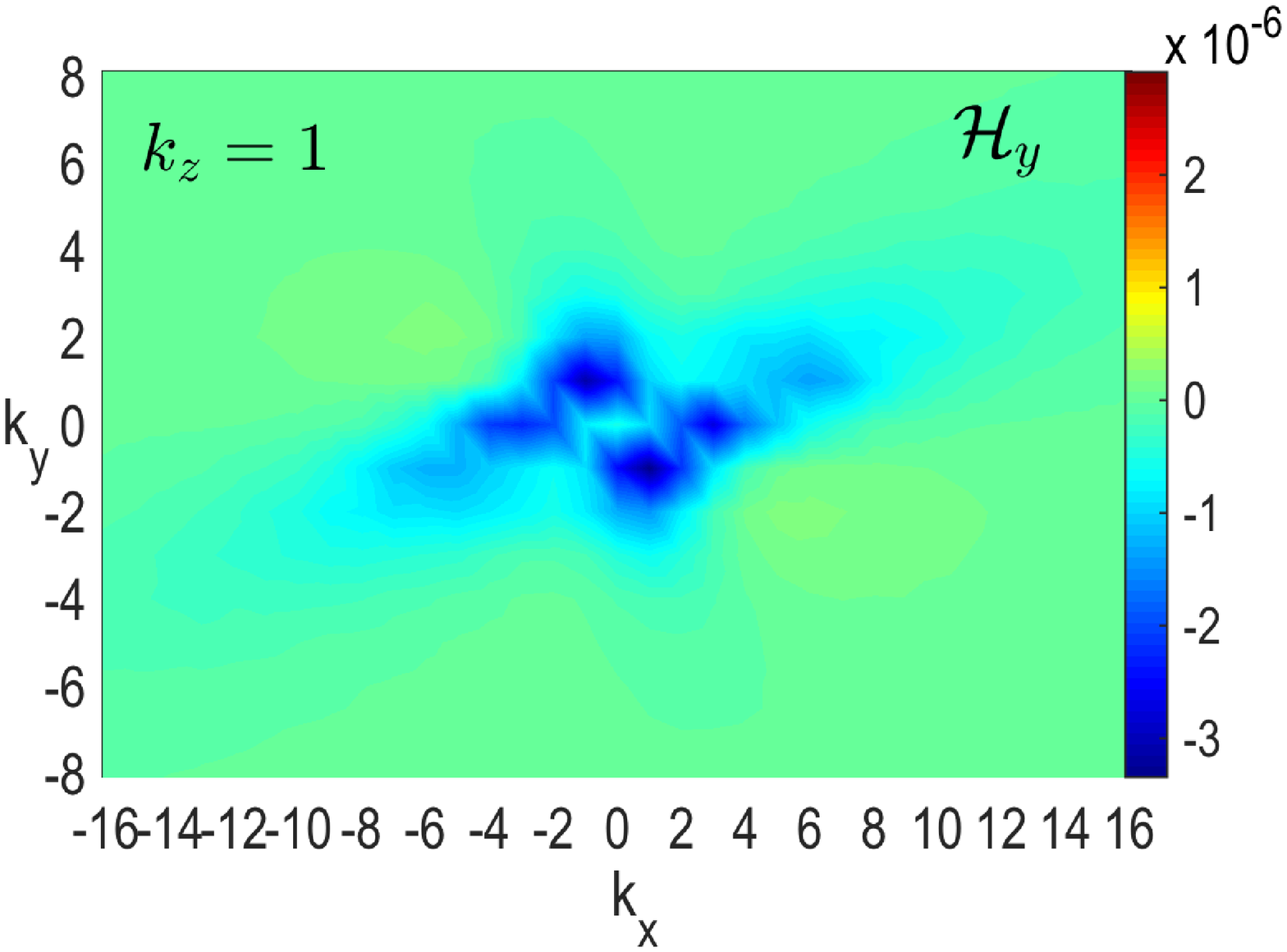}
\includegraphics[width=0.32\textwidth, height=0.2\textwidth]{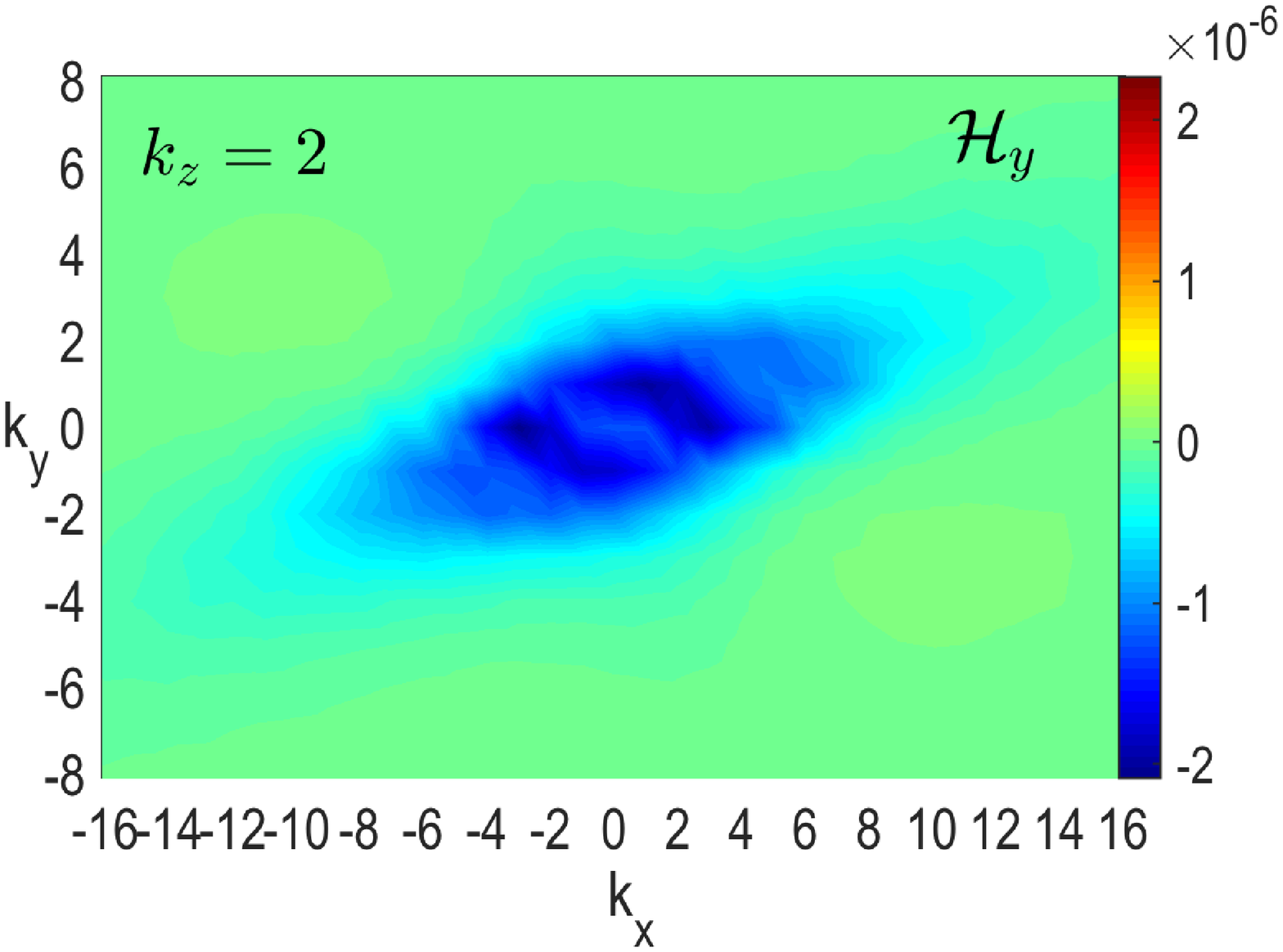}
\includegraphics[width=0.32\textwidth, height=0.2\textwidth]{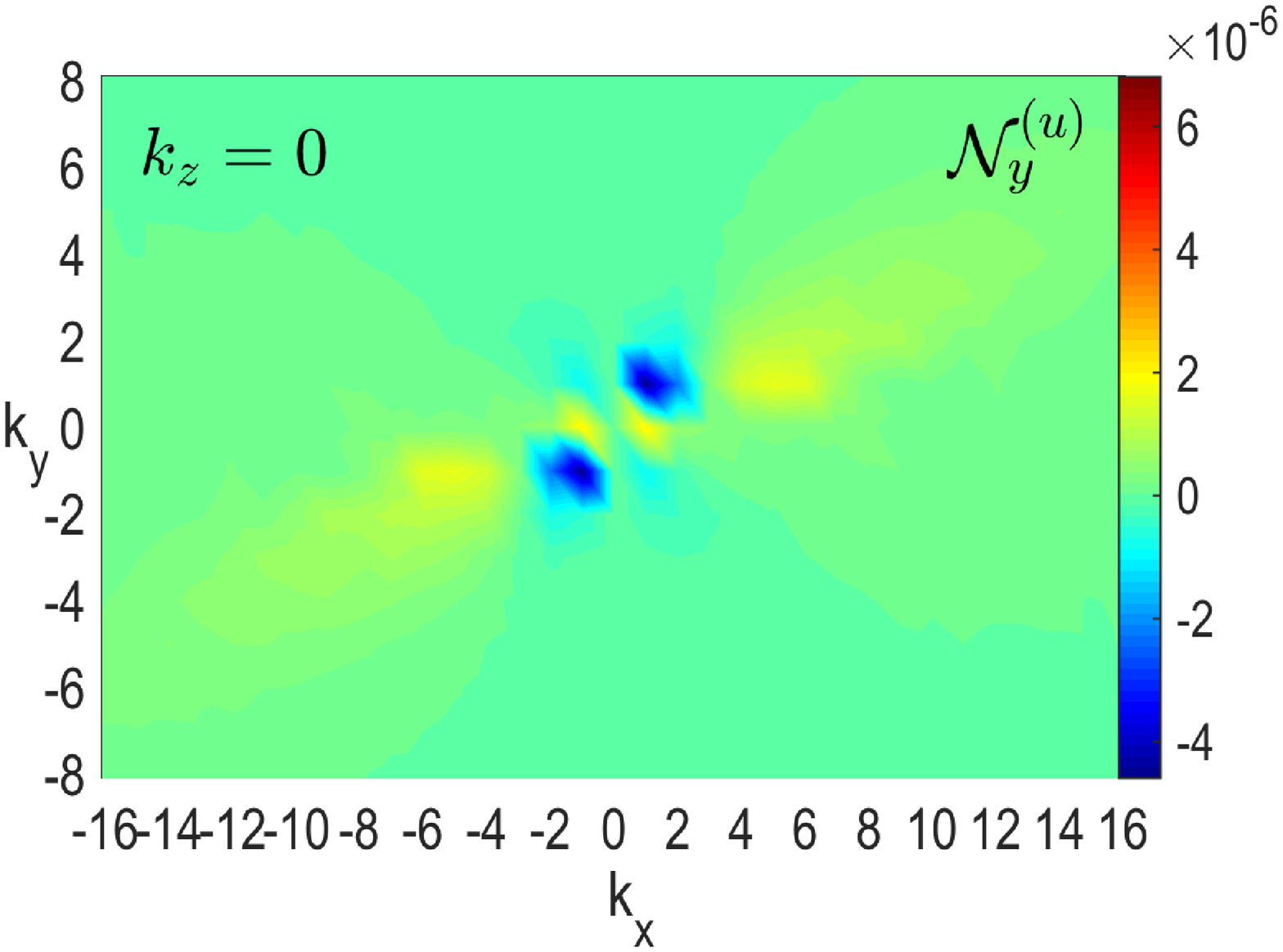}
\includegraphics[width=0.32\textwidth, height=0.2\textwidth]{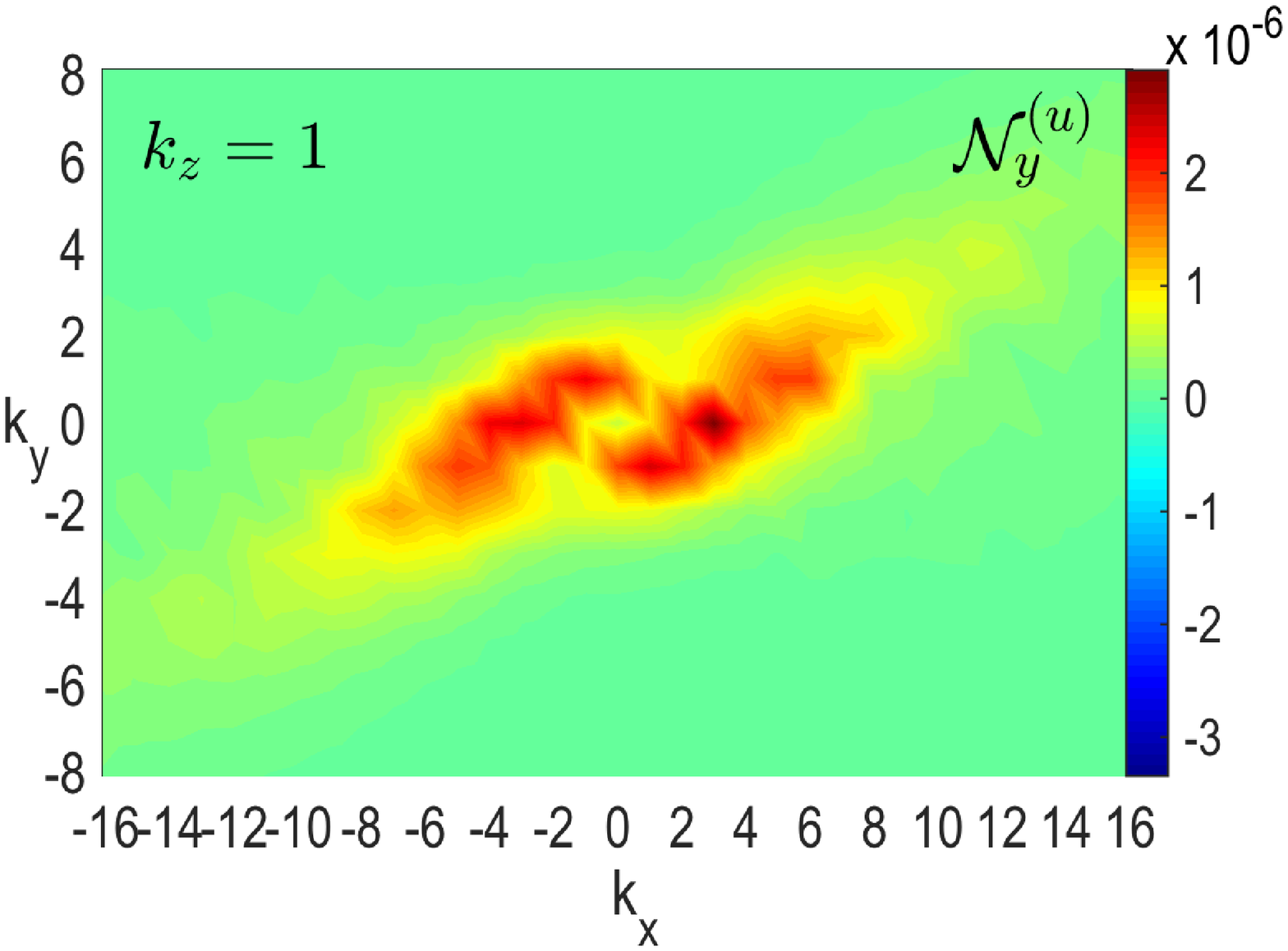}
\includegraphics[width=0.32\textwidth, height=0.2\textwidth]{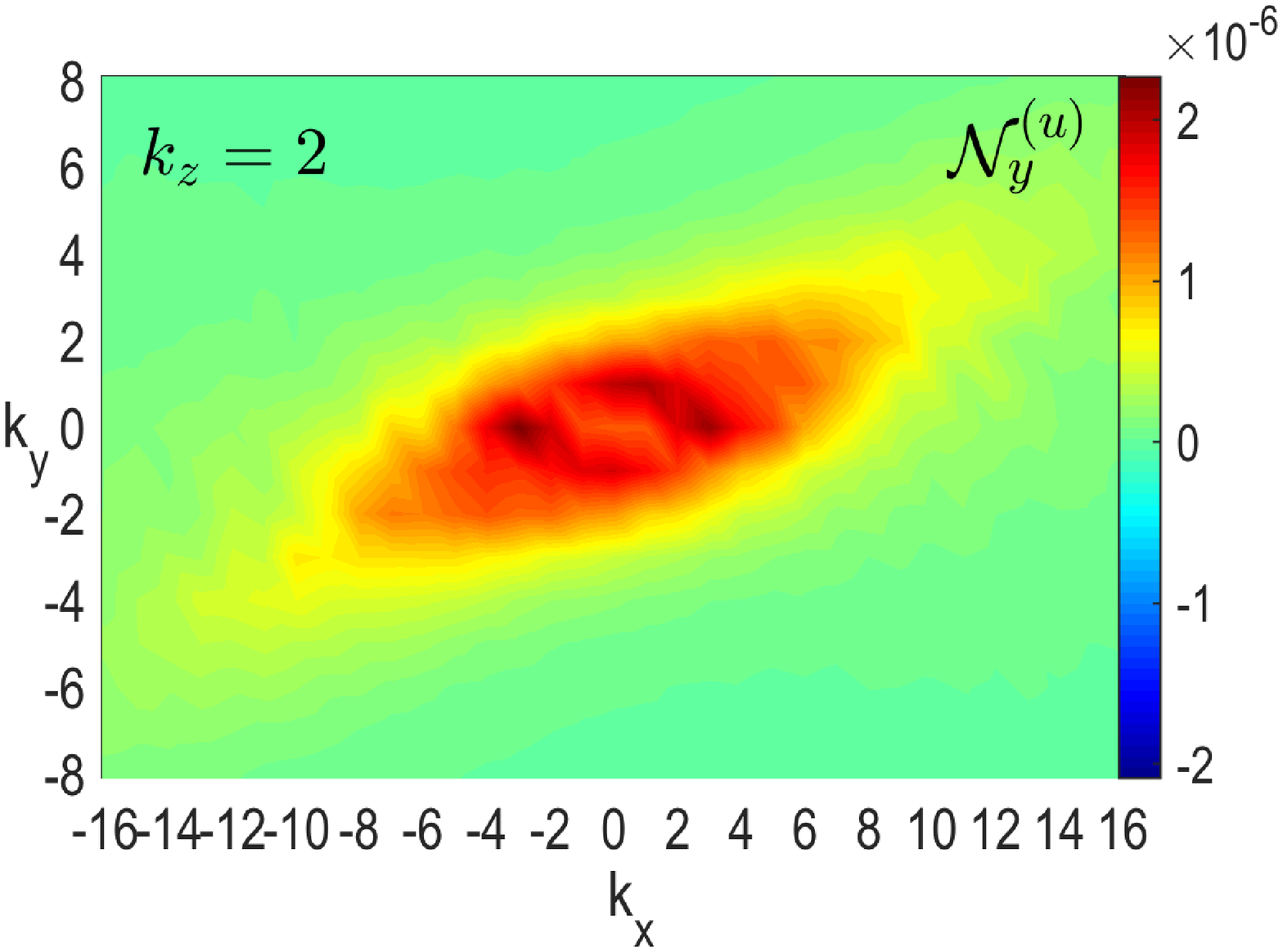}
\caption{Spectra of the azimuthal velocity $|\bar{u}_y|$ with the governing linear  ${\cal H}_y$ and nonlinear ${\cal N}^{(u)}_y$ terms. The  azimuthal velocity spectrum is quite dominated by the mode ${\bf k}_{zf}=(\pm 1, 0, 0)$, which represents the large-scale zonal flow in physical space. In contrast to the radial velocity, at $k_z \geq 1$, in the vital area, the azimuthal velocity is driven by the nonlinear term ${\cal
N}_y^{(u)}$ and drained by the linear term ${\cal H}_y$. At $k_z=0$,
the most important is the zonal flow ${\bf k}_{zf}$ mode for which
${\cal H}_y({\bf k}_{zf})=0$ and ${\cal N}^{(u)}_y({\bf k}_{zf})>0$,
so the excitation and maintenance of this mode is only due to the nonlinear term.}\label{fig:spectral_uy}
\end{figure*}

\begin{figure}
\includegraphics[width=1.2\columnwidth]{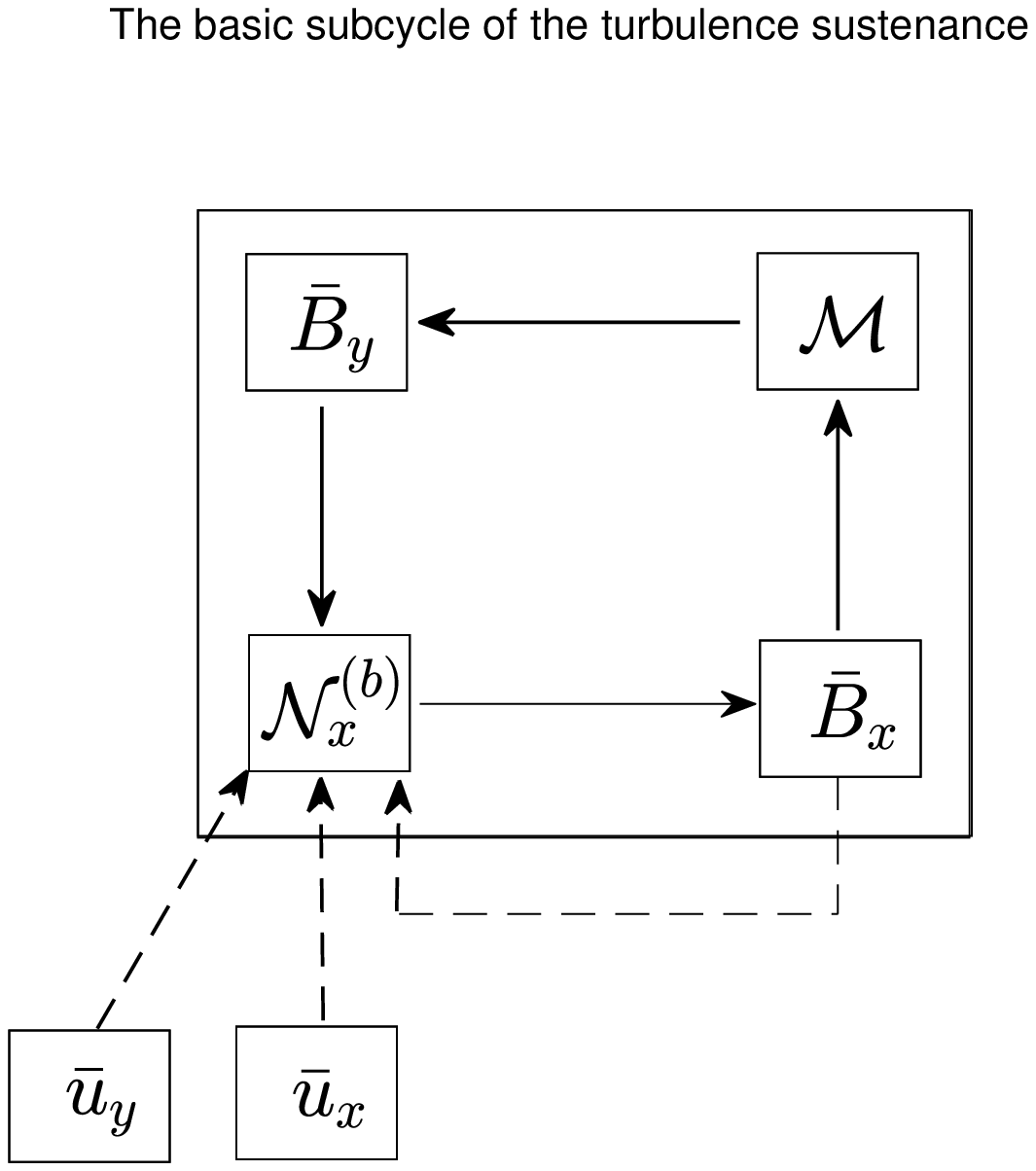}
\caption{Sketch of the core subcycle of the self-sustaining scheme of the zero net flux MRI-turbulence, consisting of the following stages (the solid arrows): (i) regeneration of $\bar{B}_x$ by the nonlinearity ${\cal N}_x^{(b)}$ via the transverse cascade process, (ii) the nonmodal MRI growth process, which produces the positive Maxwell stress ${\cal M}$ from $\bar{B}_x$ and in this way amplifies the azimuthal field energy $|\bar{B}_y|^2/2$ and finally (iii) feeding of $\bar{B}_y$ back into ${\cal N}_x^{(b)}$. Although the other quantities $\bar{u}_x$, $\bar{u}_y$, $\bar{b}_x$ also contribute to ${\cal N}_x^{(b)}$ (dashed arrows, $\bar{u}_z$ and $\bar{b}_z$ are small and not shown), they are
outside the basic subcycle.}\label{fig:sketch}
\end{figure}

The active modes are distributed quite anisotropically in
$(k_x,k_y)$-slice, being mostly concentrated on the $k_x/k_y>0$ side due to the shear. For ${\rm Pm}=4$ these modes occupy a broader range of radial wavenumbers $|k_x|\leq 11$ but narrower range of
azimuthal ones $|k_y|\leq 3$. As noted above, these active modes have small vertical wavenumbers, $|k_z|\lesssim 2$. Since this area of ${\bf k}-$space encompasses the active modes that most contribute to the turbulence dynamics, we refer to it as \emph{the vital area} of the turbulence. For ${\rm Pm}=1$ the radial and azimuthal
extents of the vital area are somewhat smaller, $|k_x|\leq 10$,
$|k_y|\leq 2$. We will see below (Section  \ref{sec:Pmdependence}) that this is because in the decaying turbulence at ${\rm Pm}=1$, in contrast to the sustained one at ${\rm Pm}=4$, the active modes gain less energy due to MRI and the nonlinear transverse cascade, while their energy is rapidly lost to higher wavenumber modes primarily via direct cascade.

The mode ${\bf k}_d=(0,0,\pm 1)$ ($(0,0,\pm 2\pi/L_z)$ in dimensional units) in Figure \ref{fig:modes} bears the large-scale dynamo field \citep[][``d'' stands for ``dynamo'']{Lesur_Ogilvie08,Herault_etal11, Shi_etal16}. It somewhat stands out among other active modes in that it has the magnetic energy larger than 50\% threshold during the longest period of time. However, it is also seen in these plots that other nearby active modes retain similar energies for a comparable time intervals. We demonstrate in Appendix B that the nonlinear interaction between the ${\bf k}_d$ mode and these active modes with comparable amplitudes, govern the dynamics of the large-scale field. For this reason and for the sake of generality, as distinct from the self-sustaining schemes of \citet{Herault_etal11,Riols_etal15,Riols_etal17}, we prefer not to separate out the dynamo ${\bf k}_d$ mode in the dynamics, but treat it on an equal footing with other modes in Fourier space.

Finally, note that the number of the active modes in Figure
\ref{fig:modes} is fairly large -- 112 for ${\rm Pm}=4$ and 59
for ${\rm Pm}=1$ (double these, if we also take into account negative $k_z$). This indicates that the self-sustaining process of zero net flux MRI-turbulence is quite complex, involving a broad range of scales and, therefore, should not be treated within simplified low-order models of the self-sustaining processes. Those models, involving much smaller number of modes, do not account for all the essential nonlinear mode interactions that lie at the heart of the self-sustaining process of MRI-turbulence.

\subsection{Dynamical balances}

We investigate now the spectral dynamics of the turbulence in $(k_x,k_y)$-slices for the same first few smallest vertical wavenumbers at ${\rm Pm}=4$. Herewith we characterize the dynamical balances individually for the radial and azimuthal velocity and magnetic field components. Understanding these dynamical balances allows us to formulate a main self-sustaining scheme for the zero net flux MRI-turbulence. Figures \ref{fig:spectral_bx}-\ref{fig:spectral_uy} present the time-averaged spectra of $|\bar{B}_x|$, $|\bar{B}_y|$, $|\bar{u}_x|$ and $|\bar{u}_y|$ as well as the corresponding main governing linear, ${\cal M}$, ${\cal H}_x, {\cal H}_y$, and nonlinear ${\cal
N}_x^{(b)}$,  ${\cal N}_y^{(b)}$, ${\cal N}_x^{(u)}$, ${\cal
N}_y^{(u)}$ dynamical terms in $(k_x,k_y)$-slices at $k_z=0,1,2$.
These $(k_x,k_y)$-sections of the spectral dynamics contain the full information on the linear and nonlinear dynamical processes and their interplay, ensuring the turbulence sustenance. The ranges of the radial and azimuthal wavenumbers are taken as $|k_x|\leq 16$, $|k_y|\leq 8$, so as to fully encompass the vital area defined above -- the location of the core of the sustaining process. We first describe the dynamical balances depicted in each figure. Here, we do not show the negative definite viscous, ${\cal D}_i^{(u)}$, and resistive, ${\cal D}_i^{(b)}$, terms, since they always oppose the sustenance. Moreover, these terms are negligible in the vital area compared to the above dynamical terms.

As is typical of shear flows, a key common feature of all these spectra of
the magnetic field and velocity as well as the dynamical terms
depicted in Figures \ref{fig:spectral_bx}-\ref{fig:spectral_uy} is
that they all display a common type of anisotropy in Fourier space
due to the shear, i.e., exhibit a strong variation over polar
angle in $(k_x,k_y)$-slices with inclination preferably on the $k_x/k_y>0$ side \citep[e.g.,][Paper I]{Lesur_Longaretti11, Mamatsashvili_etal14, Murphy_Pessah15}. The anisotropic linear terms ${\cal M}$, ${\cal H}_x, {\cal H}_y$, as discussed above, characterize the nonmodal growth of zero net flux MRI. These terms act as a source when they are positive (red and yellow areas) and as a sink when negative (blue areas). For the nonlinear processes described by ${\cal N}^{(b)}_i$ and ${\cal N}^{(u)}_i$ this spectral anisotropy evidently cannot be fitted into the framework of classical (spherically symmetric) forms of nonlinear -- direct and inverse -- cascades.

We now describe the spectra of the magnetic field and velocity
components as well as the dynamical balances for them. We start with the magnetic field spectra.
\begin{enumerate}[$\bullet$]

\item
{\bf The radial field} spectrum, $|\bar{B}_x|$, is plotted in Figure \ref{fig:spectral_bx} together with the governing nonlinear transfer
term ${\cal N}_x^{(b)}$, which is mainly responsible for its
generation and amplification. The power of the anisotropic spectrum
of the radial field is mainly concentrated in non-axisymmetric
($k_y\neq 0$) modes with $k_x/k_y>0$. The mode with $|k_x|=3, |k_y|=k_z=1$ has the largest amplitude, although the nearby non-axisymmetric modes at $k_z=0$ and $2$ have comparable amplitudes. This  anisotropy of $|\bar{B}_x|$ is a consequence of the anisotropic structure of ${\cal N}^{(b)}_x$, whose main notable effect is the transverse (i.e., over wavevector polar angle, $\varphi={\rm sin}^{-1}[k_y/(k_x^2+k_y^2)^{1/2}]$) redistribution/transfer of power in $(k_x,k_y)$-slice for all considered $k_z=0,1$ and $2$. As seen in the bottom plots of Figure \ref{fig:spectral_bx}, this term transfers the spectral energy of radial field, $|\bar{B}_x|^2/2$, from the area of ``giver'' wavenumbers where it is negative ${\cal N}_x^{(b)}<0$ (blue) into the area of ``receiver'' wavenumbers where it is positive ${\cal N}_x^{(b)}>0$ (red and yellow), as well as among different components of the velocities and magnetic field. The structure of these two areas strongly depends on polar angle in $(k_x,k_y)$-slices. The yellow and red areas are essential for the sustenance process, since they comprise the wavenumbers at which ${\cal N}_x^{(b)}$ continually regenerates the radial field as a result of nonlinear interactions between other pairs of wavenumbers forming a triad with these ones. These modes, being thereby replenished in the red and yellow areas, are non-axisymmetric with $k_x/k_y>0$. In other words, one can call these red and yellow areas together the growth area. The linear drift (the first rhs term of Equation \ref{eq:bxk2}), on the other hand, advects $|\bar{B}_x|$ of an individual non-axisymmetric mode (shearing wave with given $k_y$ and $k_z$) along $k_x$-axis from the growth (red and yellow) area back into the blue area where ${\cal N}_x^{(b)}<0$ and, therefore, $|\bar{B}_x|$ decreases while still remaining significant in this blue area. This interplay between the nonlinear transfer and linear drift establishes the specific anisotropic spectrum of $|\bar{B}_x|$ that overlaps with both -- red/yellow and blue -- areas, as is illustrated by the contours drawn on ${\cal N}_x^{(b)}$ in Figure \ref{fig:spectral_bx}. In the quasi-steady state, the drift and the transverse cascade balance each other in the vital area, forming a closed cycle. Thus, the linear drift in Fourier space gives the dynamical processes transient feature and therefore the permanent regeneration due to the nonlinear transverse cascade (exhibited by ${\cal N}_x^{(b)}$) plays a key role in the self-sustaining dynamics of the turbulence.

\item
{\bf The azimuthal field} spectrum, $|\bar{B}_y|$, is shown in Figure \ref{fig:spectral_by} together with the governing Maxwell stress, ${\cal
M}$, and the nonlinear transfer term ${\cal N}^{(b)}_y$. We have seen already in Figure \ref{fig:modes} that the dynamo mode ${\bf k}_d$ stands out among other active modes. Here, we also observe that, on
average in time, it carries the largest azimuthal field energy among
the other modes. Nearby energetic non-axisymmetric active
modes have $|k_y|=1$ \citep[see also][]{Herault_etal11,Riols_etal15,Riols_etal17}. The Maxwell
stress, ${\cal M}$, is positive in $(k_x,k_y)$-slices at
$k_y/k_x>0$ and dominant in the vital area (red and yellow), where it supplies (injects) energy in the azimuthal field (including that of the ${\bf k}_d$ mode, see also Appendix B). As a result, $|\bar{B}_y|$ undergoes nonmodal MRI growth. By contrast, the nonlinear transfer term is mainly negative, ${\cal N}^{(b)}_y<0$, in this area (blue), draining  the azimuthal field energy there and transferring it to large wavenumbers, where this term is positive but small (light green area), as well as to different components. Thus, the azimuthal field of larger wavenumber modes is supplied as a result of nonlinear transfers from the smaller wavenumber modes in the vital area, but not from the Maxwell stress directly, which is negligible there.\footnote{These larger wavenumber modes are what \cite{Riols_etal17} refer to as ``slaved modes''.} Note also that ${\cal M}$ and ${\cal N}^{(b)}_y$ have nearly similar absolute values and shapes in Fourier space, implying that in the quasi-steady state, the energy injection into and its nonlinear ``removal'' from the active modes in the vital area are approximately in balance -- the action of the linear drift for $|\bar{B}_y|$ (the first rhs term of  Equation \ref{eq:byk2}) plays only a minor role in this balance. However, because of this drift, the structure of the $|\bar{B}_y|$ spectrum in the quasi-steady state, mainly resulting from the action of the Maxwell stress and the nonlinear transfer term, is a bit more inclined towards the $k_x$-axis than ${\cal M}$, as seen from the contours of $|\bar{B}_y|$ drawn on this term in Figure \ref{fig:spectral_by}.

\item
{\bf The radial velocity} spectrum, $|\bar{u}_x|$, is shown in Figure \ref{fig:spectral_ux} together with its governing linear, ${\cal H}_x$, and nonlinear transfer, ${\cal N}^{(u)}_x$, terms. $|\bar{u}_x|$ reaches higher values at $k_z=0$, with a maximum at $k_x=0, |k_y|=1$. At $k_z\geq 1$, it is maintained by mostly positive ${\cal H}_x>0$, at the expense of the mean flow, and drained by the negative ${\cal N}^{(u)}_x<0$. At
$k_z=0$, ${\cal H}_x$ can be positive (red areas at $k_x/k_y<0$) and
negative (blue areas at $k_x/k_y>0$) in $(k_x,k_y)$-slice, similar to that in the 2D hydrodynamic case \citep{Chagelishvili_etal03,Johnson_Gammie05,Horton_etal10}. This
implies that the injected $|\bar{u}_x|$ with $k_x/k_y < 0$, undergoing
the linear drift along $k_x$-axis, achieves a maximum at $k_x = 0$ and then enters the area $k_x/k_y>0$ where it falls off due to the negative values of ${\cal H}_x$ and ${\cal N}^{(u)}_x$.

\item
{\bf The azimuthal velocity} spectrum, $|\bar{u}_y|$, is shown in Figure \ref{fig:spectral_uy} together with the governing linear, ${\cal
H}_y$, and nonlinear transfer, ${\cal N}^{(u)}_y$, terms. Note that the $|\bar{u}_y|$ spectrum is markedly dominated by the ${\bf k}_{zf}=(\pm 1, 0, 0)$ mode, which, as discussed above, corresponds to the zonal flow excited in MRI-turbulence. Just this mode gives rise to the sharp peak of $\widehat{|\bar{u}_y|^2}$ at $k_z=0$ (Figure \ref{fig:integrated in plane-ub}). ${\cal N}^{(u)}_y$ is always positive at $k_z\geq 1$, supplying the azimuthal velocity, while at $k_z=0$ this term can be positive or negative. At $k_z=0$, the linear term ${\cal H}_y$ also can be positive or negative, while it is always negative for $k_z=1,2$, draining the azimuthal velocity. In any case, for the dominant ${\bf k}_{zf}$ mode, ${\cal H}_y({\bf k}_{zf})=0$, ${\cal N}^{(u)}_y({\bf k}_{zf})>0$, implying that the generation and maintenance of the zonal flow is exclusively due to nonlinear transfers. Note also that the radial velocity at $k_z=0$ is several times smaller than the azimuthal velocity of the zonal flow, $|\bar{u}_x(0,1,0)|<|\bar{u}_y({\bf k}_{zf})|$, as it follows from the comparison of Figures \ref{fig:spectral_ux} and \ref{fig:spectral_uy}.
\end{enumerate}

\subsection{The core subcycle of the self-sustaining process}

Having described the dynamical balances in Fourier space in the sustained case with ${\rm Pm}=4$, we are now in a position to construct on their basis the main self-sustaining scheme/cycle of the zero net flux MRI-turbulence. Since the magnetic energy and Maxwell stress are
much larger and hence more important than the kinetic energy and
Reynolds stress, the sustenance should be primarily magnetically
driven. Specifically, the core of this process should involve the
most energy-carrying radial and azimuthal field components at those
wavenumbers where they are appreciable, i.e., in the vital area. In the presence of both shear and rotation these two components undergo amplification due to the nonmodal MRI process, mediated by the Maxwell stress. The nonlinear transfers, involving also the rest of the velocity and magnetic field components, ensure continual regeneration of the radial field, which initiates the MRI growth. \emph{Thus, interplay of the linear nonmodal MRI growth and the constructive nonlinear feedback due to the transverse cascade in fact determines the self-sustaining process of the MRI-turbulence in the zero net flux case}. The situation here is analogous to that of the MRI-turbulence in the nonzero net azimuthal field case studied in Paper I, except that here there are no linear kinetic-magnetic exchange terms contributing to MRI. The detailed nonlinear interactions between velocity and magnetic field components at many different wavenumbers are described in general by rather complex terms under integrals in expressions (\ref{eq:App-Nuij}) and (\ref{eq:App-Fxk})-(\ref{eq:App-Fyk}). Since the number of active modes is quite large (Figure \ref{fig:modes}), this makes the entire self-sustaining process not amenable to a vivid schematization.
Nevertheless, assuming all the details of the nonlinear mode-to-mode
interactions are encapsulated in the nonlinear magnetic transfer
terms ${\cal N}^{(b)}_x$ and ${\cal N}^{(b)}_y$ (Figures \ref{fig:spectral_bx} and \ref{fig:spectral_by}), we  can clearly disentangle the \emph{core subcycle} of the turbulence sustenance from the overall dynamical picture.

This core subcycle is sketched in Figure \ref{fig:sketch} (solid
arrows inside the rectangle) and can be characterized as follows. One
can start each such a cycle with the regeneration/seeding of the
radial field $\bar{B}_x$ by nonlinear term ${\cal N}_x^{(b)}$ via
the transverse cascade process at those wavenumbers where this term is positive (red and yellow areas in Figure \ref{fig:spectral_bx}). These modes with newly regenerated radial field, in turn, initiate the linear nonmodal MRI process due to the shear (via shear-proportional linear term in Equation \ref{eq:App-byk}), which produces positive Maxwell stress, ${\cal M}>0$, that, in turn, amplifies the azimuthal field energy $|\bar{B}_y|^2/2$. This is in agreement with the noticeable correlation between the distributions of $|\bar{B}_x|$ and ${\cal M}$ in $(k_x,k_y)$-slices seen by comparing Figures \ref{fig:spectral_bx} and \ref{fig:spectral_by}. This growth is higher for non-axiymmetric modes and thus is of a transient type because of their drift along the $k_x$-axis. Just because of this transient nature of the MRI growth (and hence energy extraction from the flow), the continual  seeding of the radial field due to the nonlinear transverse cascade is crucial for the turbulence self-sustenance. As these modes drift, they also cross the area in Fourier space where the nonlinear term ${\cal N}_y^{(b)}$ is negative (blue areas in Figure \ref{fig:spectral_by}) and therefore drains the azimuthal field energy of these modes. Eventually, due to the linear drift process, the non-axisymmetric modes move away from the amplification red area, where the Maxwell stress ${\cal M}$ is appreciable (Figure \ref{fig:spectral_by})  and decay due to resistivity. Since this azimuthal field is the dominant field component, it gives a main contribution -- positive feedback -- to ${\cal N}_x^{(b)}$, which, in turn, produces a new seed radial field, thereby closing the cycle. This self-sustenance scheme of MRI-turbulence, underlying the dynamics of active modes, naturally determines the behavior of the axisymmetric ${\bf k}_d$ dynamo mode too (see Appendix B).

We have outlined above a central part of the full self-sustaining scheme
at ${\rm Pm}=4$, in which, in principle, also the velocity
components and the vertical field participate. They, however, contribute through the nonlinear term, ${\cal N}_x^{(b)}$, which is key to the turbulence sustenance, but still it is determined primarily by the dominant $\bar{B}_y$. In Figure \ref{fig:sketch}, the dashed arrows denote the contributions from these quantities extrinsic to the core subcycle.

\begin{figure*}[t]
\includegraphics[width=0.32\textwidth]{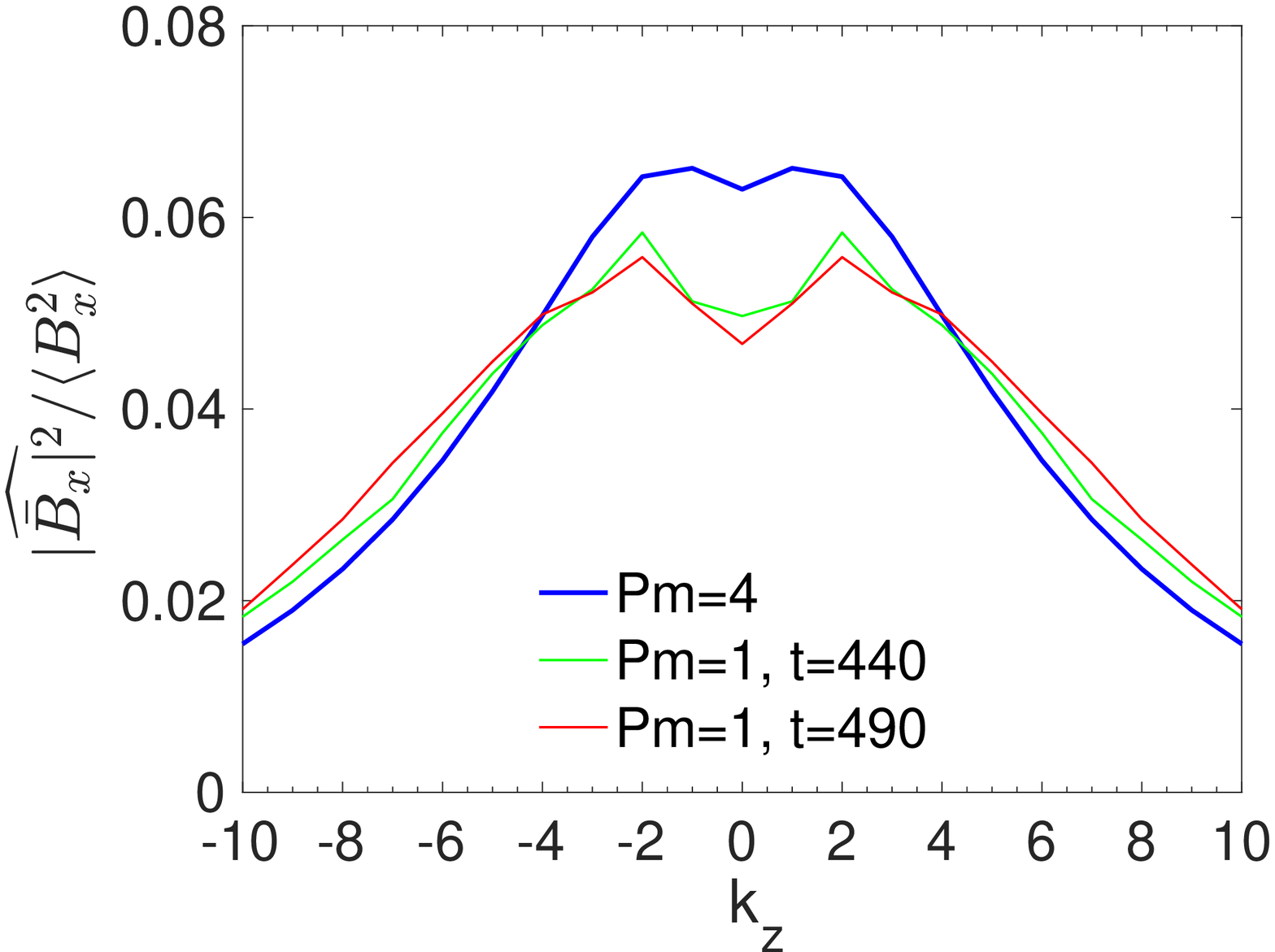}
\includegraphics[width=0.32\textwidth]{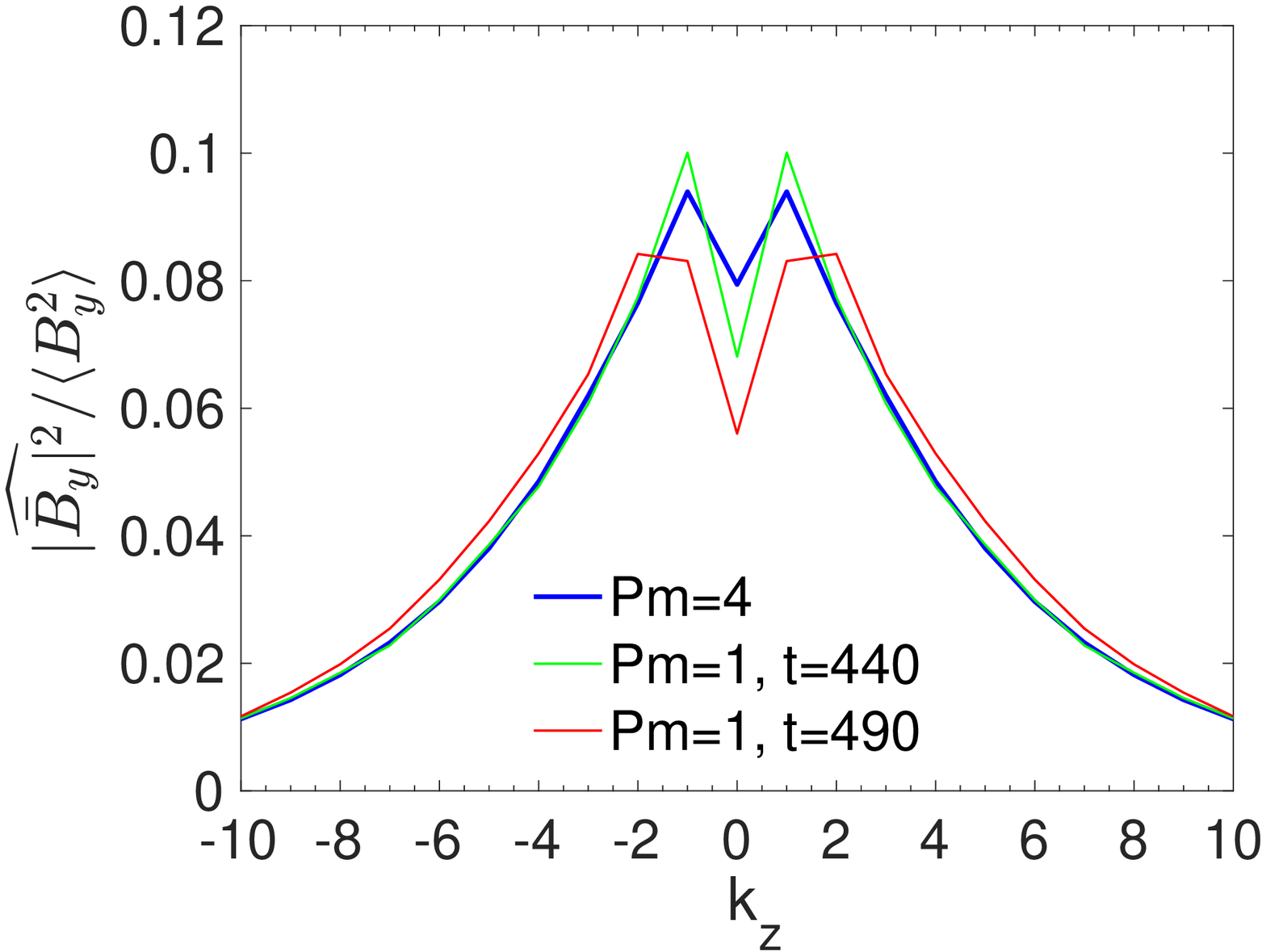}
\includegraphics[width=0.32\textwidth]{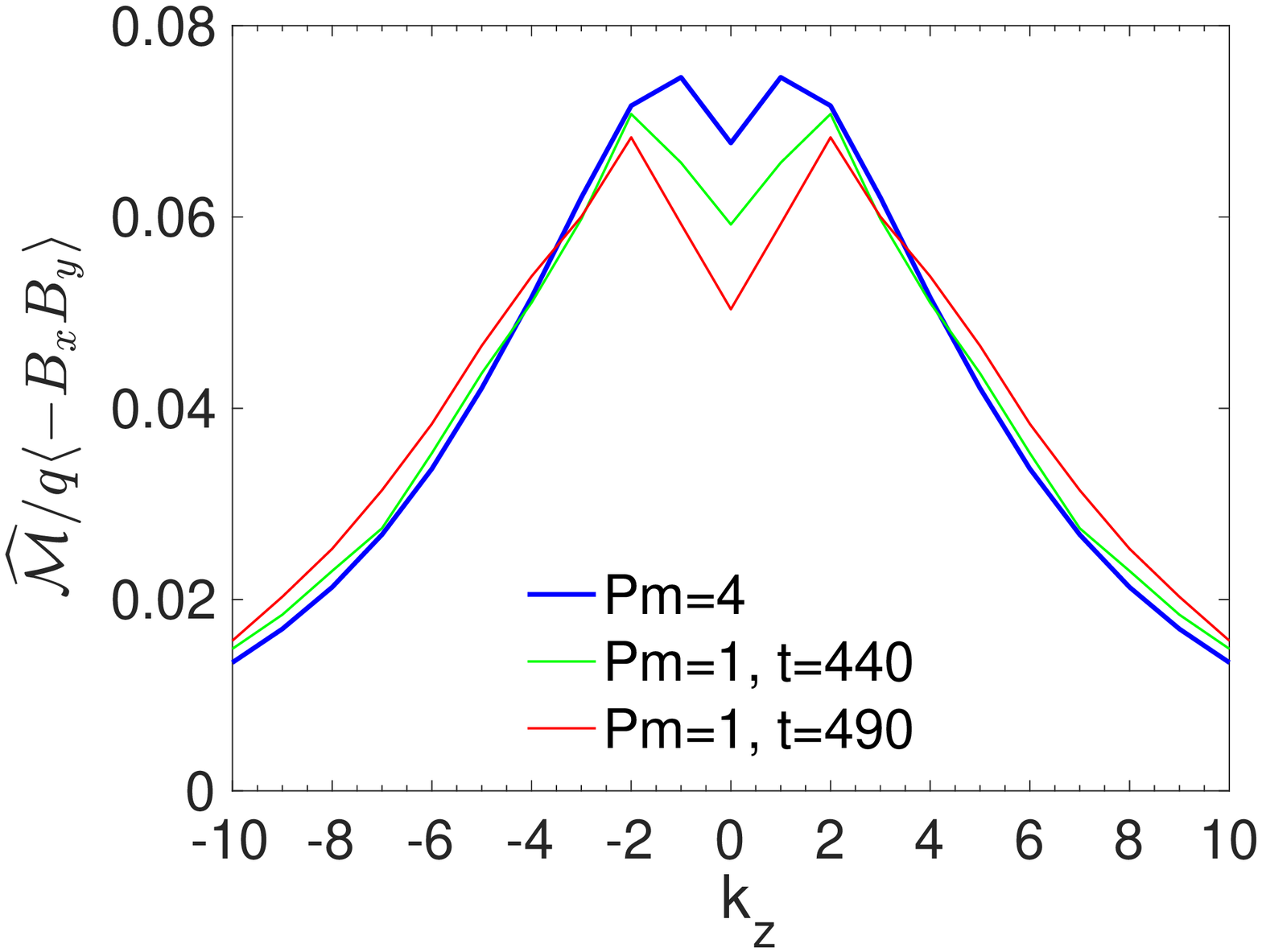}
\caption{Spectra of the radial and azimuthal magnetic field as well as the Maxwell stress integrated in $(k_x,k_y)$-slices vs. $k_z$ and normalized by their respective total values, $\langle B_x^2\rangle$, $\langle B_y^2\rangle$, $q\langle -B_xB_y\rangle$. They are averaged in time over the quasi-steady state for ${\rm Pm}=4$, and in the vicinity of two time moments $t=440$ and $t=490$ in the decaying case for ${\rm Pm}=1$ (see text). In the latter case, all these normalized spectra decrease with time mostly at small $|k_z|\leq 3$, but slightly increase at larger $|k_z|$.} \label{fig:bx_by_M_vs_kz_Pm4_1}
\end{figure*}

\begin{figure*}
\includegraphics[width=0.247\textwidth]{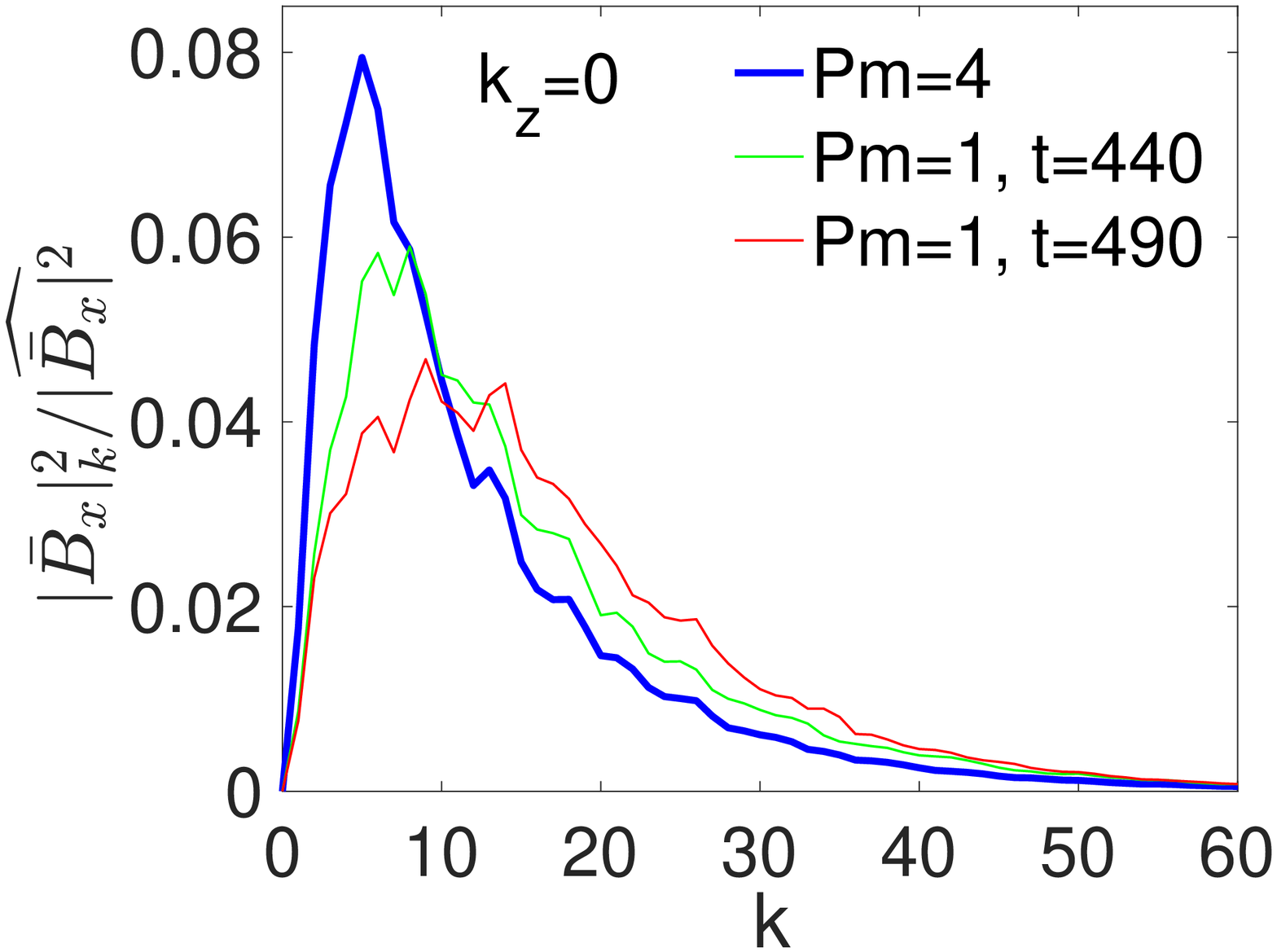}
\includegraphics[width=0.247\textwidth]{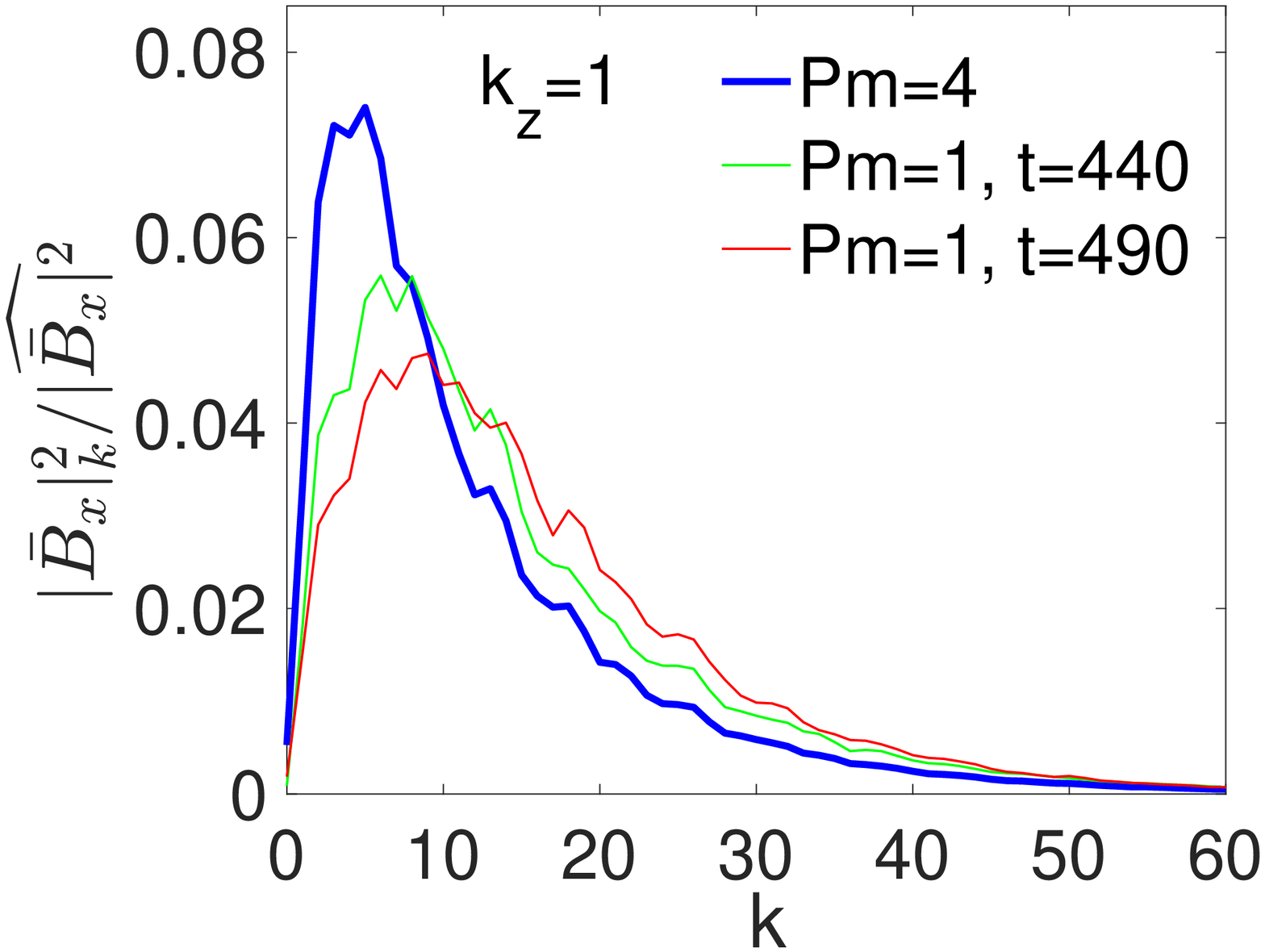}
\includegraphics[width=0.247\textwidth]{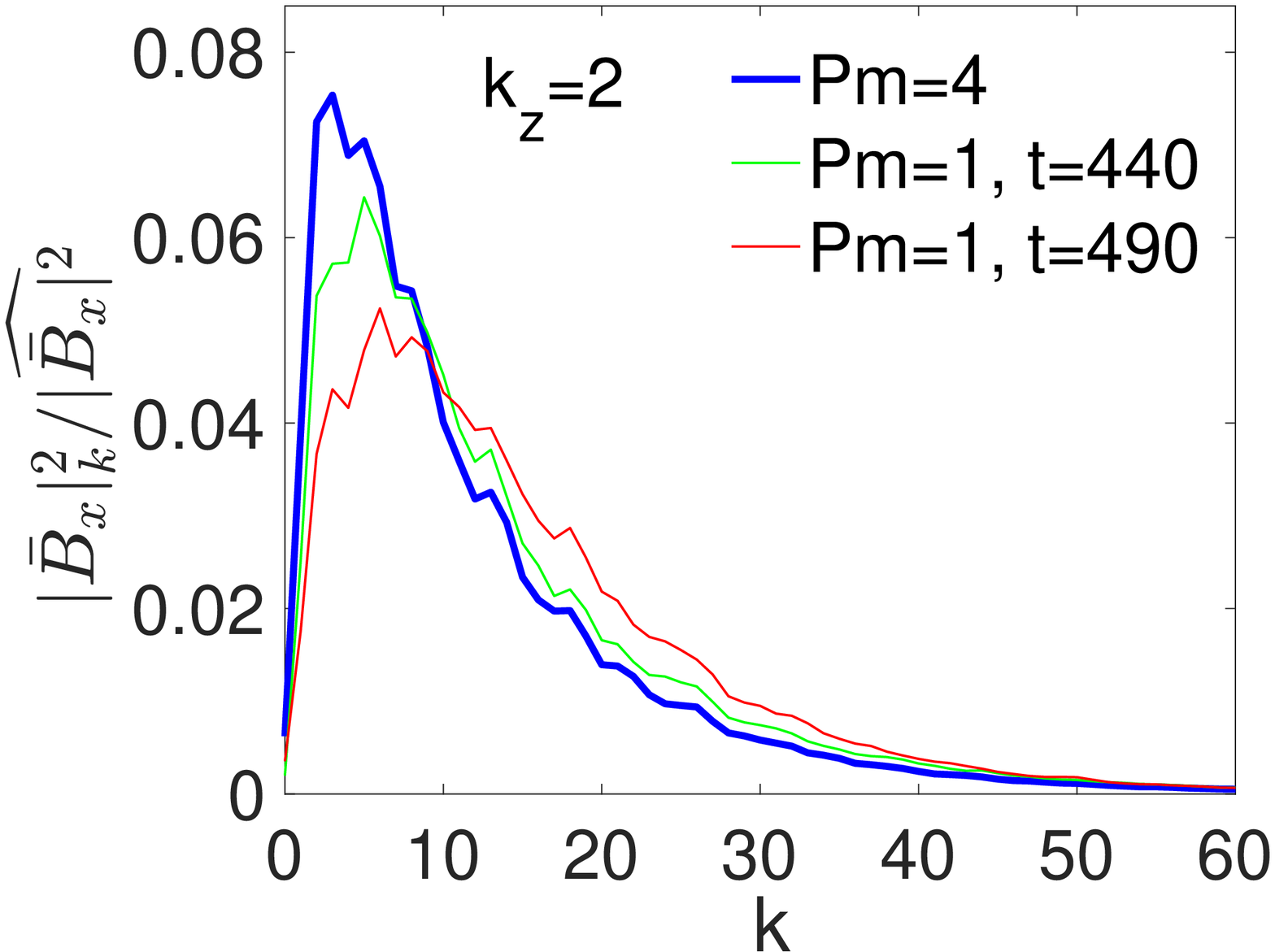}
\includegraphics[width=0.247\textwidth]{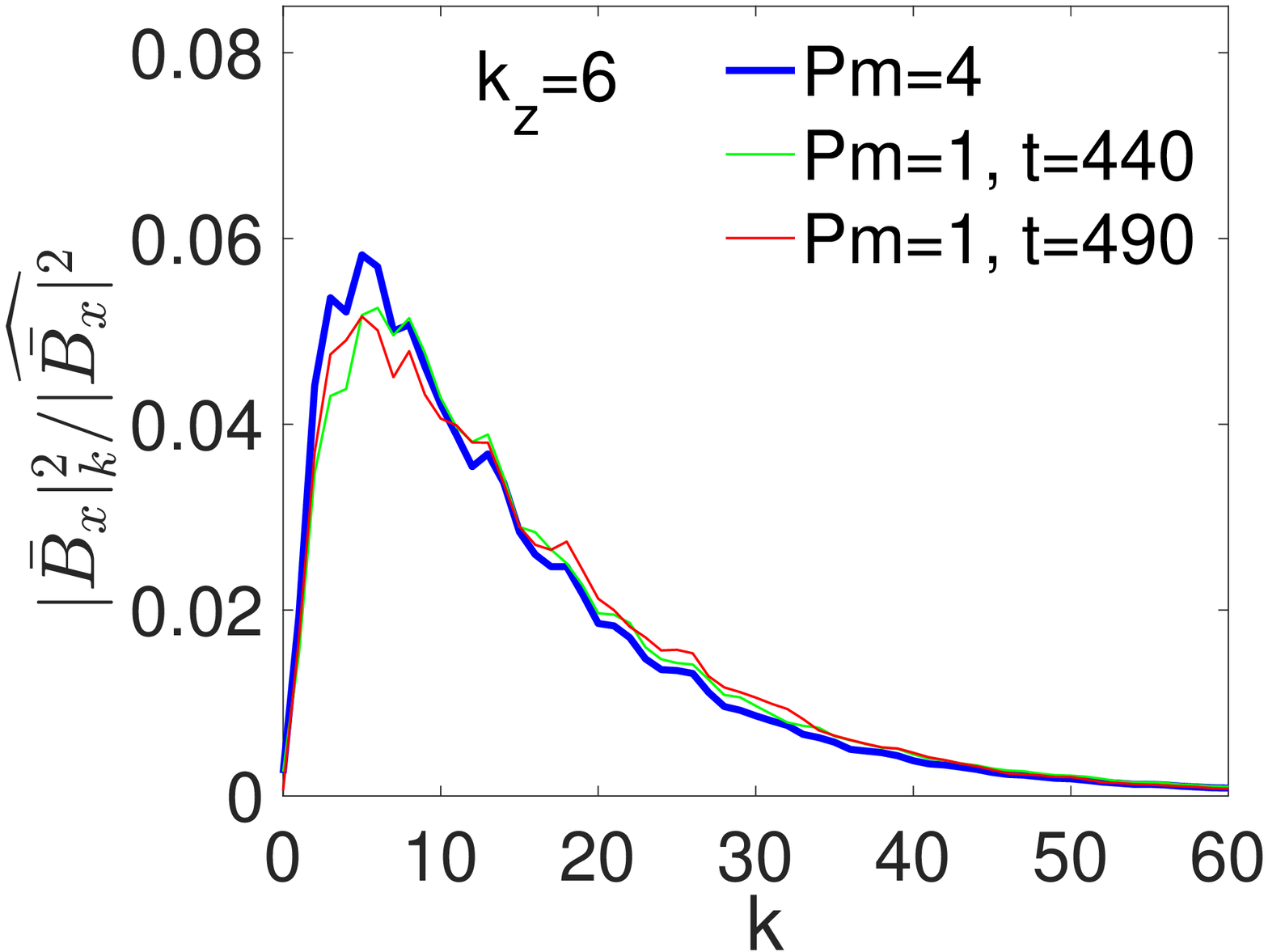}
\includegraphics[width=0.247\textwidth]{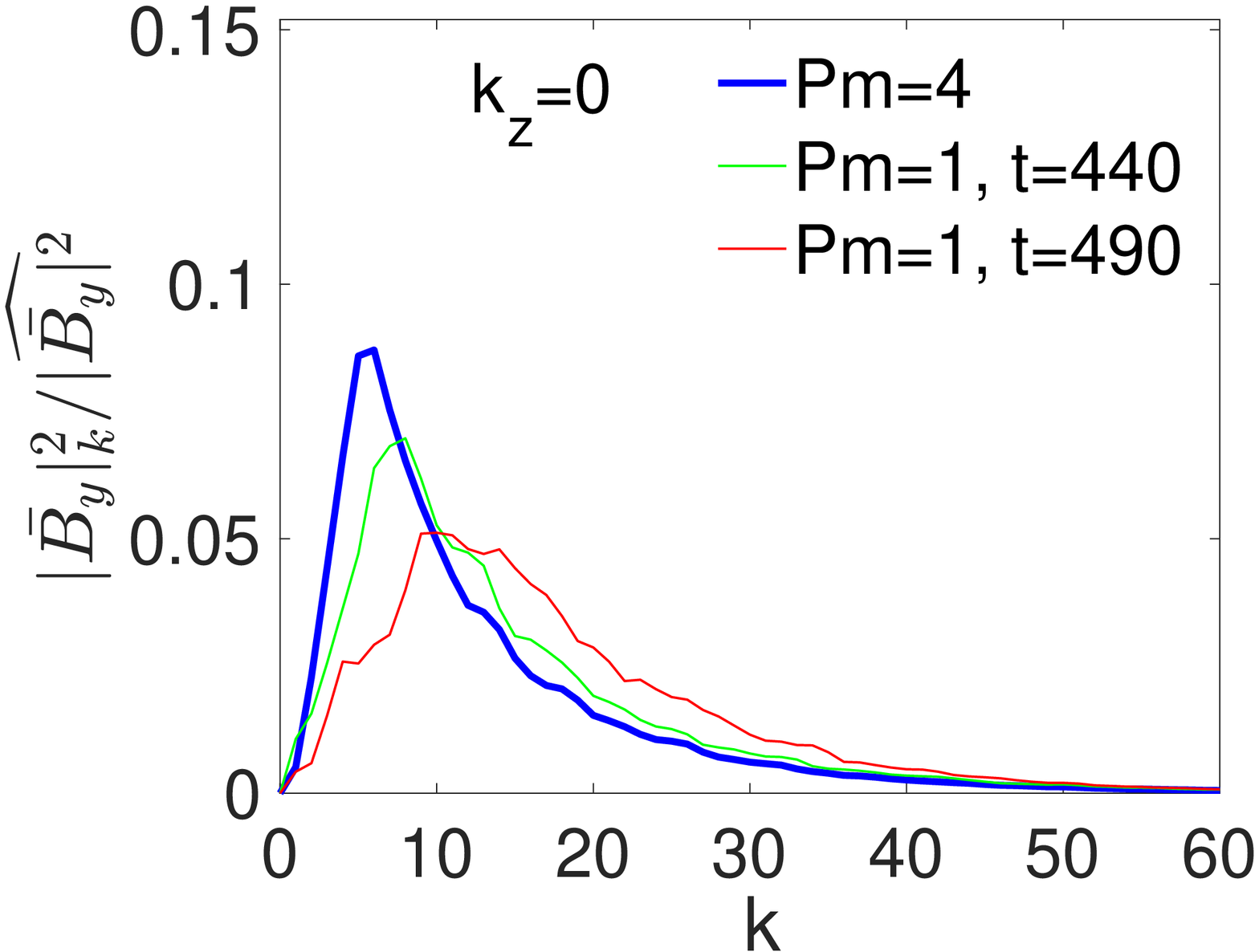}
\includegraphics[width=0.247\textwidth]{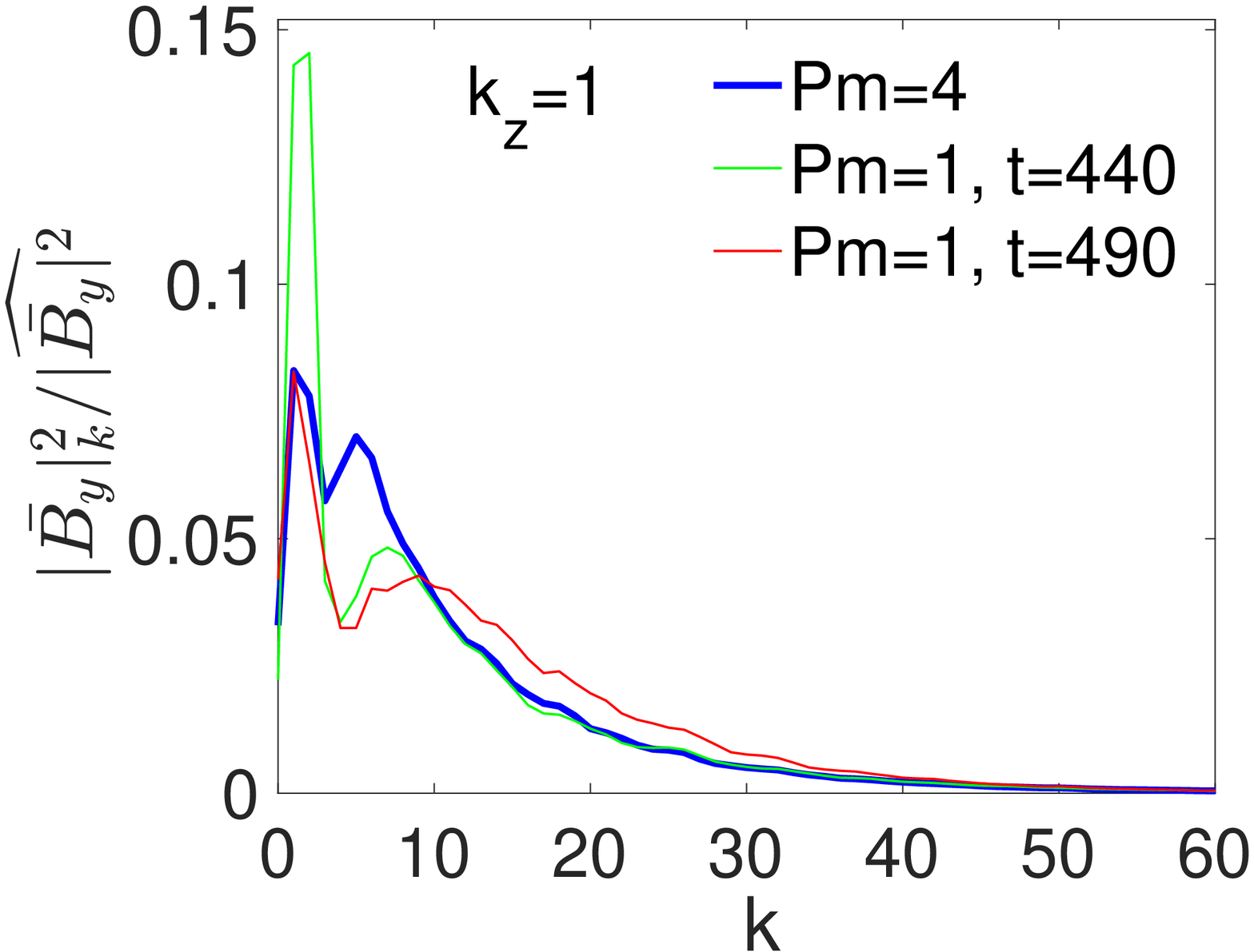}
\includegraphics[width=0.247\textwidth]{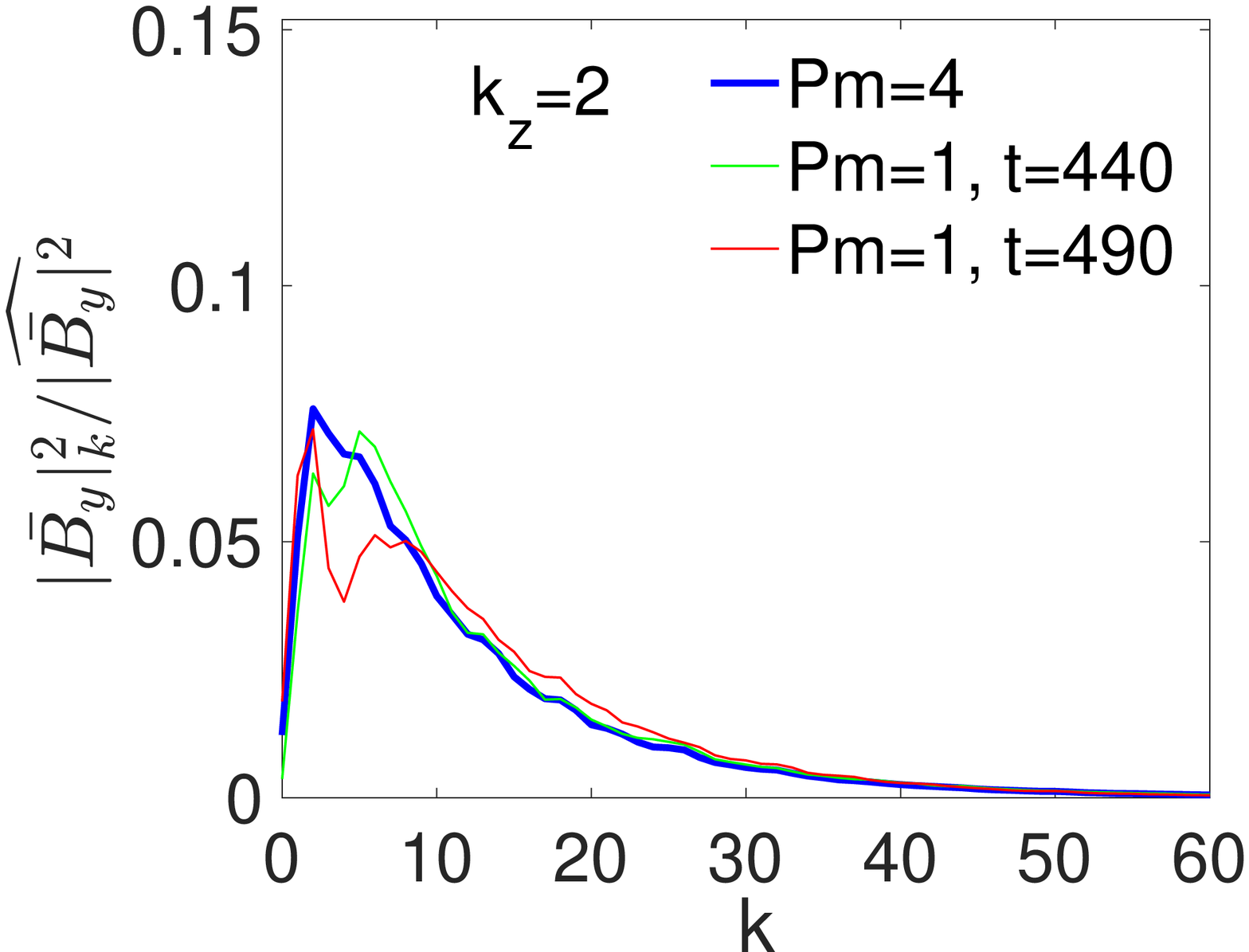}
\includegraphics[width=0.247\textwidth]{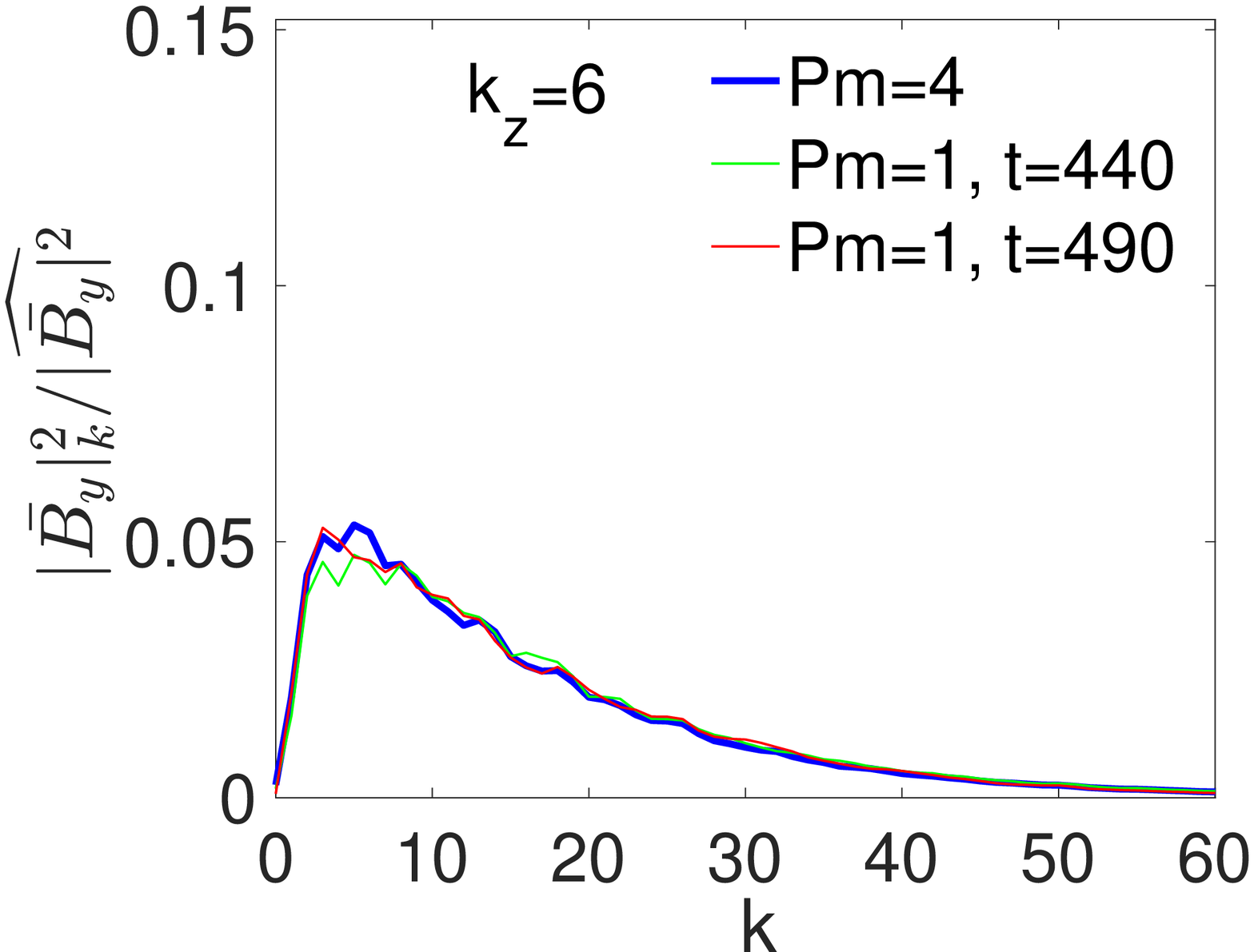}
\includegraphics[width=0.247\textwidth]{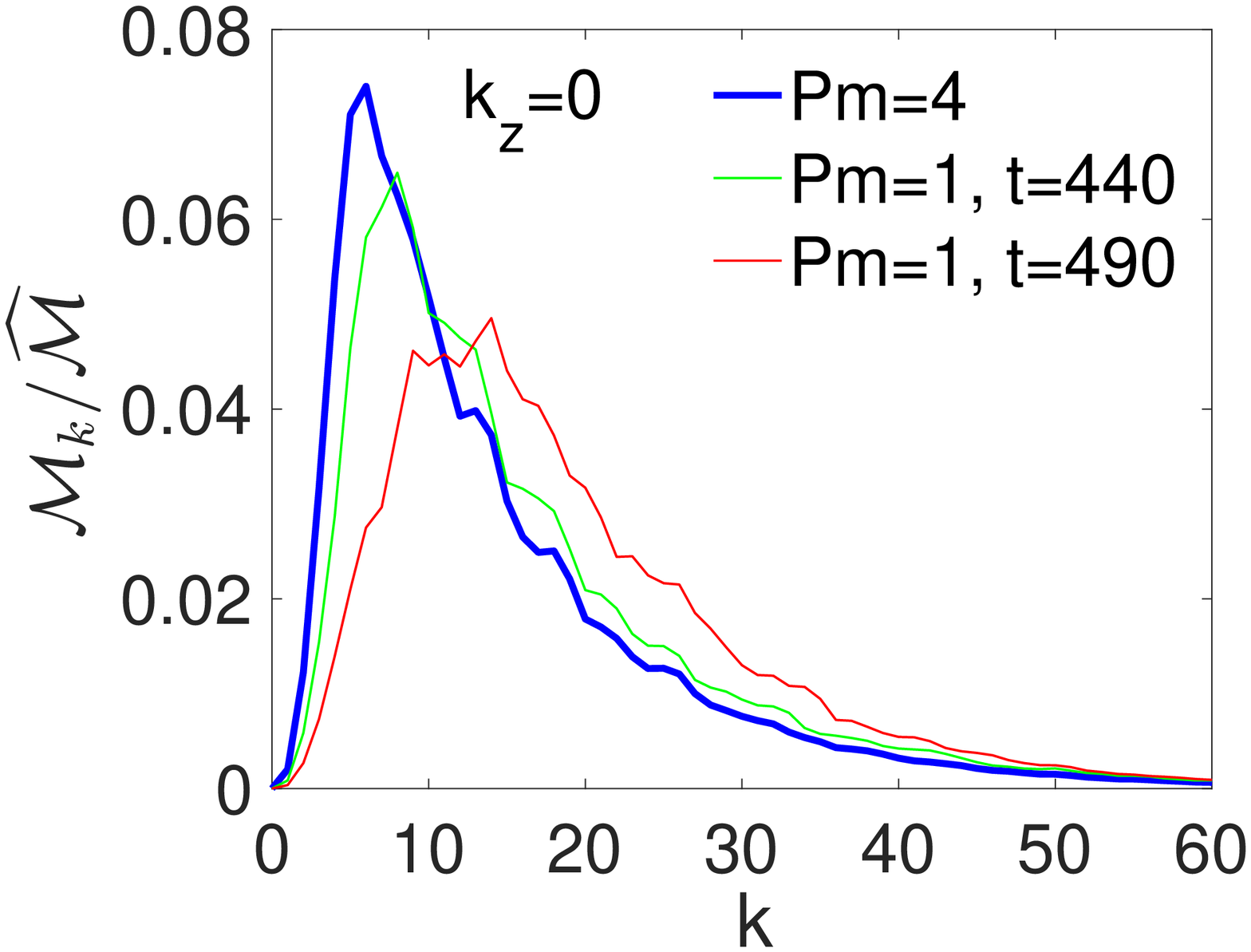}
\includegraphics[width=0.247\textwidth]{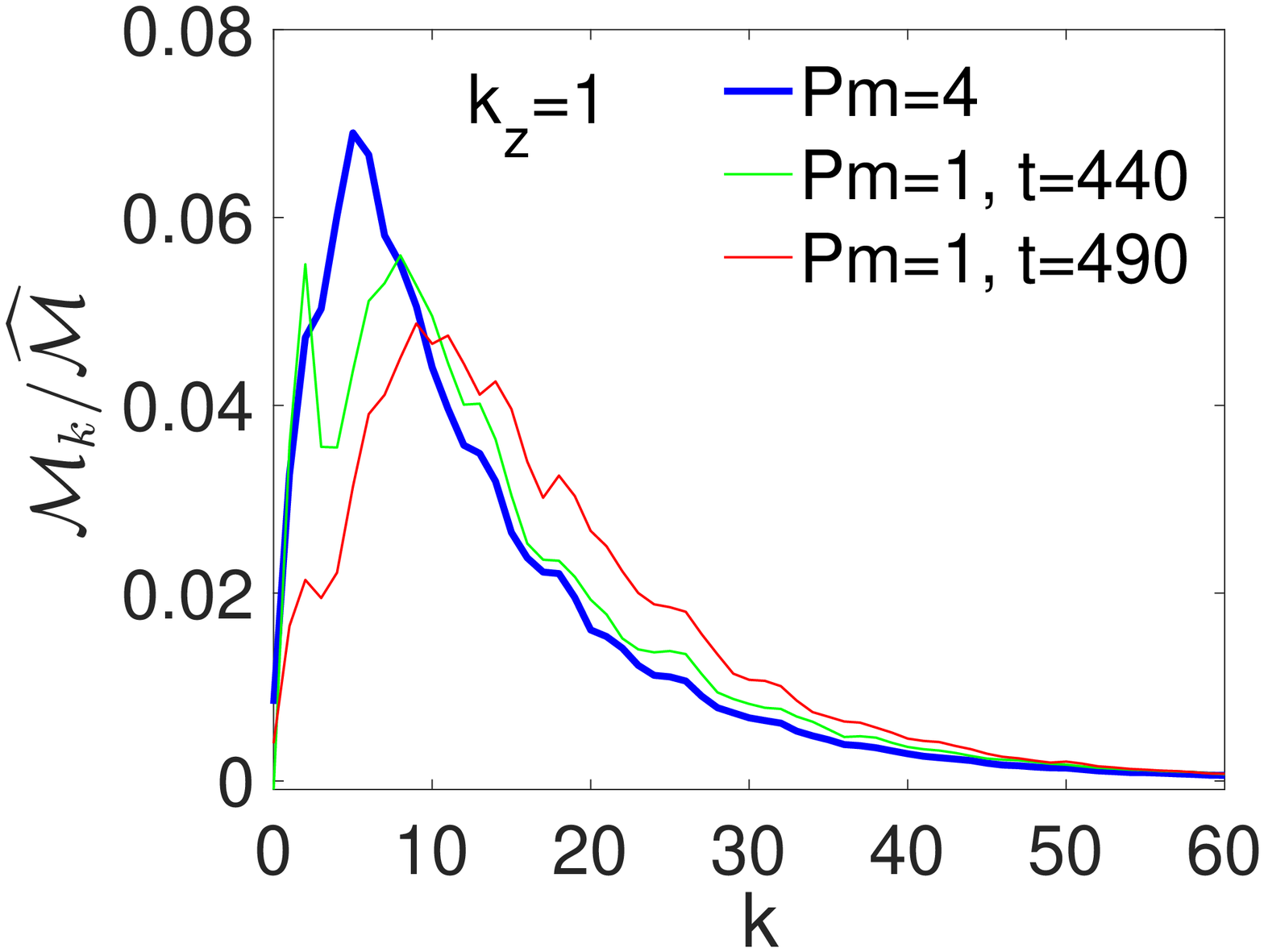}
\includegraphics[width=0.247\textwidth]{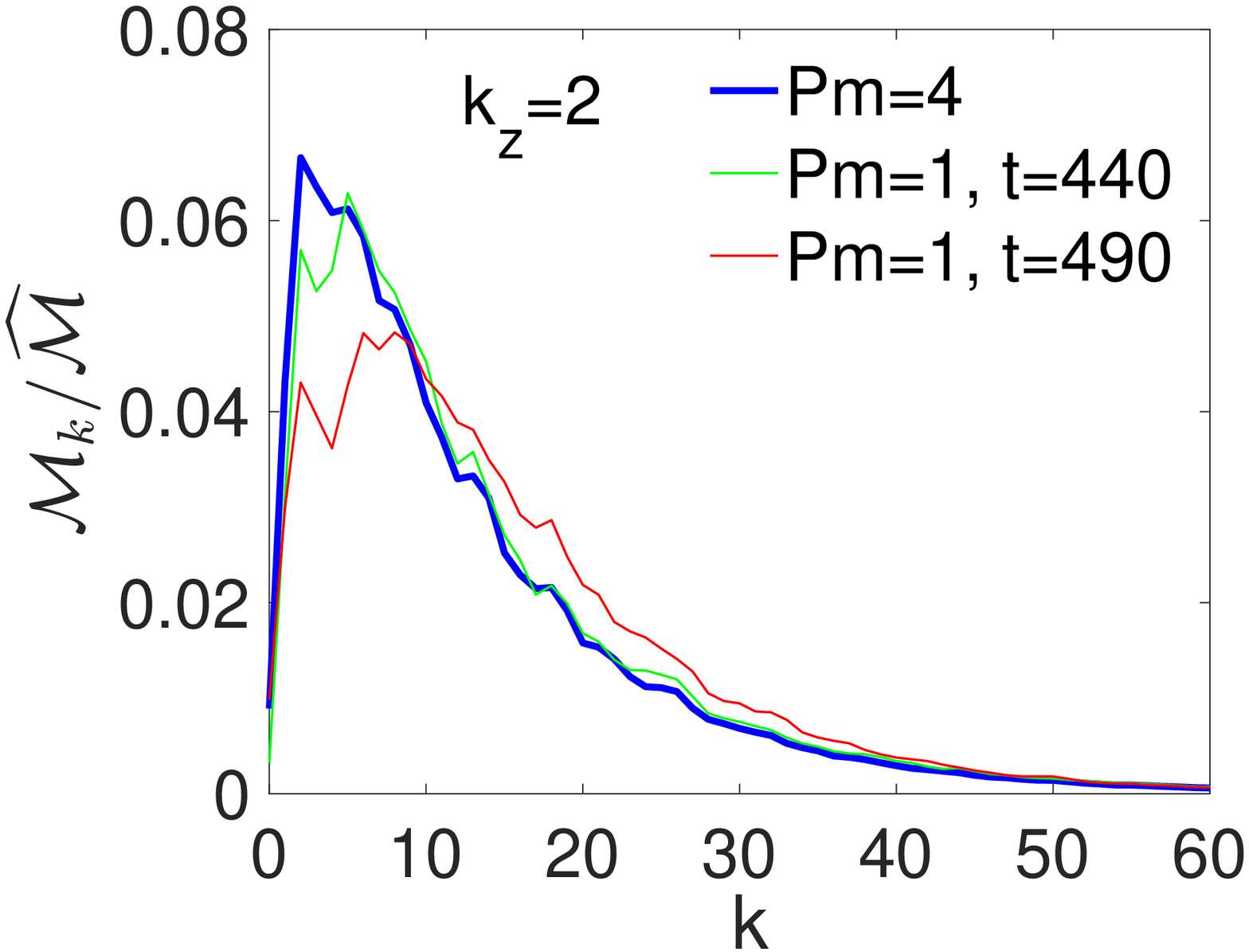}
\includegraphics[width=0.247\textwidth]{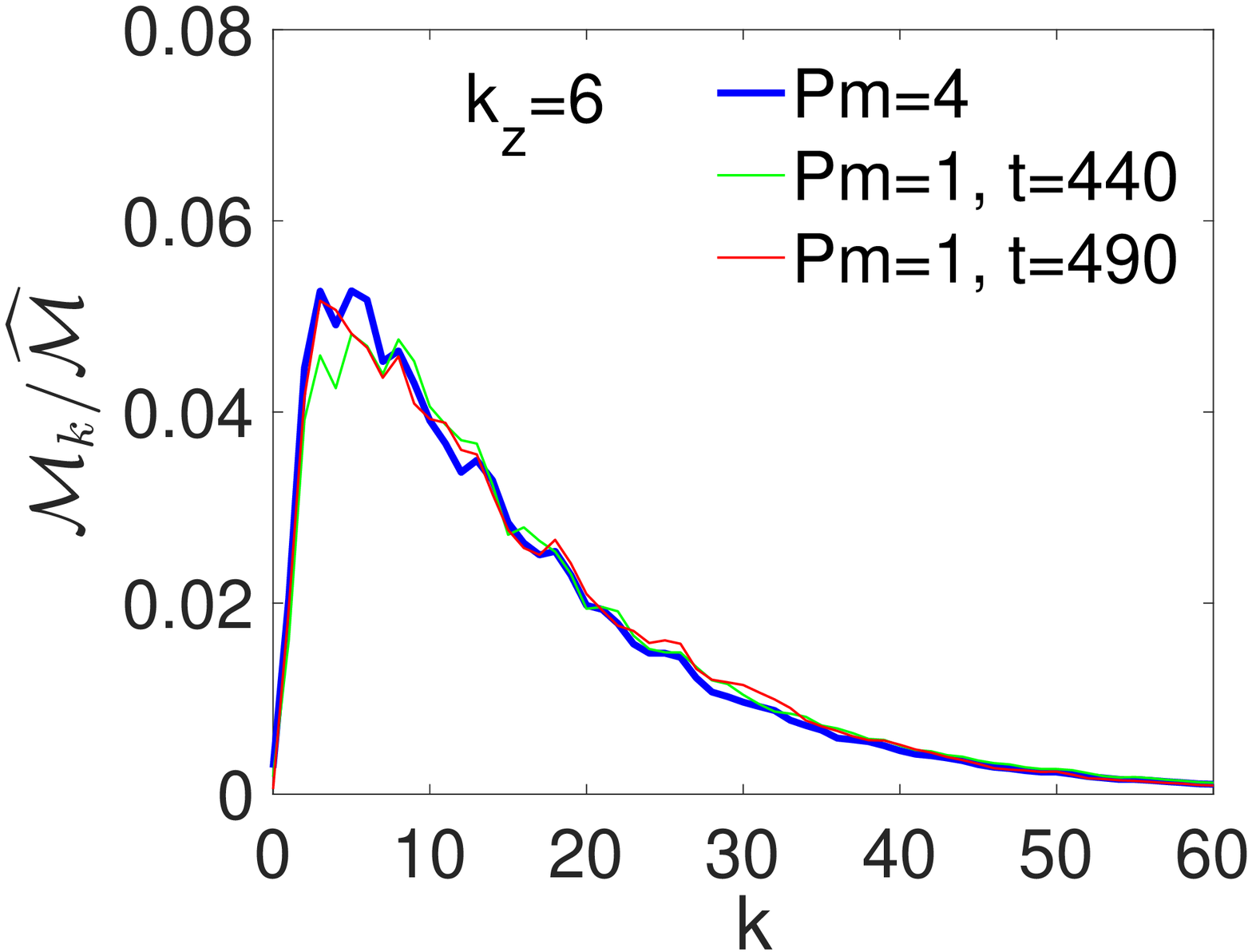}
\caption{Spectra of the radial and azimuthal
magnetic field as well as the Maxwell stress ringed-averaged in $(k_x,k_y)$-slices vs. $k=(k_x^2+k_y^2)^{1/2}$ and normalized by respective, integrated in these slices, values $\widehat{|\bar{B}_x|^2}$, $\widehat{|\bar{B}_y|^2}$, $\widehat{\cal M}$, at different $k_z=0,1,2$ and larger $k_z=6$ for $Pm=4$ and $Pm=1$ (time-averages are done in the same manner as in Figure \ref{fig:bx_by_M_vs_kz_Pm4_1}). At $Pm=1$, all these spectra decrease with time in the vital area $k\lesssim k_0= 10$, but increase at larger $k\gtrsim k_0$ mainly for $k_z=0,1,2$, so there is a flux -- direct cascade -- of these quantities out of the vital area towards larger wavenumbers. However, at $k_z=1$, there is additionally some accumulation of power in the large-scale azimuthal field at $k=0$. At higher, $k_z=6$, however, the shape of these normalized spectra changes neither with time nor $Pm$.}\label{fig:bx_by_M_vs_k_Pm4_1}
\end{figure*}

\begin{figure*}[t]
\includegraphics[width=0.247\textwidth]{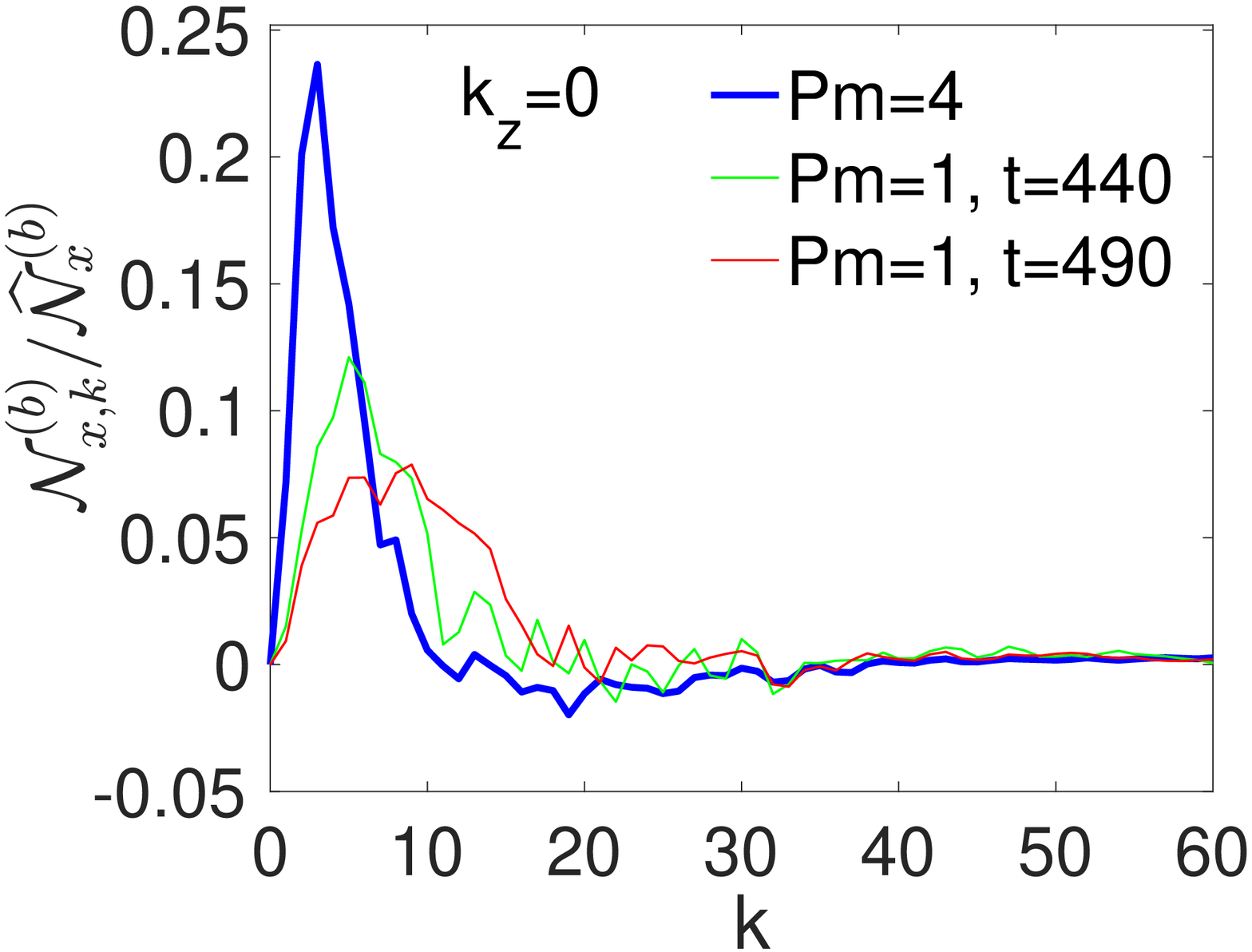}
\includegraphics[width=0.247\textwidth]{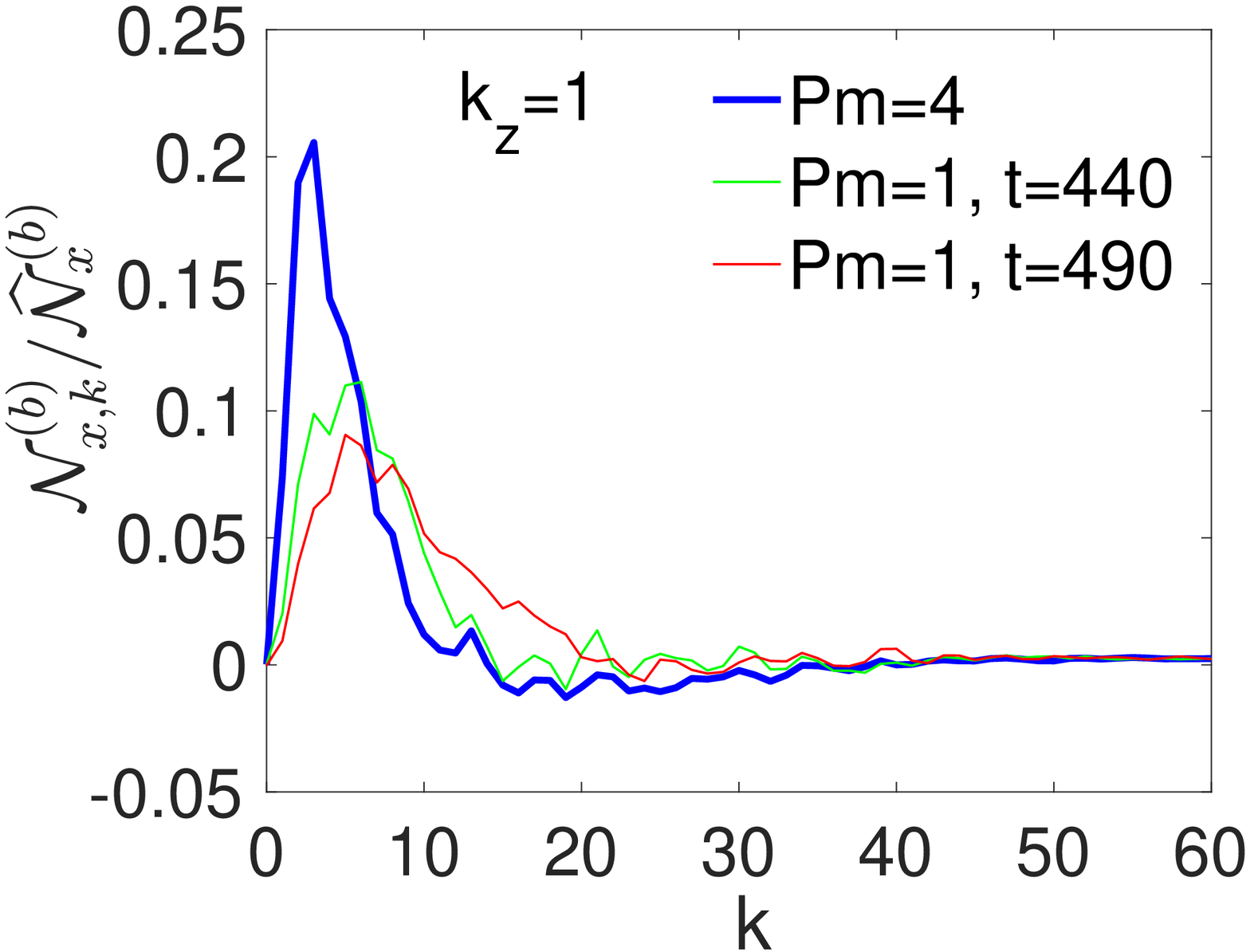}
\includegraphics[width=0.247\textwidth]{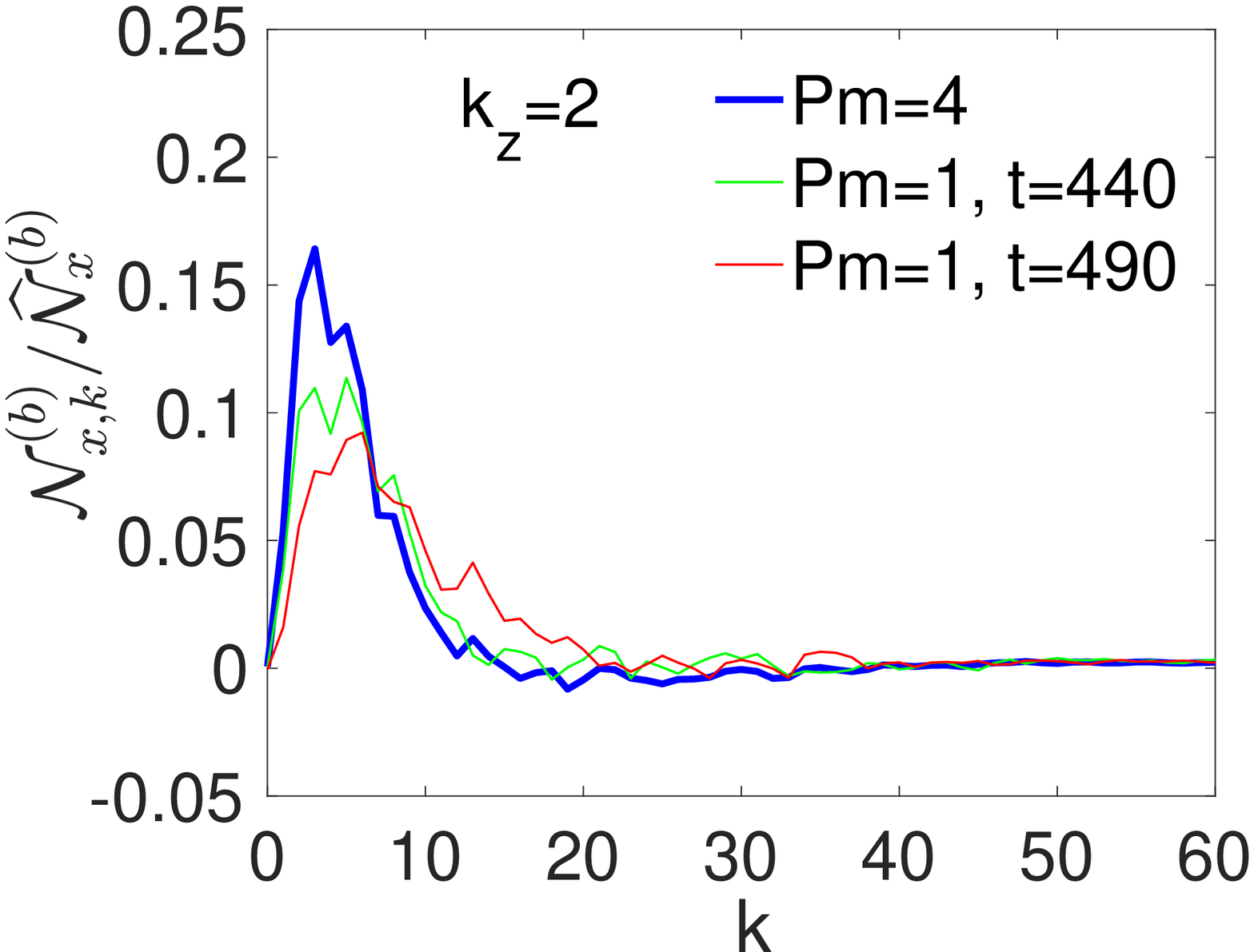}
\includegraphics[width=0.247\textwidth]{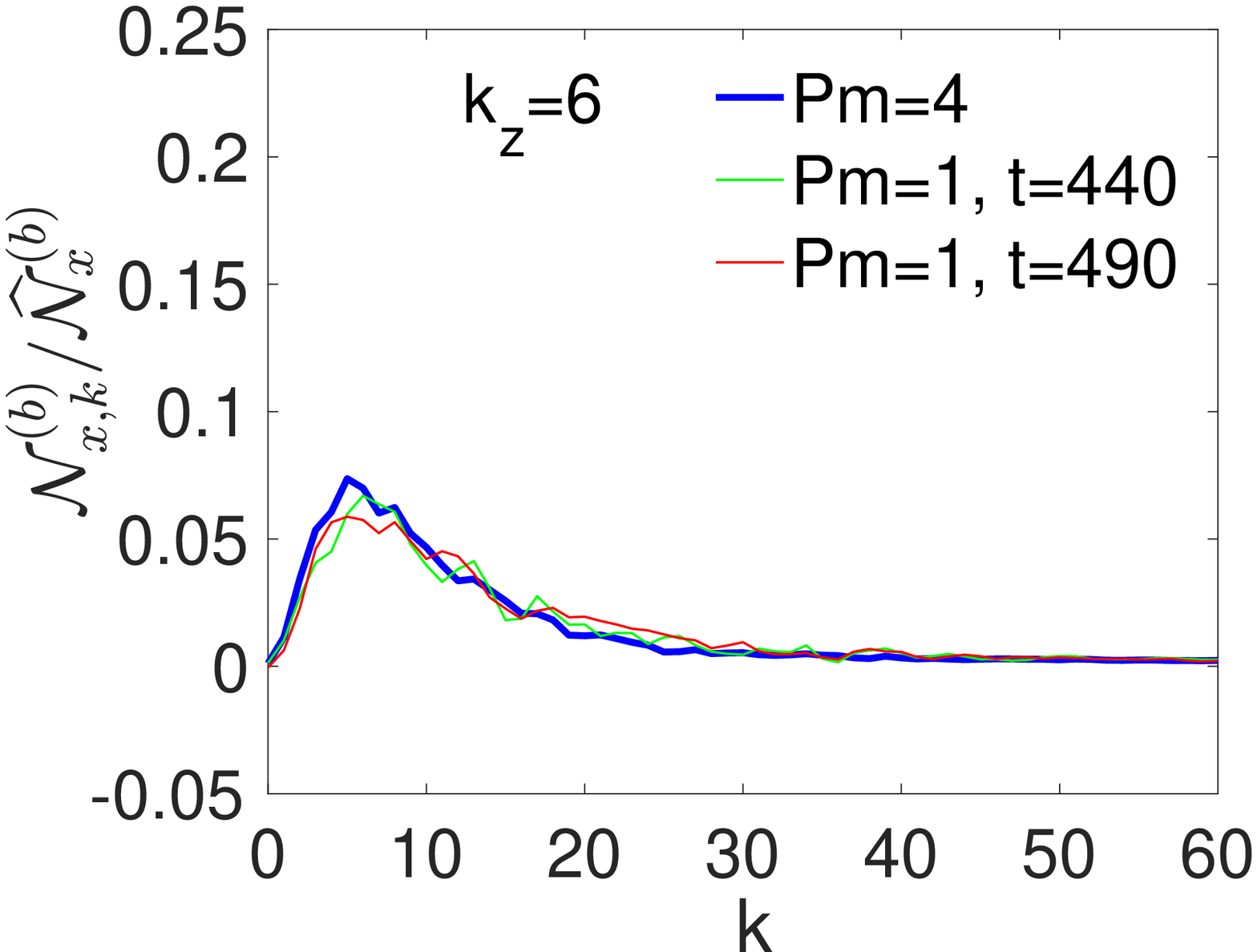}
\caption{Spectrum of the nonlinear term ${\cal N}^{(b)}_x$ ringed averaged in $(k_x,k_y)$-slices vs. $k=(k_x^2+k_y^2)^{1/2}$ and normalized by its integrated in these slices value $\widehat{\cal N}_x^{(b)}$ at different $k_z=0,1,2$ and larger $k_z=6$ for $Pm=4$ and $Pm=1$ (time-averages are done in the same manner as in Figure \ref{fig:bx_by_M_vs_kz_Pm4_1}). At ${\rm Pm}=1$, the action of this term decreases with time in the vital area $k\lesssim k_0=10$, but increases at larger $k\gtrsim k_0$ mainly at $k_z=0,1,2$. By contrast, at ${\rm Pm}=4$ it is more concentrated at smaller $k$ in the vital area. This indicates an enhanced direct cascade along $k$ with decreasing ${\rm Pm}$ at smaller $k_z$. At larger $k_z=6$, however, the shape of the normalized spectrum of the nonlinear term changes neither time nor $Pm$.}\label{fig:Nbx_vs_k_Pm4_1}
\end{figure*}

The proposed self-sustaining scheme share some similarities with the nonlinear 3D MRI-dynamo cyclic solutions reported by \citet{Herault_etal11,Riols_etal15,Riols_etal17}, despite the fact that these dynamo states are more regular in space and time (not fully turbulent), with much reduced number of active modes. A detailed comparison of our self-sustaining scheme with that proposed in those papers is discussed in subsection 6.1.

\section{Dependence of the turbulence sustenance and dynamics on {\rm P\MakeLowercase{m}}} \label{sec:Pmdependence}

In this section, we explore the effects of reducing ${\rm Pm}$ on the nonlinear transfers and hence on magnetic field spectrum, which eventually lead to the decay of the turbulence (Figure \ref{fig:evol_diff_Pm}). We do a  comparative analysis by juxtaposing spectral quantities at ${\rm Pm}=4$, when turbulence is sustained, and at ${\rm Pm}=1$, when it decays fastest, and characterizing the differences between the turbulence dynamics in Fourier space in these two cases. Such an analysis is at first glance analogous to that carried out by  \citet{Lesur_Longaretti11} for nonzero net vertical flux case at different ${\rm Pm}$, who, however, used spherical shell-averaging technique, overlooking the spectral anisotropy and hence the transverse cascade, whose weakening with ${\rm Pm}$, as shown below, is a main cause for the decay of the turbulence. As seen in Figure \ref{fig:evol_diff_Pm}, the turbulence slowly decays from the moment ($t=400$) at which ${\rm Pm}$ is abruptly lowered from 4 to 1. Due to this, it is not possible to do time-averaging of the spectra at ${\rm
Pm}=1$ over an entire evolution, as done for the quasi-steady state
at ${\rm Pm}=4$ in the previous section. In this case, we choose
instead two time moments: $t=440$ in about the middle and $t=490$
near the end of the decay (see Figure \ref{fig:evol_diff_Pm}) and
average the spectral quantities over a short time interval around
these moments in order to filter noise and make spectral quantities
smoother. Since the radial field and its regeneration process are the most important ingredients in the turbulence sustenance, in this section we focus mainly on the behavior of its spectrum and nonlinear term $ {\cal N}_x^{(b)}$ at different ${\rm Pm}$, though we also consider the behavior of the azimuthal field and the Maxwell stress.

Finally, in the decaying turbulence, all the spectra of the magnetic field and dynamical terms decrease in time. This makes it somewhat difficult to see how the shape of these spectral quantities changes with time at ${\rm Pm}=1$ as well as with respect to the quasi-steady ones at ${\rm Pm}=4$. This is important, because changes in the form of spectra with time can give clues on transfer directions in Fourier space when ${\rm Pm}$ is varied. To circumvent this, we compare below normalized spectra that better highlight the differences in the magnetic field and dynamical terms' spectra in the sustained and decaying cases. So, below we describe how the spectra of the main quantities change with decreasing ${\rm Pm}$ and clarify its physics by examining the behavior of the nonlinear transfers.

\begin{figure*}
\includegraphics[width=0.32\textwidth,height=0.2\textwidth]{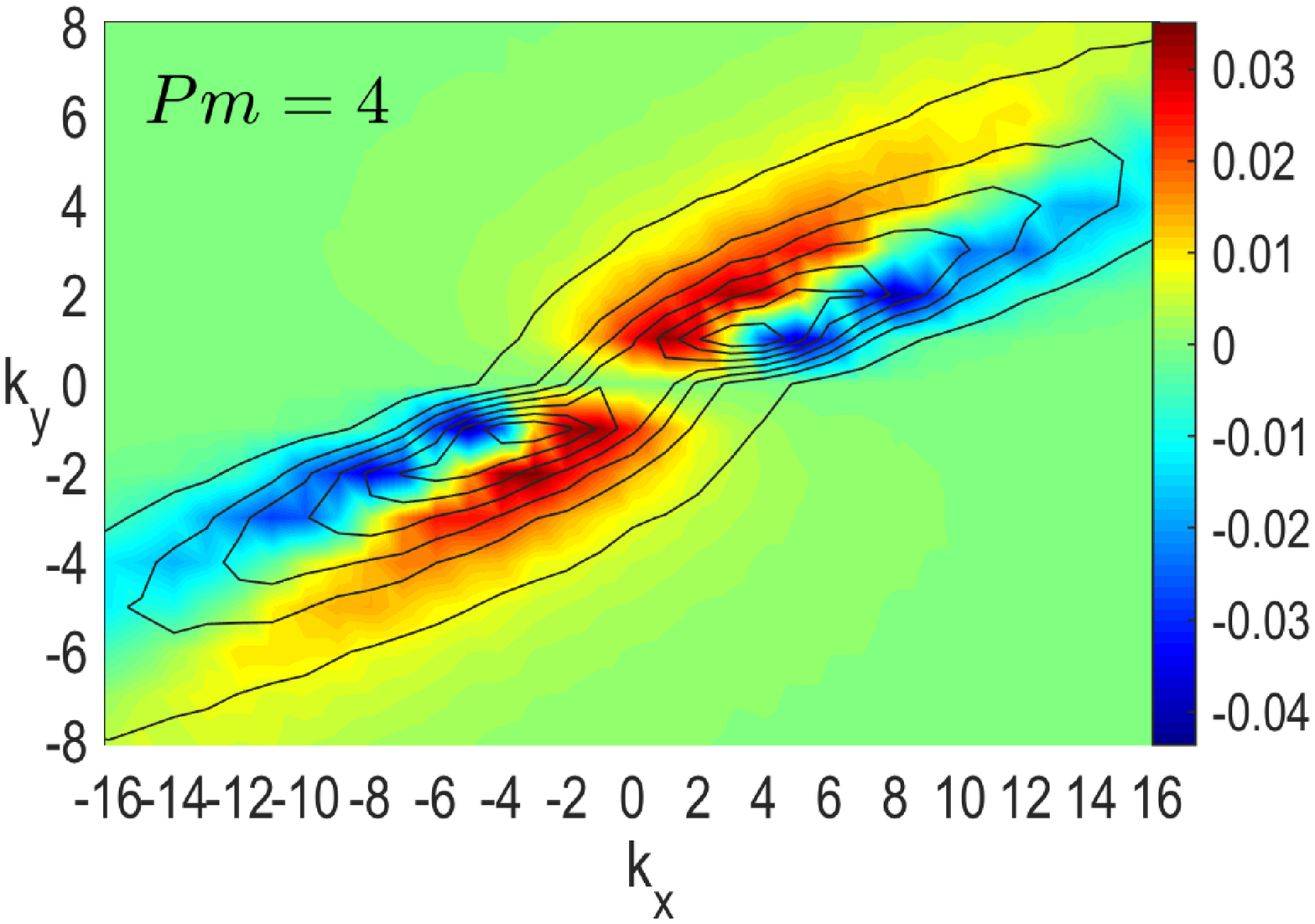}
\includegraphics[width=0.32\textwidth,height=0.2\textwidth]{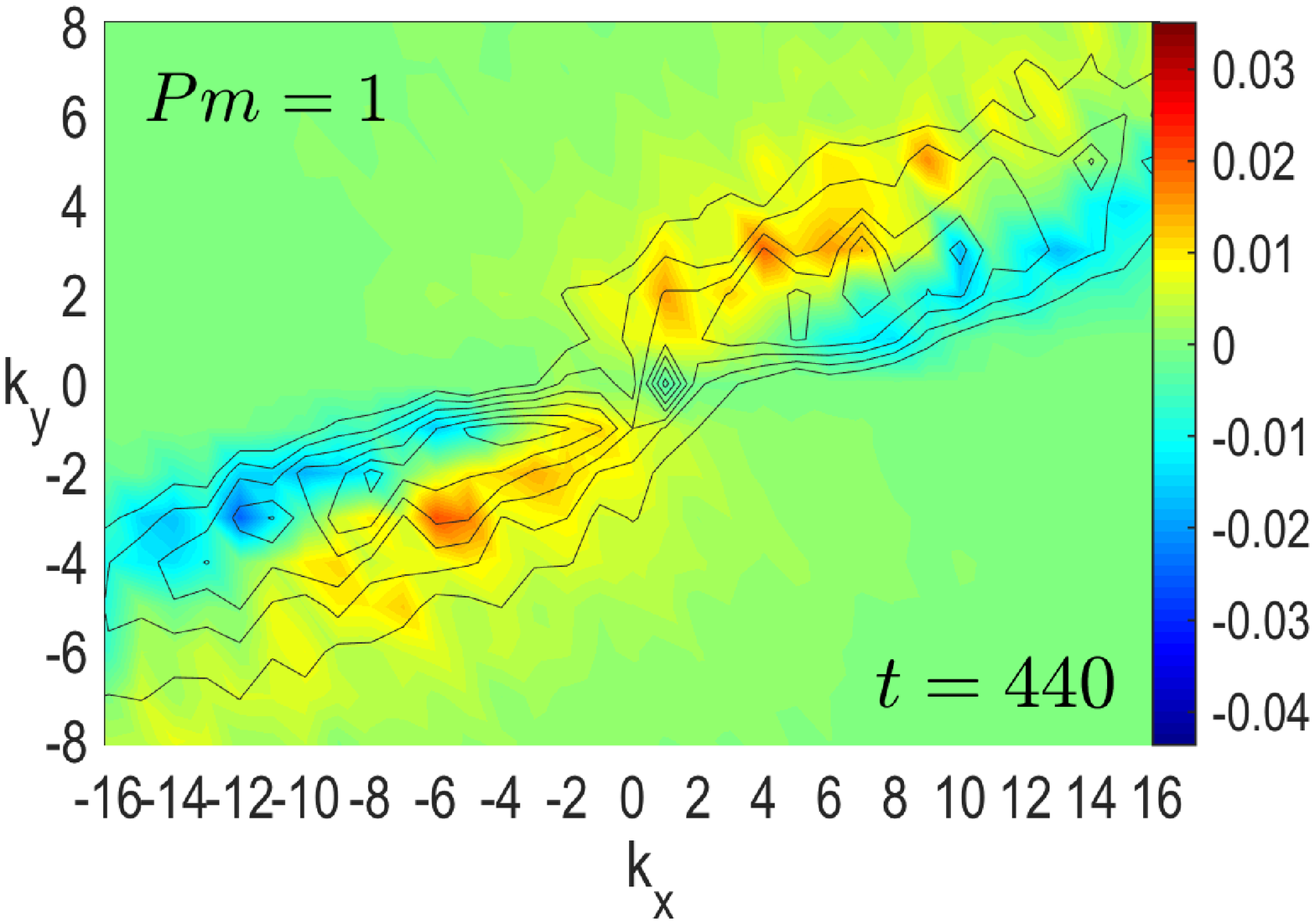}
\includegraphics[width=0.32\textwidth,height=0.2\textwidth]{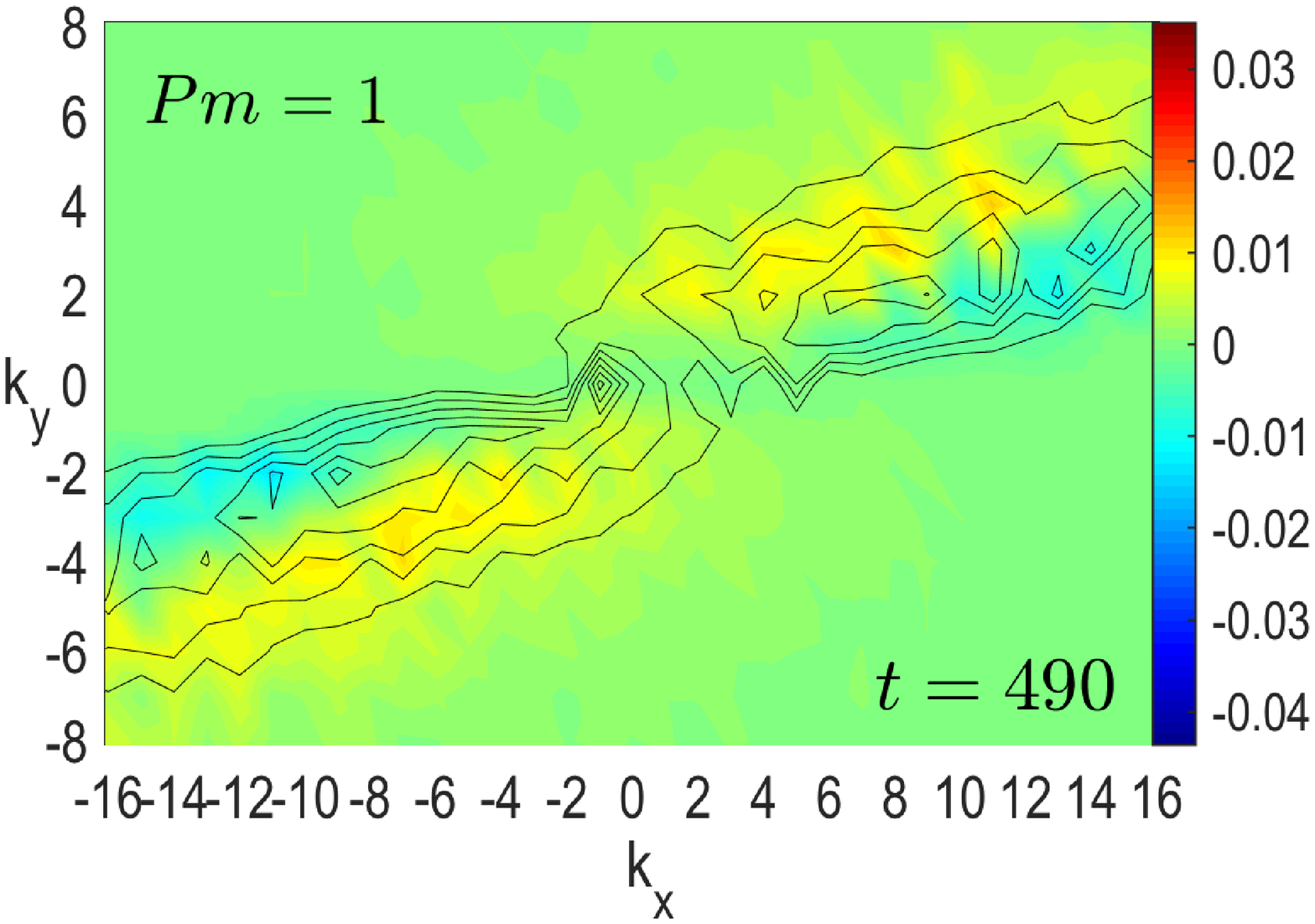}
\caption{Normalized spectrum of the nonlinear term for the radial field, ${\cal N}_x^{(b)}/\hat{\cal N}^{(b)}_x$, in $(k_x,k_y)$-slice at $k_z=1$
for $Pm=4$ (left) and $Pm=1$, for two moments $t=440$ (middle) and
$t=490$ (right). The contours of the corresponding $|\bar{B}_ x|$ are also drawn on this spectrum. In contrast to the sustained $Pm=4$ case, at $Pm=1$ this nonlinear term is less concentrated in the vital area $|k_x|\lesssim k_0=10, |k_y|\lesssim 3$ and spreads out with time to larger $k_x$ and $k_y$, decreasing in amplitude. The color intensity of red, yellow and blue areas also fades away, implying the reduction of the transverse cascade efficiency.}\label{fig:Nbx_vs_kxky_Pm4_1}
\end{figure*}

\begin{figure*}[t!]
\centering
\includegraphics[width=0.32\textwidth,height=0.2\textwidth]{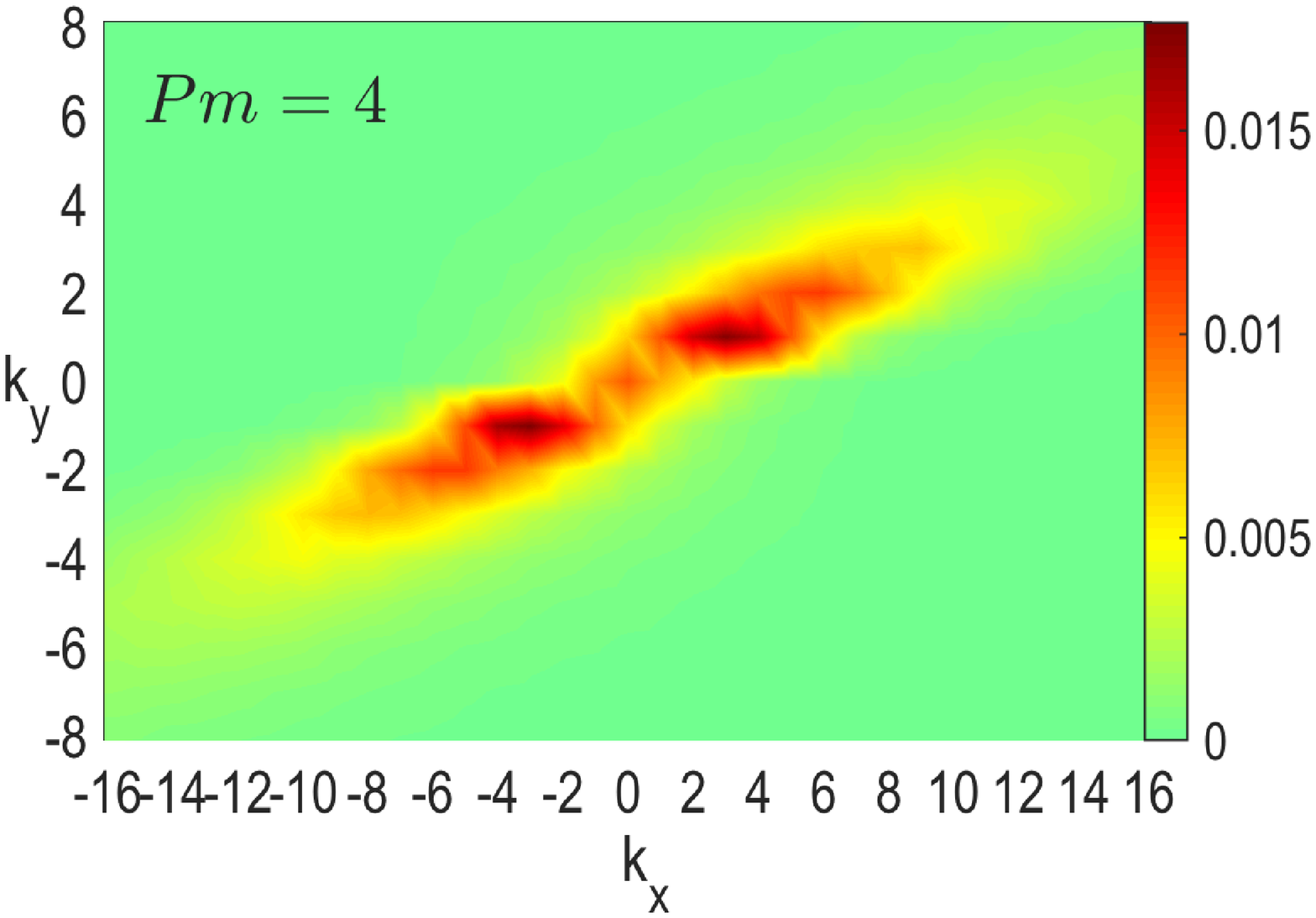}
\includegraphics[width=0.32\textwidth,height=0.2\textwidth]{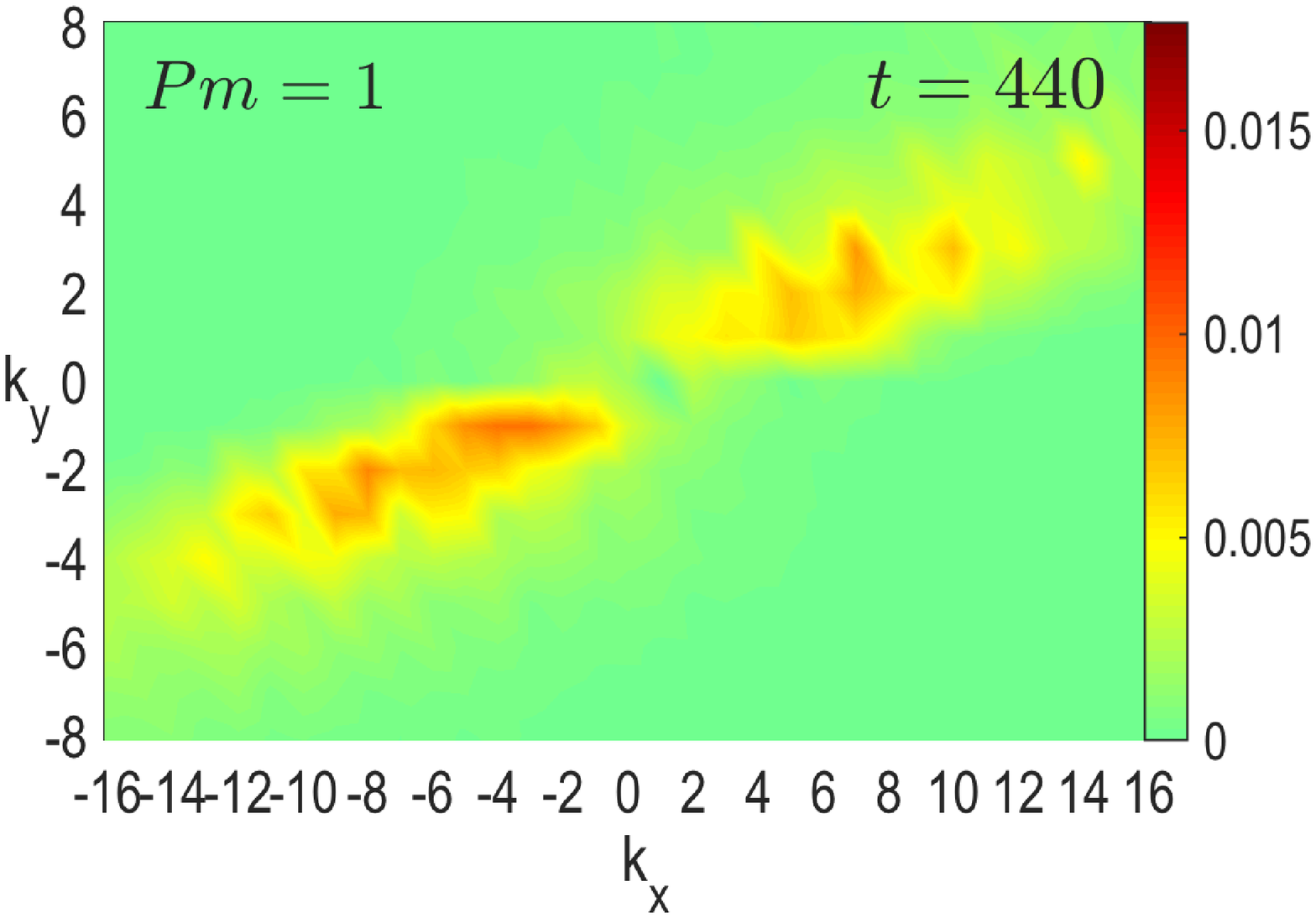}
\includegraphics[width=0.32\textwidth,height=0.2\textwidth]{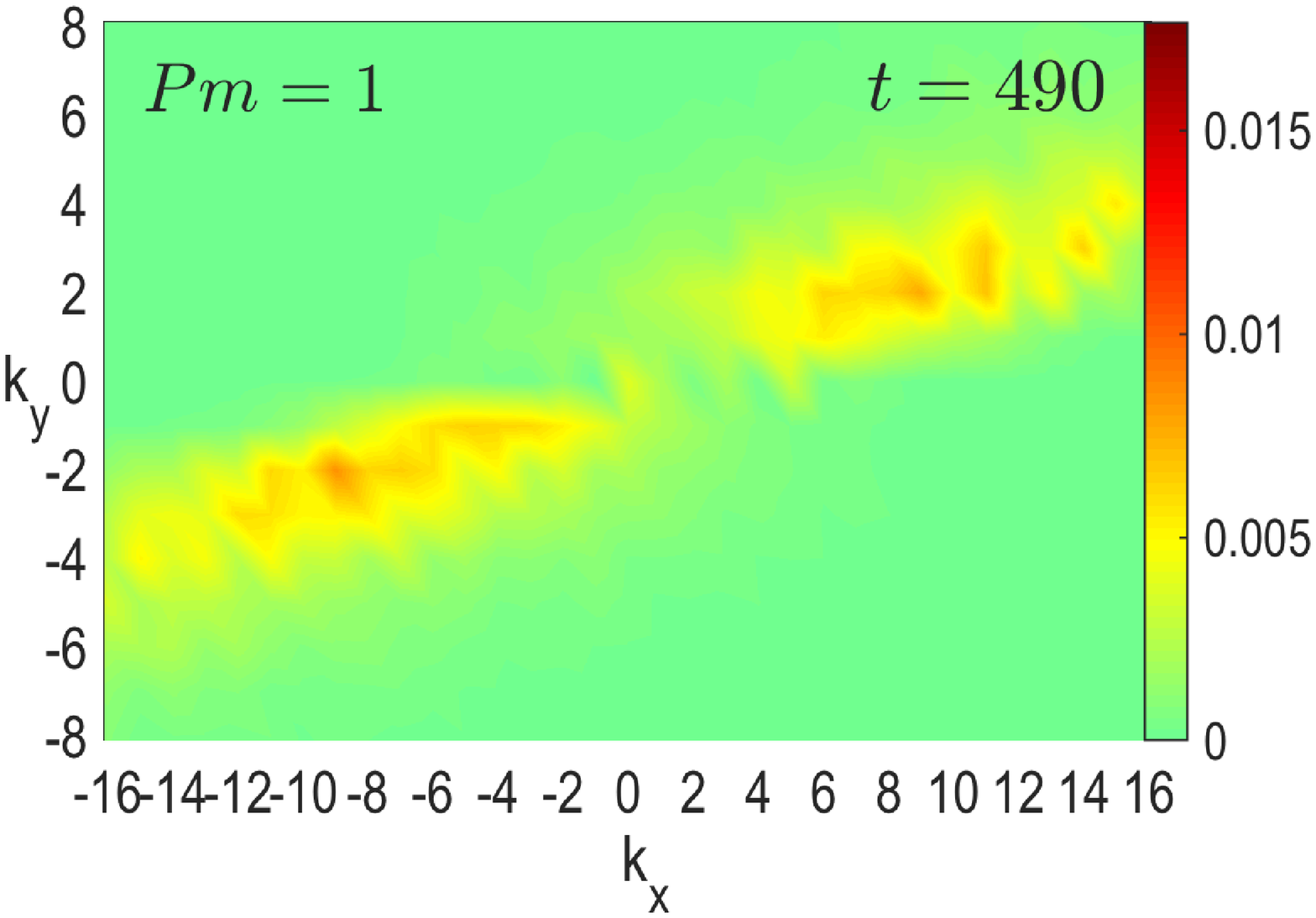}
\includegraphics[width=0.32\textwidth,height=0.2\textwidth]{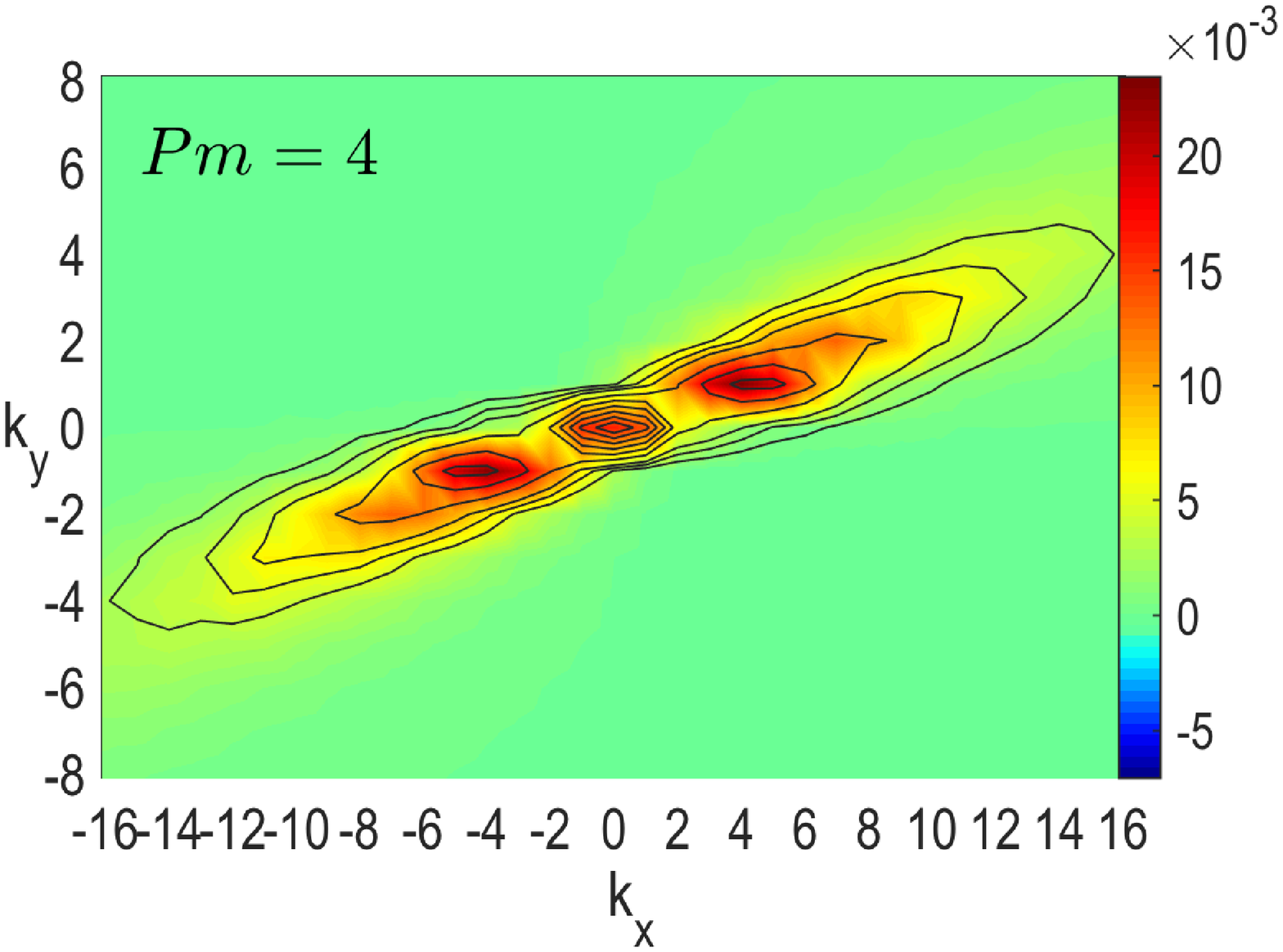}
\includegraphics[width=0.32\textwidth,height=0.2\textwidth]{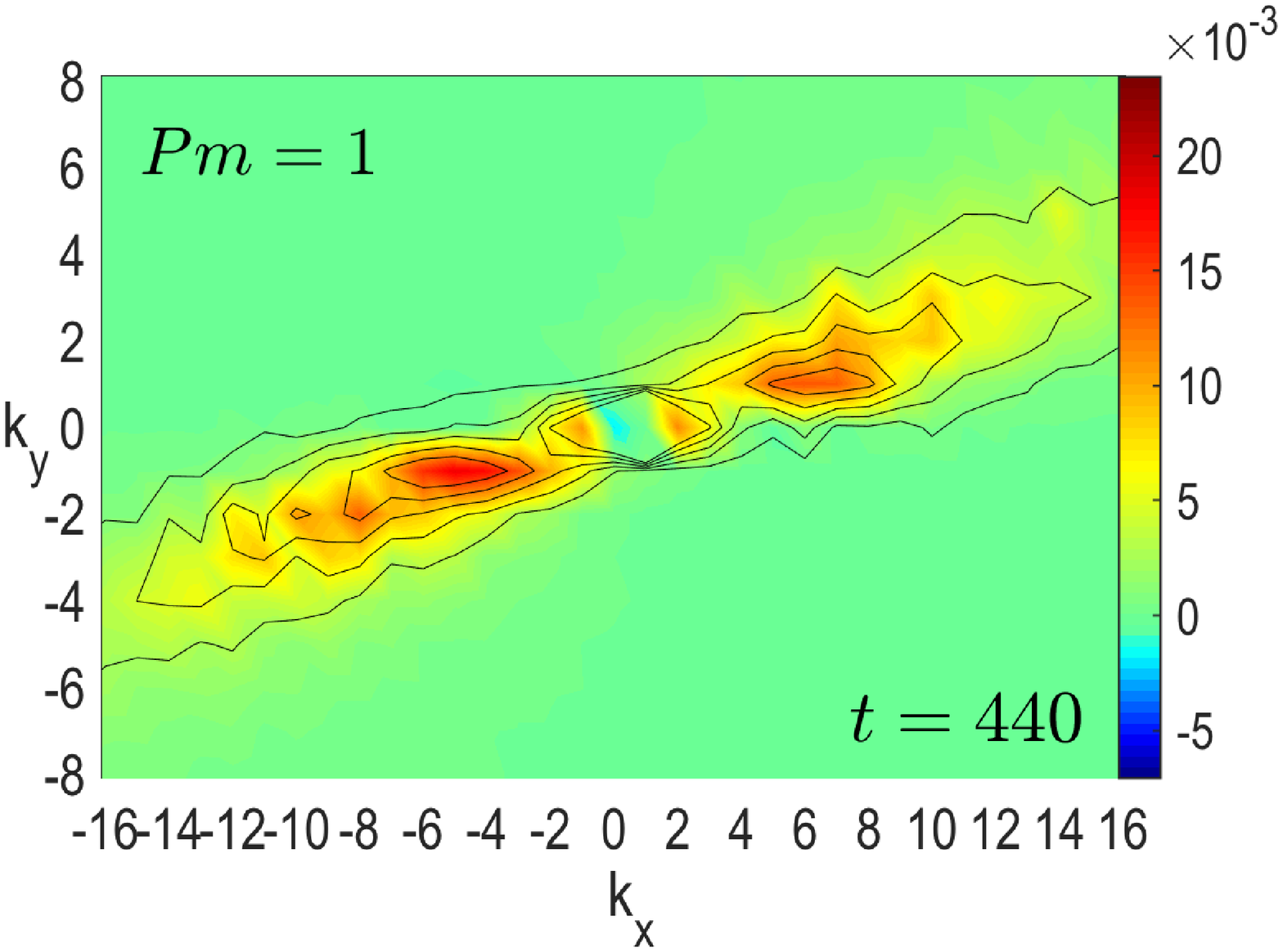}
\includegraphics[width=0.32\textwidth,height=0.2\textwidth]{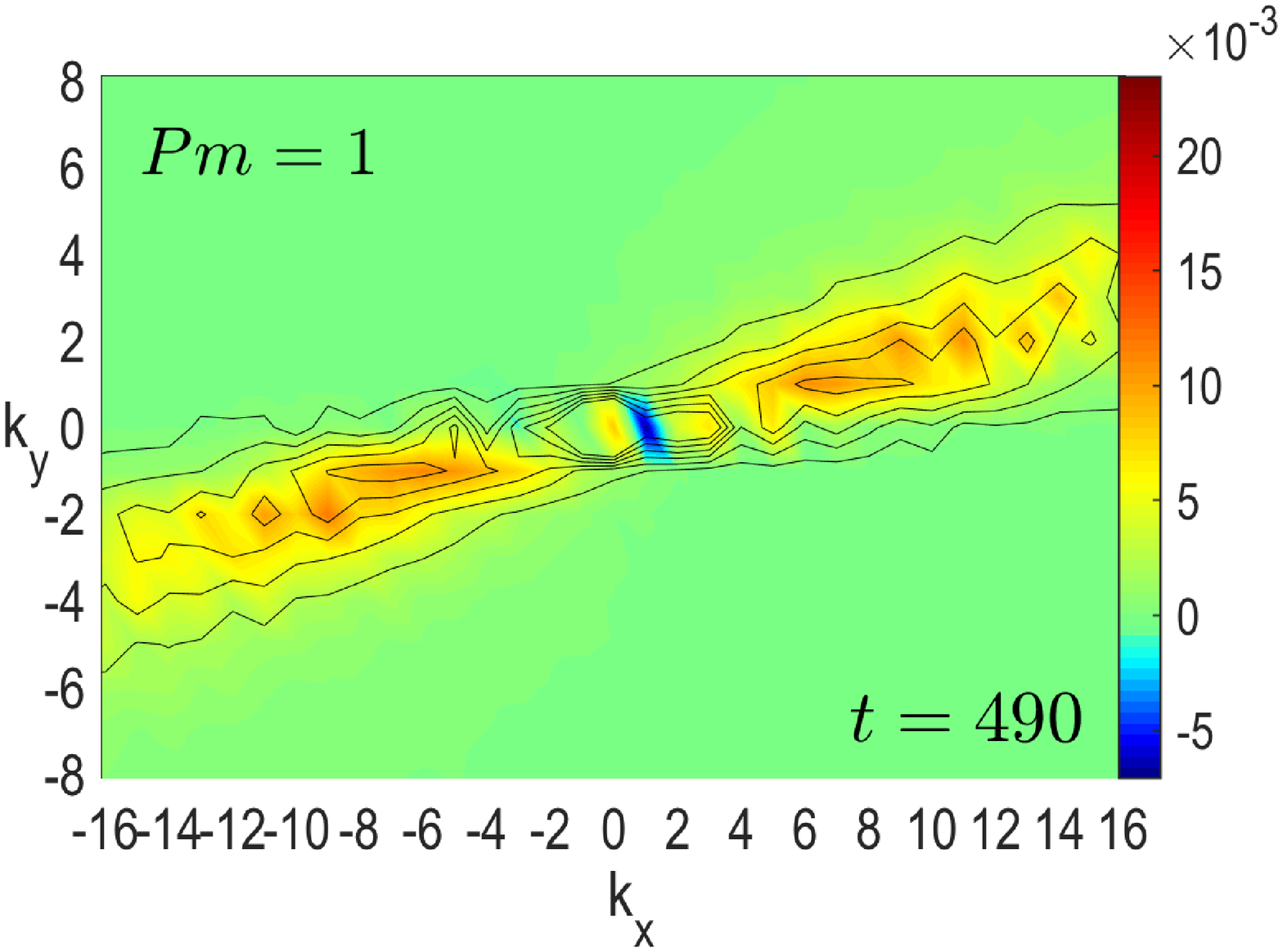}
\caption{Normalized spectra of the radial field, $|\bar{B}_x|^2/\widehat{|\bar{B}|_x^2}$ (top row), and Maxwell stress, ${\cal M}/\widehat{\cal M}$ (bottom row), in $(k_x,k_y)$-slice at $k_z=1$ for $Pm=4$ (left) and $Pm=1$ at two moments $t=440$ (middle) and $t=490$ (right). The contours of the corresponding azimuthal field spectrum, $|\bar{B}_ y|$, are drawn on the Maxwell stress spectrum at the lower row. The large-scale mode ${\bf k}=(\pm 1, 0, 1)$ giving large contribution in the Maxwell stress at $Pm=1$ has been artificially faded in the middle and right plots in order to better highlight the dynamics of other modes in $(k_x,k_y)$-slice. As a consequence of the action of the nonlinear term (Figure \ref{fig:Nbx_vs_kxky_Pm4_1}), at $Pm=1$ these normalized spectra of the magnetic field and Maxwell stress also spread to larger wavenumbers with time as well as decrease in magnitude.}\label{fig:bx_by_M_vs_kxky_Pm4_1}
\end{figure*}

\begin{figure}
\includegraphics[width=\columnwidth]{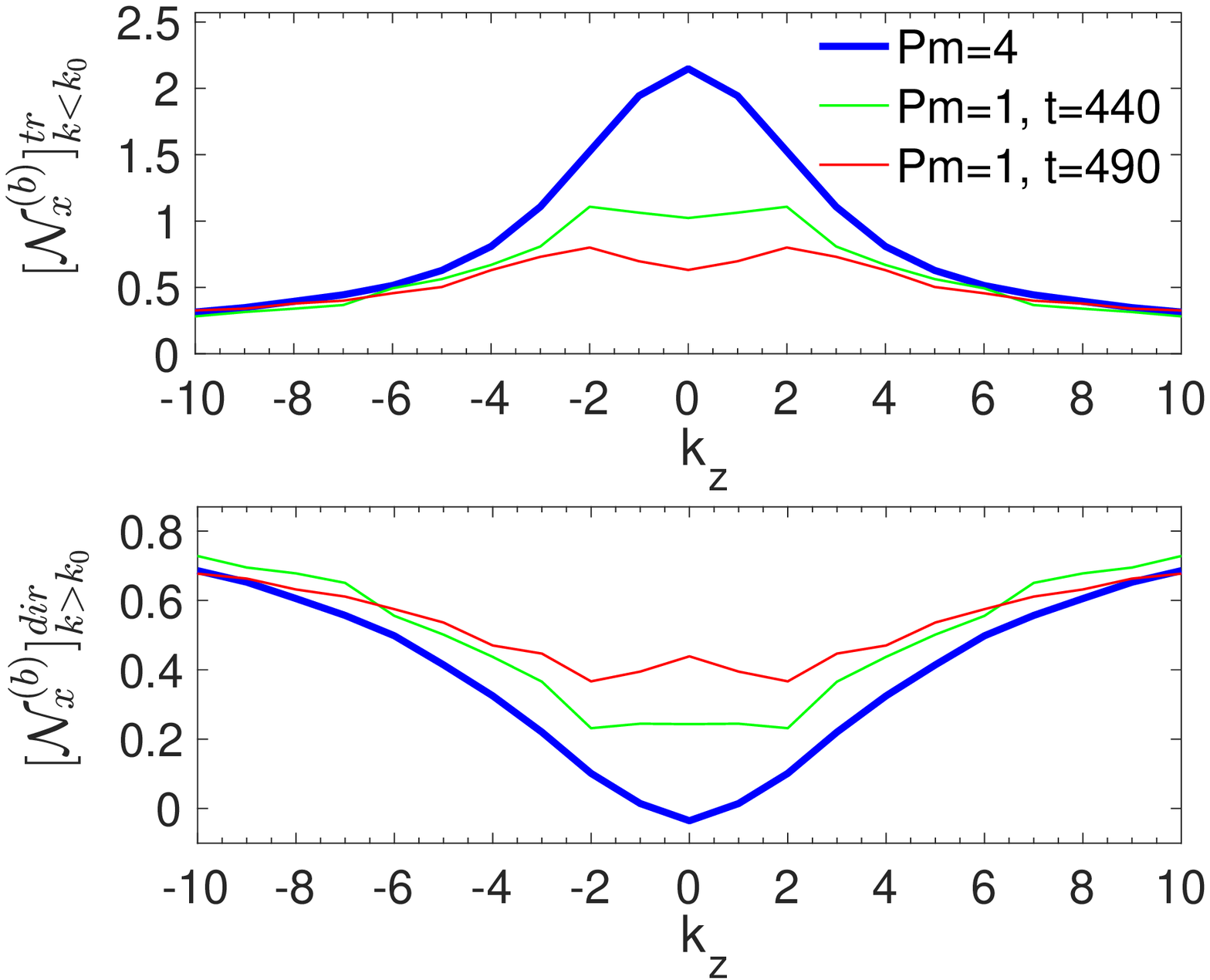}
\caption{The quantities $[{\cal N}_{x}^{(b)}]_{k < k_0}^{tr}$ (top) and $[{\cal N}_{x}^{(b)}]_{k > k_0}^{dir}$ (bottom), defined, respectively, within and outside the vital area with radius $k_0= 10$, are represented here as a function of $k_z$ for ${\rm Pm}=4$ and ${\rm Pm}=1$. The upper plot shows that at ${\rm Pm}=4$, the transverse cascade is the dominant nonlinear process in the vital area, but at ${\rm Pm}=1$ its efficiency decreases with time. The bottom plot, on the other hand, shows that the transfer of power to larger wavenumbers due to the direct cascade intensifies with time at ${\rm Pm}=1$ in contrast to the ${\rm Pm}=4$ case.}\label{fig:Nbx_va_out_Pm4_1}
\end{figure}

\subsection{Spectra at different ${\rm Pm}$}

We start with comparing the spectra of the radial field, azimuthal field and Maxwell stress integrated in $(k_x,k_y)$-slice, which are shown in
Figure \ref{fig:bx_by_M_vs_kz_Pm4_1} at ${\rm Pm}=4$ and for the above two moments at ${\rm Pm}=1$. These $k_z$-spectra are normalized by the corresponding values integrated over $k_z$, i.e., by
$\langle B_x^2\rangle$, $\langle B_y^2\rangle$, $q\langle -
B_xB_y\rangle$, respectively. These plots show how the power is
redistributed along $k_z$ in the process of the turbulence decay with time at ${\rm Pm}=1$ compared to the sustained case at ${\rm Pm}=4$. The relative power in the radial and azimuthal components and hence the Maxwell stress mainly decreases with time at small $|k_z|\leq 3$, with a slight increase at larger $|k_z|$. Thus, with decreasing ${\rm Pm}$, the magnetic energy is taken from the modes of the vital area by the nonlinear terms and is distributed among higher-$k_z$ modes.

Let us now see how the spectra of these quantities are redistributed in $(k_x,k_y)$-slices at different $k_z$. First we consider the normalized spectra of these quantities averaged over rings of constant wavenumber magnitude $k\equiv (k_x^2+k_y^2)^{1/2}$ in $(k_x,k_y)$-slices,\footnote{This notation for the magnitude of wavevector in $(k_x,k_y)$-slices, ${\bf k}=(k_x,k_y)$, is used everywhere below in this section.} $|\bar{B}_{x,y}|^2_k=\int_0^{2\pi}|\bar{B}_{x,y}|^2kd\varphi$,
${\cal M}_k=\int_0^{2\pi}{\cal M}_kkd\varphi$, where $\varphi$ is the polar angle of wavevectors in $(k_x,k_y)$-slices. These ring-averaged spectra are shown in Figure \ref{fig:bx_by_M_vs_k_Pm4_1} for the first few $k_z=0,1,2$ and larger $k_z=6$. Here, each spectral quantity is normalized by its integrated values over $(k_x,k_y)$-slices at a given $k_z$, i.e., by
$\widehat{|\bar{B}_x|^2}$, $\widehat{|\bar{B}_y|^2}$, $\widehat{\cal
M}$, respectively. At ${\rm Pm}=1$ and small $k_z$, the normalized spectra of these quantities decrease with time for $k \lesssim k_0=10$ ($k_0$ is approximately the radial extent of the vital area in Figure \ref{fig:modes}), but increase for larger $k \gtrsim k_0$, with the maximum also decreasing and moving towards higher $k$. We demonstrate below that this behavior is a consequence of the enhancement  of the role of the direct cascade and decrease of the role of the transverse cascade when ${\rm Pm}$ is reduced from 4 to 1. It is also seen in the azimuthal field spectrum at ${\rm Pm}=1$ that there is additionally some temporal accumulation of power in the dynamo mode $k=0, k_z=1$ with time.

This tendency of the enhanced transfer of the spectral magnetic energy to larger wavenumbers with lowering ${\rm Pm}$ in fact is related to the specific action of its governing nonlinear term ${\cal N}_x^{(b)}$. Figure \ref{fig:Nbx_vs_k_Pm4_1} depicts this nonlinear term ring-averaged in $(k_x,k_y)$-slices,
\[
{\cal N}_{x,k}^{(b)}=\int_0^{2\pi}{\cal N}_{x}^{(b)}kd\varphi,
\]
and normalized by its integrated value for each $k_z=0,1,2$ and $6$, i.e., $\widehat{\cal N}_x^{(b)}=\int {\cal N}_{x,k}^{(b)}dk$, at the same times $t=440$ and $t=490$ at ${\rm Pm}=1$ and its time-averaged value at ${\rm Pm}=4$. The ring-averaging hides dependence of the spectrum on polar angle in $(k_x,k_y)$-slice and, hence, ${\cal N}_{x,k}^{(b)}$ describes the transfer of the radial field only along wavevector  ${\bf k}=(k_x,k_y)$, i.e., the direct/inverse cascade.

Figure \ref{fig:Nbx_vs_k_Pm4_1} shows that at ${\rm Pm}=1$ and small $k_z=0,1,2$, the normalized ${\cal N}_{x,k}^{(b)}/\widehat{\cal N}_x^{(b)}$ decreases with time at $k \lesssim k_0$ in the vital area, but increases at $k \gtrsim k_0$, that is, it supplies more power to modes with $k$ higher than $k_0$ and less power to modes with $k$ smaller than $k_0$. As a result, the ring-averaged spectrum of the radial field exhibits a similar behavior with time, as we have seen above in Figure \ref{fig:bx_by_M_vs_k_Pm4_1}. We interpret this as the transfer of power from the vital area to larger $k$ due to the intensified direct cascade, caused by the decrease of ${\rm Pm}$, that is, increase of ${\rm Re}$ (at fixed ${\rm Rm}$), as if the latter opens up a ``channel'' of magnetic energy flux towards small-scale modes. For the Maxwell stress spectrum, this transfer implies that the energy injection into the turbulence also shifts to higher $k$. By contrast, at ${\rm Pm}=4$, the nonlinear term, and hence the spectral magnetic field and the Maxwell stress, are concentrated within the vital area, implying that the role of the energy transfer to higher wavenumber modes outside the vital area is not so important. On the other hand, at higher $k_z$, say, $k_z=6$, the normalized spectra of ${\cal N}_{x,k}^{(b)}$ as well as the spectral radial and azimuthal fields and the Maxwell stress no longer vary either with time or ${\rm Pm}$, as seen in Figures \ref{fig:bx_by_M_vs_k_Pm4_1} and \ref{fig:Nbx_vs_k_Pm4_1}. This is also consistent with Figure \ref{fig:bx_by_M_vs_kz_Pm4_1}, indicating that most of the changes in the $k_z$-spectra of these quantities with time and ${\rm Pm}$ occur at lower $k_z$.

Figures \ref{fig:bx_by_M_vs_k_Pm4_1} and \ref{fig:Nbx_vs_k_Pm4_1} give a first indication of increasing transfers of the ring-averaged spectral magnet energy and Maxwell stress towards larger wavenumbers with lowering ${\rm Pm}$. Figures \ref{fig:Nbx_vs_kxky_Pm4_1} and \ref{fig:bx_by_M_vs_kxky_Pm4_1} give a fuller picture of how these transfers proceed in $(k_x,k_y)$-slices with time as a result of the variation of the nonlinear term. At ${\rm Pm}=1$, the action of ${\cal N}_x^{(b)}$ shifts from small to large $k_x$ and $k_y$ (on the $k_x/k_y>0$ side) as the turbulence decays, spreading over a broader range of wavenumbers, while its normalized value, ${\cal N}_x^{(b)}/\widehat{\cal N}_x^{(b)}$, gradually decreases (Figure  \ref{fig:Nbx_vs_kxky_Pm4_1}). The intensity (color saturation) of positive (red and yellow) and negative (blue) areas of ${\cal N}_x^{(b)}$ fades away in the vital area, indicating the weakening efficiency of the nonlinear transverse cascade with time there relative to the direct cascade. By contrast, in the sustained turbulence at ${\rm Pm}=4$, the transverse cascade is more pronounced and efficient, being concentrated mainly at small wavenumbers in the vital area.

This behavior of the nonlinear transfer terms with time at ${\rm Pm}=1$ induces similar redistribution of the spectra of the radial field and Maxwell stress in $(k_x,k_y)$-slices, as seen in Figure \ref{fig:bx_by_M_vs_kxky_Pm4_1}. This figure shows the normalized spectral radial field, $|\bar{B}_x|^2/\widehat{|\bar{B}_x|^2}$, and the Maxwell stress, ${\cal M}/\widehat{\cal M}$, which appear also to shift and spread towards higher $k_x$ and $k_y$.

\subsection{MRI-turbulence decay at low ${\rm Pm}$}

To better characterize  the above discussed nonlinear transfers and,
especially, the relative role of the transverse and direct cascades
with decreasing ${\rm Pm}$, we introduce a measure of the intensity
of the transverse cascade in the vital area. It is defined as the ratio of
the sum of the absolute values of the nonlinear transfer term within
the vital area $k<k_0$  to the total value $\widehat{\cal N}_x^{(b)}$ in $(k_x,k_y)$-slices (for a given $k_z$), 
 \[
 [{\cal N}_{x}^{(b)}]_{k < k_0}^{tr} = \frac{1}{\widehat{\cal N}_x^{(b)}}\sum_{k < k_0}| {\cal N}_{x}^{(b)}|.
 \]
Note that in this expression, we take the absolute value of ${\cal
N}_x^{(b)}$, because the transverse cascade is characterized by its
positive and negative areas. These would otherwise nearly cancel
each other had we taken simply the sum of ${\cal N}_x^{(b)}$ over
the wavenumbers. Physically, $[{\cal N}_{x}^{(b)}]_{k < k_0}^{tr}$ describes the role of the regeneration process of the radial field by the transverse cascade inside $k < k_0$ with respect to the total value
$\widehat{\cal N}_x^{(b)}$ at the same $k_z$. From the top plot of
Figure \ref{fig:Nbx_va_out_Pm4_1} it is seen that this ratio is
largest at ${\rm Pm}=4$ and decreases with time at ${\rm Pm}=1$,
indicating a drop of intensity of the transverse cascade 
for active mode wavenumbers in the vital area.

The second process competing with the transverse cascade is the
direct cascade of power from the active modes inside the vital area
to larger wavenumber modes outside it. Following
\cite{Lesur_Longaretti11}, we can characterize this by the ratio of the
total energy gain for all the modes with wavenumber magnitudes $k>k_0$ which lie outside the vital area and $\widehat{\cal N}_x^{(b)}$ at given $k_z$,
 \[
 [{\cal N}_{x}^{(b)}]_{k > k_0}^{dir} = \frac{1}{\widehat{\cal N}_x^{(b)}}\sum_{k > k_0} {\cal N}_{x}^{(b)}.
 \]
This measure of the direct cascade is plotted in the lower panel of
Figure \ref{fig:Nbx_va_out_Pm4_1}. It is smallest on average in time
in the quasi-steady turbulence for ${\rm Pm}=4$ and increases with
time at ${\rm Pm}=1$, indicating increasing intensity of the direct
cascade for all $k_z$.

From the above analysis, we can interpret the decay of zero net flux
MRI-turbulence with decreasing ${\rm Pm}$ as follows. At ${\rm
Pm}=4$, the transverse cascade concentrated in the vital area is
strong enough to supply the active modes there and counteract the
effect of the direct cascade, which acts to transfer magnetic energy to higher wavenumber modes outside it. However, as seen from Figure
\ref{fig:Nbx_va_out_Pm4_1}, the decrease of ${\rm Pm}$ initiates the topological rearrangement of the nonlinear cascade processes:\\
-- It weakens the transverse cascade in the vital area, that is,
weakens the angular transfer of modes in Fourier space and thus replenishment of MRI growing modes, which is essential for the turbulence sustenance, and\\
-- It intensifies the direct cascade, that is, transfer of modes from small to high wavenumbers that acts as a sort of nonlinear (turbulent) diffusion for the active modes \citep[see also][]{Riols_etal15,Riols_etal17}. This transfer (leakage) of the magnetic energy from the vital area towards large wavenumber modes, in turn, leads to enhanced resistive dissipation of these modes. 

Thus, the role of the nonlinear cascades in the turbulence decay is indirect -- the topological rearrangement of the nonlinear cascade processes modifies the spectral distribution of the magnetic field, thereby intensifying the resistive dissipation at large wavenumbers and weakening the regeneration of MRI growing modes (mediated by the transverse cascade) in the vital area, i.e., weakening the energy supply to the turbulence. The diminishing of the transverse cascade and intensification of the direct cascade for the magnetic field ultimately results in the drop of the turbulence level (at ${\rm Pm}=3$) or its decay (at ${\rm Pm}=1$ and 2). In the decaying case, the reduced efficiency of the regeneration of MRI growing modes cannot counteract the increased resistive dissipation, making the self-sustaining scheme incapable of maintaining the turbulence.

\section{Summary and discussion}
\label{sec:Conclusion}

In this paper, we investigated the nature of MRI-turbulence --
dynamical balances, self-sustenance and dependence on magnetic
Prandtl number  -- in the unstratified shearing box approximation of
Keplerian disks with zero net magnetic flux. New insights into these
processes were gained by first performing simulations of the
turbulence at different ${\rm Pm}=1-4$  and then carrying out a
detailed analysis of the individual linear and nonlinear dynamical
terms in the main MHD equations and their interplay in Fourier
(\textbf{k}-)space. We found, in agreement with other related
studies, that in standard boxes at ${\rm Pm}\gtrsim 1$, MRI-turbulence is sustained, while at ${\rm Pm}\lesssim 1$ it decays on a time scale of hundred orbits.

As distinct from previous studies, we did the spectral analysis in a
general way, without doing spherical shell-averaging in Fourier space and thereby capturing the spectral anisotropy of nonlinear processes -- the transverse cascade -- due to the flow shear. In the net zero flux case, there are no large-scale purely exponentially growing MRI modes, so MRI is of transient, or nonmodal type. This transient growth itself is ``flawed'' in the sense that it lasts during a limited (dynamical/orbital) time and therefor needs to be supported by nonlinear positive feedback for the long-term maintenance of the turbulence, which is provided just by the transverse cascade. In other words, the nonlinear transverse cascade lies at the heart of the zero net flux MRI-turbulence, ensuring its self-sustenance. This type of new cascade generic to shear flows is fundamentally (topologically) different from usual classical (inverse/direct) nonlinear cascades. So, the conventional characterization of nonlinear MHD cascade processes in strongly sheared Keplerian flows in terms of direct and inverse cascades only \citep[e.g., using shell-averaging][]{Fromang_Papaloizou07,Simon_etal09,Lesur_Longaretti11}, which ignores the shear-induced spectral anisotropy and the
resulting transverse cascade, is incomplete.

Specifically, the main results of this paper are 
\begin{enumerate}[$\bullet$]
\item
{\it Self-sustenance scheme}
\\
In the sustained case ${\rm Pm}=4$, we established that the zero net flux MRI-turbulence is sustained by the interplay of the linear nonmodal growth of MRI and the nonlinear transverse cascade in Fourier space. Although this interplay is generally quite intricate, we were able to isolate a core subcycle of the turbulence sustenance from it (Figure
\ref{fig:sketch}). The radial and azimuthal components of the magnetic field play the main role in this process, which proceeds as follows. The radial field is generated/seeded and maintained by the nonlinear term ${\cal N}^{(b)}_x$ through the transverse cascade process. The azimuthal field is then produced from the radial field through the linear nonmodal MRI process. The azimuthal field, in turn, largely contributes to the production of the nonlinear source (feedback) ${\cal N}^{(b)}_x$ for the radial field, thereby closing the self-sustaining cycle. This self-sustaining dynamics of the turbulence mainly occupies a small wavenumber area of Fourier space, the so-called vital area, and involves a sizable number ($\gtrsim 100$) of large-scale active modes. For relatively large ${\rm Pm}=4$, the transverse cascade is able to replenish the radial field in the vital area sufficiently faster than the direct cascade transfers it to larger wavenumber modes and hence the
sustenance is possible. As typical of MRI-turbulence, large-scale zonal flow is generated with the dominant azimuthal velocity.

\item
{\it Dependence on ${\rm Pm}$}
\\
We also performed a comparative analysis of the sustained (at ${\rm
Pm}=4$) and decaying (at ${\rm Pm}=1$) cases, exploring the specific
differences between the nonlinear transfers and, consequently, of the magnetic field and Maxwell stress spectra in Fourier space. We showed that the essence of these differences is that decreasing ${\rm Pm}$ leads to the structural/topological rearrangement of the
nonlinear transfers, such that in contrast to the sustained case,
now the direct cascade prevails over the transverse one and shifts the magnetic field spectrum to higher wavenumbers. In other words, at ${\rm Pm}=1$, the action of the nonlinear transverse cascade, and particularly the associated regeneration of MRI growing modes, weakens in the vital area (top plot of Figure \ref{fig:Nbx_va_out_Pm4_1}). It can no longer oppose the increased action of the direct cascade, which more rapidly transfers magnetic energy from small wavenumber modes in the vital area to higher wavenumber ones outside it (lower plot of Figure \ref{fig:Nbx_va_out_Pm4_1}). This intensified transfer of the magnetic field to the small-scale modes, in turn, enhances resistive dissipation of these modes relative to the regeneration of MRI growing modes due to the transverse cascade. This enhancement of resistive dissipation in conjunction with the reduction of MRI growth, ultimately, makes the self-sustaining scheme ineffective of the turbulence sustenance and leads to its decay. Thus, the decrease of ${\rm Pm}$ results in weakening the transverse cascade at small wavenumbers (large scales) and a simultaneous intensification of the direct cascade to large wavenumbers, which, in turn, leads to enhanced resistive dissipation of the magnetic field at small scales. However, as to why this course of events takes place still needs to be better understood.
\end{enumerate}

\subsection{Connection to related studies}

The self-sustenance and dynamics of 3D nonlinear periodic MRI-dynamo solutions in the presence of zero net magnetic flux and their dependence on viscous and resistive dissipation were studied
previously by \citet{Herault_etal11,Riols_etal15,Riols_etal17} in
Fourier space. Therefore, making connections with their works is in
order. We note from the beginning that these papers were
focused on smaller Reynolds numbers, smaller numerical resolution,
larger box aspect ratios, $L_y/L_x, L_z/L_x$, etc. than used here. For comparison, we took much higher numerical resolution $(N_x,N_y,N_z)=(512,512,128)$ and ${\rm Rm}=1.2\times 10^4$ in the standard box $(L_x,L_y,L_z)=(3,3,1)$. These limiting factors in the above studies likely resulted in the resistive dissipation penetrating into
the vital area, as we call it, and hence reducing the number of
active modes participating in the self-sustaining dynamics to only first few large-scale non-axisymmetric and two axisymmetric ones. These two axisymmetric modes are the ${\bf k}_d=(0,0,\pm 2\pi/L_z)$ dynamo mode, which carries a slowly varying in time, large-scale azimuthal magnetic field, and ${\bf k}_m=(\pm 2\pi/L_x,0,\pm 2\pi/L_z)$ mode, which carries a radially modulated magnetic field. Overall, the nonlinear MRI-dynamo state considered in those papers is not fully turbulent, but instead spatially and temporally more regular with a scarce spectrum of active modes contributing to the dynamics, in contrast to our case of the fully developed MRI-turbulence.

A self-sustaining scheme of this zero net flux MRI-dynamo state was worked out in \citet{Herault_etal11}. In this scheme, the role of the nonlinearity is twofold:\\
1. to maintain the large-scale axisymmetric radial field of the ${\bf k}_d$ dynamo mode via nonlinear scattering of non-axisymmetric MRI-amplified modes (shearing waves). The azimuthal dynamo field, in turn, is produced by shear-induced stretching of this radial field. \\
2. to regenerate (seed) leading non-axisymmetric modes due to the nonlinear scattering of trailing non-axisymmetric modes with the ${\bf k}_m$ mode. These leading modes are then capable of undergoing transient MRI growth against the background of the large-scale azimuthal  axisymmetric dynamo field.  

The next step of zero net magnetic flux MRI study was performed in \citet{Riols_etal15} considering again low-order model (azimuthally very elongated box $(L_x,L_y,L_z)=(0.7,20,2)$, low numerical resolution $(N_x,N_y,N_z)=(48,48,72)$ and $(128, 128, 96)$, moderate ${\rm Re}, {\rm Rm}=100-3000$) and focusing on the mechanism of decrease/disappearance of the MRI-dynamo states with decreasing  ${\rm Pm}$. To explain this behavior, \citet{Riols_etal15} split the nonlinear terms in the magnetic field equation into positive induction and negative advection parts. It was shown that there is an enhanced nonlinear magnetic diffusion due to transfer of energy from small to large wavenumbers (direct cascade) with decreasing ${\rm Pm}$ that results in turbulence decay.\footnote{It should be mentioned that \citet{Riols_etal15, Riols_etal17} sum the spectra of the dynamical terms along the radial wavenumber $k_x$, thereby missing out essential shear-induced anisotropy of the nonlinear processes in Fourier space and hence the transverse cascade.}

At the last step, \citet{Riols_etal17} increased ${\rm Re}$ and ${\rm Rm}$ and took moderate aspect ratios, $(L_x,L_y,L_z)=(0.7,6,2), (0.5,2,1)$. In this case, although the dynamical picture is more complex (``chimera'' MRI-dynamo states) than that in the previous low-order cases, it is still not yet completely turbulent -- the vital area contains only a few largest-scale dominant active modes with $|k_x|\leq 4, |k_y|,|k_z| \leq 1$. In that study, using the same approach, the roles/interplay of the induction and advection terms was analyzed in the decay of magnetic perturbations with decreasing ${\rm Pm}$ given this more complex nature of the MRI-dynamo states.

As distinct from the works of \citet{Herault_etal11,Riols_etal15,Riols_etal17}, we deal with much
larger number of active modes, i.e., with  fully developed
MRI-turbulence and propose a general scheme of the turbulence
self-sustenance. Our simulations indicate that in
this turbulent state not only the large-scale axisymmetric azimuthal 
field, but also a certain, quite extensive set of non-axisymmetric
modes undergo appreciable nonmodal/transient MRI growth without mediation of this axisymmetric field. The existence of a nonlinear feedback provided by transversal/angular redistribution of the
non-axisymmetric modes in Fourier space is a key factor for the
turbulence self-sustenance, as it replenishes new non-axisymmetric
modes experiencing the nonmodal amplification. This angular redistribution of the multitude of the non-axisymmetric modes can be
structurally/topologically classified as a nonlinear transverse
cascade. In conclusion, the nonlinear transverse cascade is naturally inherent in the turbulence dynamics (bottom plots of Figure \ref{fig:spectral_bx}) in shear flows and therefore is an integral part of our scheme.

Due to its general nature, the transverse cascade also comprises a
particular nonlinear feedback process central in the self-sustenance scheme of \citet{Herault_etal11} -- replenishment of new leading non-axisymmetric modes via scattering of trailing ones by the ${\bf k}_m$ mode. Our study indicates that, in contrast to those low-order models, this particular nonlinear process is not the one and only process resupplying new active modes in zero net flux MRI-turbulence. A broad class of other such ``constructive'' nonlinear triad interactions are collectively accounted for in the transverse cascade process. Besides, our approach gives us insight into the topological changes of the nonlinear cascades in Fourier space with lowering ${\rm Pm}$, which lead to the reduction/disappearance of the turbulence in standard  ($L_z/L_x\lesssim 1$) boxes. 

The representation of the nonlinear magnetic term into induction and advection parts, as done in \citet{Riols_etal15,Riols_etal17}, is useful and interesting in its own right. However, dealing with fully developed turbulence (i.e., with much larger number of active
modes), instead of the explicit decomposition of the nonlinear term into induction and advection parts, we use another approach -- structural/topological analysis of the linear and nonlinear dynamical terms of the induction equation in Fourier space -- which we previously vindicated for hydrodynamic \citep{Horton_etal10} and MHD \citep{Mamatsashvili_etal14} 2D shear flows. In our opinion,
such a topological analysis is more general/universal, as it has
confirmed viability with regard to hydrodynamic shear flows too
\citep{Horton_etal10, Mamatsashvili_etal16}, when these induction and advection nonlinear terms of magnetic origin are absent by definition. In these papers, investigating the nonlinear terms of fluid equations in Fourier space, we elucidated, for example, the self-sustenance scheme of coherent vortical perturbations and homogeneous shear turbulence in hydrodynamic 2D and 3D constant shear flows, respectively.

%%%%%%%%%%%%%%%%%%%%%%%%%%%%%%%%%%%%%%%%%%

In related papers, \citet{Nauman_Pessah16, Nauman_Pessah18} carried out a series of zero net flux MRI-turbulence simulations for taller boxes with vertical aspect ratios $L_z/L_x=4-32$ and a wide range of ${\rm
Pm}=0.1-10$. They found sustained turbulence even at ${\rm Pm}\leq1$
for large enough $L_z/L_x \geq 4$ in contrast to standard boxes with
$L_z/L_x \sim 1$, where there is no turbulence. The main difference
between the tall and standard boxes is that the large-scale
azimuthal dynamo field is much stronger and varies slower in tall boxes \citep[see also][]{Lesur_Ogilvie08,Shi_etal16}. In this case, this mean azimuthal field contributes to and enhances the nonmodal growth of non-axisymmetric MRI modes by introducing linear kinetic-magnetic exchange terms in the momentum and induction equations. In Fourier space, this linear exchange term can be comparable to the action of the nonlinear term, i.e., transverse cascade (Paper I). As a result, the turbulence dynamics becomes more similar to that in the presence of nonzero net azimuthal field, which is known to persist at lower ${\rm Pm} \leq 1$, but with level decreasing with ${\rm Pm}$ \citep[e.g.,][]{Simon_Hawley09, Meheut_etal15}. Apparently, this mean large-scale azimuthal field helps the MRI-turbulence to persist at small ${\rm Pm}$ by enhancing its energetic supply despite the reduction of the transverse cascade efficiency with decreasing ${\rm Pm}$. However, this is only a qualitative interpretation and for a proper understanding of the sustenance of zero net flux MRI-turbulence in tall boxes, one should do a similar analysis of the turbulence dynamics in Fourier space at large $L_z/L_x \geq 4$ and different ${\rm Pm}$.

\subsection{Alternative instability mechanisms at small ${\rm Pm}$}

It is well-known by now that the existence of MRI-turbulence and
associated dynamo action is quite problematic, near to impossible at
small ${\rm Pm}\lesssim 1$ both for nonzero and zero net magnetic
fluxes, as also confirmed in the present study. On the other hand,
from global magnetized Taylor-Couette studies -- laboratory analogs
of accretion disks -- it has been long shown that at small ${\rm
Pm}\ll 1$ the helical (with both azimuthal and vertical fields) and
azimuthal (with purely azimuthal field) versions of MRI, or for
short  HMRI and AMRI can operate \citep[e.g.,]{Hollerbach_Rudiger05,Stefani_etal09,Hollerbach_etal10,Seilmayer_etal14,Kirillov_Stefani14,Mamatsashvili_Stefani16},
even down to very low ${\rm Pm}\sim 10^{-6}-10^{-5}$
relevant to cold interiors of protoplanetary disks. These instabilities, relatives of MRI, have been overlooked in shearing box studies,
because they rely on the curvature term proportional to the background azimuthal magnetic field in the linearized induction equation, which is present only in the global cylindrical geometry \citep{Pessah_etal05}. Another reason was that they require radial shear of the rotational velocity larger than the Keplerian if the azimuthal field is current-free \citep{Liu_etal06}. The last
constraint may not be realized in protoplanetary disks, and
the dominant azimuthal field can in principle deviate from the
current-free profile. In this case, as demonstrated by
\citet{Kirillov_Stefani13}, HMRI and AMRI can in fact extend to
Keplerian rotation. This opens up a new possibility that these
instabilities can be alternatives of
standard MRI in cold and dense parts (``dead zones'') of
protoplanetary disks with small ${\rm Pm} \ll 1$. They can play a central
role in driving dynamical processes and magnetic dynamo action
there, through which, in turn, they may also sustain themselves.
However, the viability of this proposition should be further
explored in global disk simulations.

\acknowledgments

This project has received funding from the European Union's Horizon
2020 research and innovation programme under the Marie Sk{\l}odowska
-- Curie Grant Agreement No. 795158 and the ERC Advanced Grant
Agreement No. 787544 as well as from the Shota Rustaveli National
Science Foundation of Georgia (SRNSFG, grant No. FR17-107). We thank Dr. F. Rincon for useful discussions and the Referee for suggestions that improved the presentation of our work.

\appendix

\section{Perturbation equations in physical and Fourier space}
\label{sec:AppendixA}

Substituting velocity and pressure perturbations, ${\bf u}={\bf
U}-{\bf U}_0$, $p=P-P_0$, about the background Keplerian shear flow
${\bf U}_0=-q\Omega x{\bf e}_y$, into the main Equations
(\ref{eq:mom})-(\ref{eq:div}) we get:
\begin{equation}\label{eq:App-ux}
\frac{Du_x}{Dt} = 2\Omega u_y- \frac{1}{\rho_0}\frac{\partial
p}{\partial x}+\frac{\partial}{\partial
x}\left(\frac{B_x^2}{4\pi\rho_0}-u_x^2\right)+\frac{\partial}{\partial
y}\left(\frac{B_xB_y}{4\pi\rho_0}-u_xu_y\right)
+\frac{\partial}{\partial
z}\left(\frac{B_xB_z}{4\pi\rho_0}-u_xu_z\right)+\nu\nabla^2u_x,
\end{equation}
\begin{equation}\label{eq:App-uy}
\frac{Du_y}{Dt} = (q-2)\Omega u_x-\frac{1}{\rho_0}\frac{\partial
p}{\partial y}+\frac{\partial}{\partial
x}\left(\frac{B_xB_y}{4\pi\rho_0}-u_xu_y\right)+\frac{\partial}{\partial
y}\left(\frac{B_y^2}{4\pi\rho_0}-u_y^2\right)+\frac{\partial}{\partial
z}\left(\frac{B_zB_y}{4\pi\rho_0}-u_zu_y\right)+\nu\nabla^2u_y
\end{equation}
\begin{equation}\label{eq:App-uz}
\frac{Du_z}{Dt} = -\frac{1}{\rho_0}\frac{\partial p}{\partial
z}+\frac{\partial}{\partial
x}\left(\frac{B_xB_z}{4\pi\rho_0}-u_xu_z\right)+\frac{\partial}{\partial
y}\left(\frac{B_yB_z}{4\pi\rho_0}-u_yu_z\right)+\frac{\partial}{\partial
z}\left(\frac{B_z^2}{4\pi\rho_0}-u_z^2\right)+\nu\nabla^2u_z
\end{equation}
\begin{equation}\label{eq:App-bx}
\frac{DB_x}{Dt}= \frac{\partial}{\partial y} \left(u_xB_y-u_yB_x
\right)-\frac{\partial}{\partial z}(u_zB_x-u_xB_z)+\eta\nabla^2B_x,
\end{equation}
\begin{equation}\label{eq:App-by}
\frac{DB_y}{Dt}= -q\Omega B_x -\frac{\partial}{\partial x}
\left(u_xB_y-u_yB_x\right)+\frac{\partial}{\partial
z}(u_yB_z-u_zB_y)+\eta\nabla^2B_y,
\end{equation}
\begin{equation}\label{eq:App-bz}
\frac{DB_z}{Dt}= \frac{\partial}{\partial x}
\left(u_zB_x-u_xB_z\right)-\frac{\partial}{\partial
y}(u_yB_z-u_zB_y)+\eta\nabla^2B_z,
\end{equation}
\begin{equation}\label{eq:App-divu}
\frac{\partial u_x}{\partial x}+\frac{\partial u_y}{\partial
y}+\frac{\partial u_z}{\partial z}=0,
\end{equation}
\begin{equation}\label{eq:App-divB}
\frac{\partial B_x}{\partial x}+\frac{\partial B_y}{\partial
y}+\frac{\partial B_z}{\partial z}=0,
\end{equation}
where $D/Dt=\partial/\partial t-q\Omega x\partial/\partial y$ is the
total derivative along the Keplerian shear flow.

Substituting decomposition (\ref{eq:Fourier}) into Equations
(\ref{eq:App-ux})-(\ref{eq:App-divB}) and taking into account the
normalization, we obtain the evolution equations for the Fourier transforms of the velocity and magnetic field components:
\begin{equation}\label{eq:App-uxk}
\left(\frac{\partial}{\partial t}+qk_y\frac{\partial}{\partial
k_x}\right)\bar{u}_x=2\bar{u}_y-{\rm i}k_x\bar{p}-\frac{k^2}{\rm Re}\bar{u}_x+{\rm i}k_x
N^{(u)}_{xx}+{\rm i}k_yN^{(u)}_{xy}+{\rm i}k_zN^{(u)}_{xz},
\end{equation}
\begin{equation}\label{eq:App-uyk}
\left(\frac{\partial}{\partial t}+qk_y\frac{\partial}{\partial
k_x}\right)\bar{u}_y=(q-2)\bar{u}_x-{\rm i}k_y\bar{p}-\frac{k^2}{\rm Re}\bar{u}_y+{\rm
i}k_xN^{(u)}_{xy}+{\rm i}k_yN^{(u)}_{yy}+{\rm i}k_zN^{(u)}_{yz},
\end{equation}
\begin{equation}\label{eq:App-uzk}
\left(\frac{\partial}{\partial t}+qk_y\frac{\partial}{\partial
k_x}\right)\bar{u}_z=-{\rm i}k_z\bar{p}-\frac{k^2}{\rm Re}\bar{u}_z+{\rm
i}k_xN^{(u)}_{xz}+{\rm i}k_yN^{(u)}_{yz}+{\rm i}k_zN^{(u)}_{zz},
\end{equation}
\begin{equation}\label{eq:App-bxk}
\left(\frac{\partial}{\partial t}+qk_y\frac{\partial}{\partial
k_x}\right)\bar{B}_x={\rm i}k_y\bar{F}_z-{\rm i}k_z\bar{F}_y-\frac{k^2}{\rm
Rm}\bar{B}_x
\end{equation}
\begin{equation}\label{eq:App-byk}
\left(\frac{\partial}{\partial t}+qk_y\frac{\partial}{\partial
k_x}\right)\bar{B}_y=-q\bar{B}_x+{\rm
i}k_z\bar{F}_x-{\rm i}k_x\bar{F}_z-\frac{k^2}{\rm Rm}\bar{B}_y
\end{equation}
\begin{equation}\label{eq:App-bzk}
\left(\frac{\partial}{\partial t}+qk_y\frac{\partial}{\partial
k_x}\right)\bar{B}_z={\rm i}k_x\bar{F}_y-{\rm i}k_y\bar{F}_x-\frac{k^2}{\rm
Rm}\bar{B}_z
\end{equation}
\begin{equation}\label{eq:App-divvk}
k_x\bar{u}_x+k_y\bar{u}_y+k_z\bar{u}_z=0,
\end{equation}
\begin{equation}\label{eq:App-divbk}
k_x\bar{B}_x+k_y\bar{B}_y+k_z\bar{B}_z=0,
\end{equation}
where $k^2=k_x^2+k_y^2+k_z^2$. These spectral equations involve the
linear and nonlinear ($N^{(u)}_{ij}({\bf k},t), \bar{F}_i({\bf
k},t)$, where $i,j=x,y,z$) terms that are the Fourier transforms of
the corresponding linear and nonlinear terms of Equations
(\ref{eq:App-ux})-(\ref{eq:App-divB}) in physical space. These spectral
nonlinear terms are given by convolutions
\begin{equation}\label{eq:App-Nuij}
N^{(u)}_{ij}({\bf k},t)=\int d^3{\bf k'}\left[\bar{B}_i({\bf
k'},t)\bar{B}_j({\bf k}-{\bf k'},t)-\bar{u}_i({\bf
k'},t)\bar{u}_j({\bf k}-{\bf k'},t)\right],
\end{equation}
and $\bar{F}_x, \bar{F}_y, \bar{F}_z$, which are the Fourier
transforms of the respective components of the perturbed
electromotive force ${\bf F}={\bf u}\times {\bf B}$,
\begin{equation}\label{eq:App-Fxk}
\bar{F}_x({\bf k},t)=\int d^3{\bf k'}\left[\bar{u}_y({\bf
k'},t)\bar{B}_z({\bf k}-{\bf k'},t)-\bar{u}_z({\bf
k'},t)\bar{B}_y({\bf k}-{\bf k'},t)\right]
\end{equation}
\begin{equation}\label{eq:App-Fyk}
\bar{F}_y({\bf k},t)=\int d^3{\bf k'}\left[\bar{u}_z({\bf
k'},t)\bar{B}_x({\bf k}-{\bf k'},t)-\bar{u}_x({\bf
k'},t)\bar{B}_z({\bf k}-{\bf k'},t)\right]
\end{equation}
\begin{equation}\label{eq:App-Fzk}
\bar{F}_z({\bf k},t)=\int d^3{\bf k'}\left[\bar{u}_x({\bf
k'},t)\bar{B}_y({\bf k}-{\bf k'},t)-\bar{u}_y({\bf
k'},t)\bar{B}_x({\bf k}-{\bf k'},t)\right],
\end{equation}
describe the effect of nonlinearity on the magnetic field
perturbations. From Equations (\ref{eq:App-uxk})-(\ref{eq:App-uzk})
and the divergence-free conditions (\ref{eq:App-divvk}) and
(\ref{eq:App-divbk}) we can express pressure
\begin{equation}\label{eq:App-pk}
\bar{p}=2{\rm i}(1-q)\frac{k_y}{k^2}\bar{u}_x-2{\rm
i}\frac{k_x}{k^2}\bar{u}_y+{\rm
i}\frac{k_z}{k^2}+\sum_{(i,j)=(x,y,z)}\frac{k_ik_j}{k^2}N^{(u)}_{ij}
\end{equation}
Inserting it back into Equations
(\ref{eq:App-uxk})-(\ref{eq:App-uzk}), we have
\begin{equation}\label{eq:App-uxk1}
\left(\frac{\partial}{\partial t}+qk_y\frac{\partial}{\partial
k_x}\right)\bar{u}_x=2\left(1-\frac{k_x^2}{k^2}\right)\bar{u}_y+
2(1-q)\frac{k_xk_y}{k^2}\bar{u}_x-\frac{k^2}{\rm Re}\bar{u}_x+Q_x,
\end{equation}
\begin{equation}\label{eq:App-uyk1}
\left(\frac{\partial}{\partial t}+qk_y\frac{\partial}{\partial
k_x}\right)\bar{u}_y=\left[q-2-2(q-1)\frac{k_y^2}{k^2}\right]\bar{u}_x-
2\frac{k_xk_y}{k^2} \bar{u}_y+\frac{k_yk_z}{k^2}-\frac{k^2}{\rm Re}\bar{u}_y+Q_y,
\end{equation}
\begin{equation}\label{eq:App-uzk1}
\left(\frac{\partial}{\partial t}+qk_y\frac{\partial}{\partial
k_x}\right)\bar{u}_z=2(1-q)\frac{k_yk_z}{k^2}\bar{u}_x-2\frac{k_xk_z}{k^2}\bar{u}_y-\frac{k^2}{\rm Re}\bar{u}_z+Q_z,
\end{equation}
where
\begin{equation}\label{eq:App-Qi}
Q_i={\rm i}\sum_jk_jN^{(u)}_{ij}-{\rm
i}k_i\sum_{m,n}\frac{k_mk_n}{k^2}N^{(u)}_{mn}, ~~~~~ i,j,m,n=x,y,z.
\end{equation}
\\
\\

\section{Dynamics of the large-scale field in the turbulent state}

In our setup, the MRI-turbulence does support the large-scale
dynamo action producing axisymmetric radial and azimuthal magnetic
fields belonging to the ${\bf k}_d=(0,0,\pm 1)$ mode with only
large-scale vertical variation. This dynamo action and the dynamics
of this mode was previously studied in unstratified zero net flux
MRI-turbulence \citep{Lesur_Ogilvie08, Herault_etal11, Shi_etal16,
Walker_Boldyrev17,Riols_etal17}, where it was shown to exhibit
either regular/quasi-periodic or more chaotic spatio-temporal
behavior, depending on the vertical aspect ratio of a box. In taller
and azimuthally more elongated boxes with aspect ratios $L_y/L_x\geq
4$ and $L_z/L_x \geq 2$ mostly considered in those papers, this
large-scale dynamo mode exhibits variations on the timescale of
several tens of orbits or more. By contrast, in the present setup
with the standard box where $L_z/L_x=1/3$, this mode and associated
large-scale dynamo field do not display any remarkable long-time quasi-periodic behavior and vary instead on shorter timescale, of the order of dynamical/orbital time.

\begin{figure}
\centering
\includegraphics[width=0.45\textwidth]{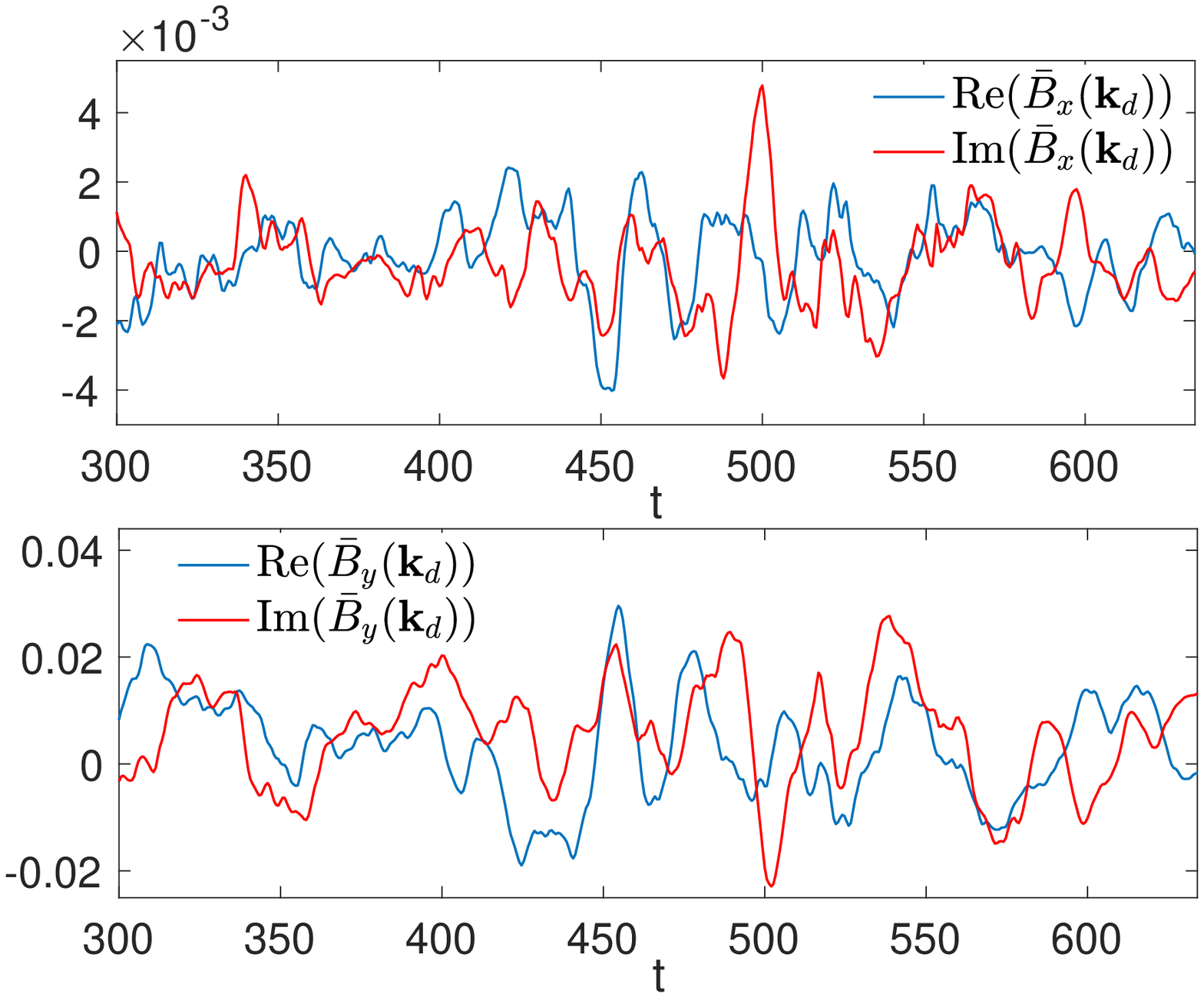}
\includegraphics[width=0.45\textwidth]{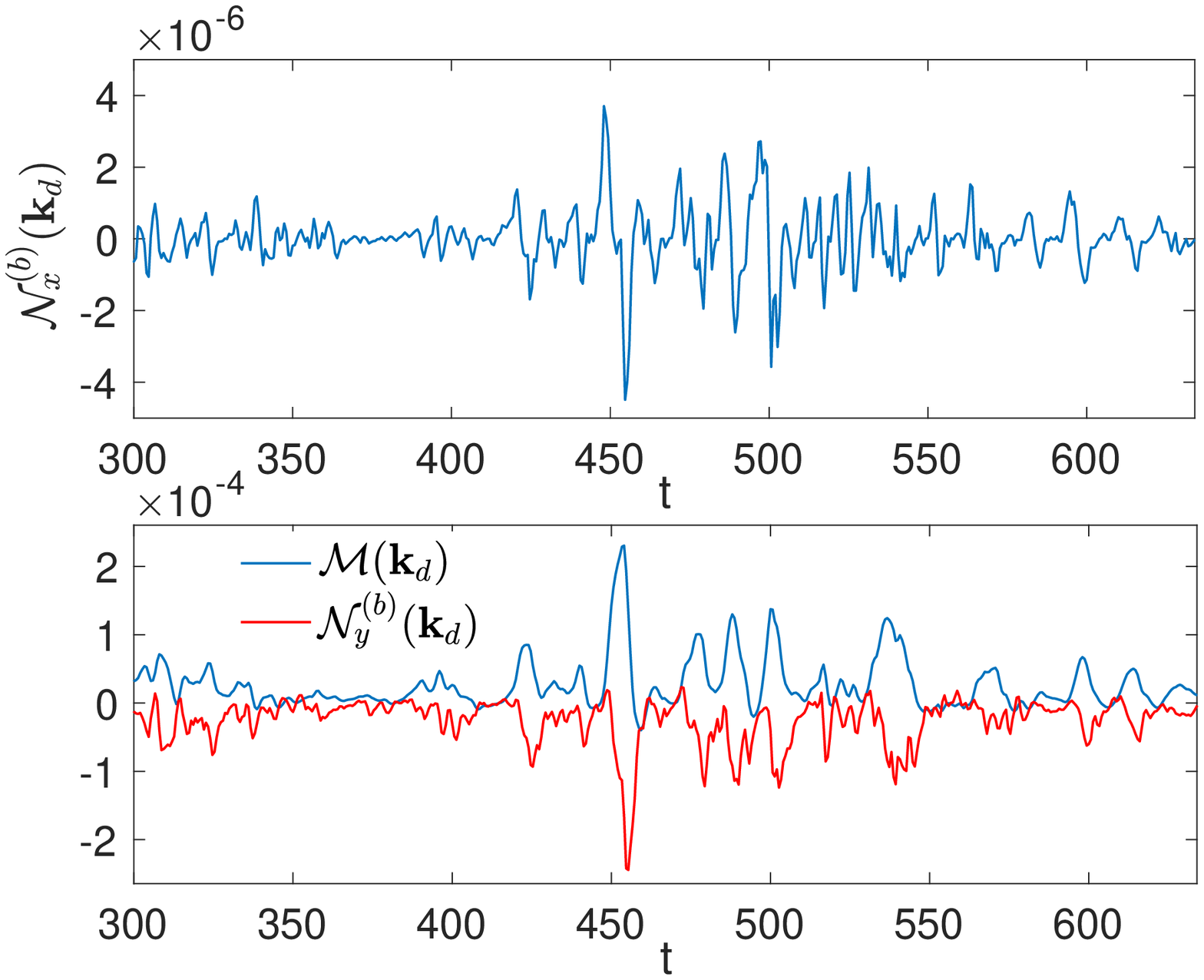}
\caption{Evolution of the real and imaginary parts of the radial,
$\bar{B}_x({\bf k}_d)$, and azimuthal, $\bar{B}_y({\bf k}_d)$,
magnetic fields of the ${\bf k}_d=(0,0,\pm 1)$ mode at ${\rm Pm}=4$
together with the corresponding governing terms, ${\cal
N}_x^{(b)}({\bf k}_d)$ as well as the Maxwell stress ${\cal M}({\bf
k}_d)$ and ${\cal N}_y^{(b)}({\bf k}_d)$. The azimuthal field is
about 10 times larger than the radial one. Both these field
components and the dynamical terms exhibit quite irregular
variations on the relatively short time-scale of $\lesssim 10$. The nonlinear term ${\cal N}_x^{(b)}({\bf k}_d)$ alternates signs and causes corresponding dips and rises of the radial field. The Maxwell stress ${\cal M}({\bf k}_d)$ is always positive, maintaining the azimuthal field, whereas the nonlinear term ${\cal N}_y^{(b)}({\bf k}_d)$ is always negative, draining the azimuthal field, i.e., acting as a turbulent magnetic dissipation for it.}\label{fig:dynamo_mode_Bxk}
\end{figure}

\begin{figure}
\centering
\includegraphics[width=0.6\textwidth, height=0.4\textwidth]{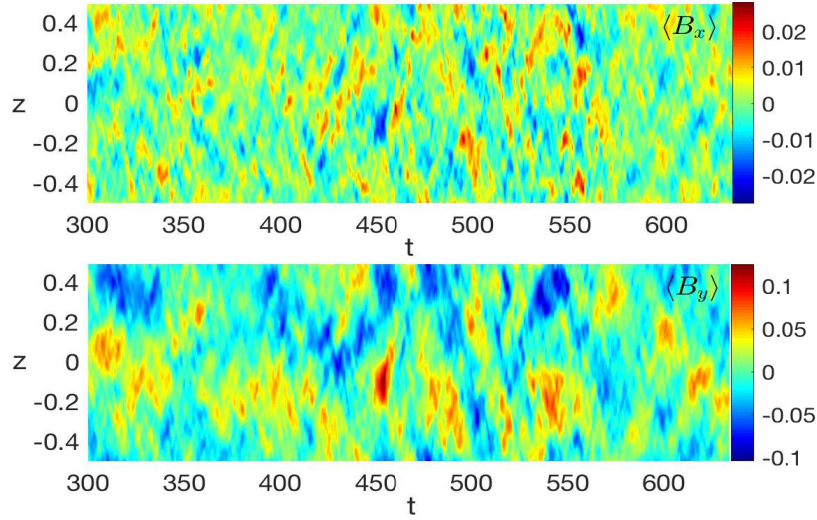}
\caption{Space-time diagrams of the radial and azimuthal magnetic fields in physical space averaged horizontally in $(x,y)$-slice at ${\rm
Pm}=4$. Their temporal behavior appears to be less regular, varying
on the similar time-scale as the ${\bf k}_d$ mode in Figure
\ref{fig:dynamo_mode_Bxk} (perhaps on even a shorter time-scale due
to the contribution of higher $k_z\geq 2$ axisymmetric modes). Consequently, the patches of positive (yellow and red) and negative (blue) values of the fields are smaller contrary to those in taller boxes.}\label{fig:dynamo_mode_Bxa}
\end{figure}

This behavior is seen in Figure \ref{fig:dynamo_mode_Bxk} depicting
the evolution of the radial $\bar{B}_x({\bf k}_d)$, and the dominant
azimuthal $\bar{B}_y({\bf k}_d)$, fields associated with the ${\bf k}_d$ mode together with driving Maxwell stress ${\cal M}({\bf k}_d)$ and
the  nonlinear terms ${\cal N}_x^{(b)}({\bf k}_d)$, ${\cal
N}_y^{(b)}({\bf k}_d)$ at ${\rm Pm}=4$. As it is often done in
MRI-turbulence studies, in Figure \ref{fig:dynamo_mode_Bxa}, we also
plot the corresponding space-time diagrams of the radial and
azimuthal fields in physical space averaged horizontally in
$(x,y)$-slice, which are dominated by the ${\bf k}_d$ mode. In these
figures, we have chosen a similar total time interval, $334$ (in
units of $\Omega^{-1}$), as used by \citet{Lesur_Ogilvie08,
Shi_etal16} in order to facilitate comparison with the related plots
of the horizontally averaged fields in those papers. It is seen that
the variation of $\bar{B}_x({\bf k}_d)$ and $\bar{B}_y({\bf k}_d)$ with time is even more irregular and non-periodic than those for taller ($L_z/L_x \geq 2$) boxes, occurring over a shorter timescale  $\lesssim 10$, nearly as fast as the characteristic time of nonmodal MRI growth of non-axisymmetric active modes. This time-scale of the mean field is consistent with the estimates of \citet{Shi_etal16} for standard boxes. Note that the radial field exhibits variations on even shorter time-scale
compared to the azimuthal field. Since the ${\bf k}_d$ mode makes
the largest contribution to the horizontally (over $x$ and $y$) averaged fields, $\langle B_x \rangle$ and $\langle B_y\rangle$, the latter
also exhibit temporal variations on a similar time-scale (Figure \ref{fig:dynamo_mode_Bxa}). As a result, the
overall pattern of large-scale field as a function of $t$ and $z$ appears to be less organized than that in taller boxes considered in
\citet{Lesur_Ogilvie08, Shi_etal16,Walker_Boldyrev17}.


The evolution of the large-scale radial and azimuthal fields is a
consequence of the specific temporal behavior of the Maxwell stress
and the nonlinear terms also shown in Figure
\ref{fig:dynamo_mode_Bxk}. ${\cal N}_x^{(b)}({\bf k}_d)$ oscillates
irregularly and rapidly, increasing the radial field when positive
and draining it when negative. The Maxwell stress ${\cal M}({\bf
k}_d)$ is always positive, acting as a main driver of the azimuthal
field, and varies with time slower than ${\cal N}_x^{(b)}({\bf
k}_d)$ does. On the other hand, ${\cal N}_y^{(b)}({\bf k}_d)$ is always negative, draining the azimuthal field. It nearly balances
the positive Maxwell term, closely following its peaks.

The above analysis shows that in the standard box $(L_x,L_y,L_z)=(3,3,1)$ used here, there is no {\it temporal} scale separation between the ${\bf k}_d$ mode and other non-axisymmetric smaller-scale modes. Besides, as we have also checked, for our setup, the contribution of the ${\bf k}_d$ mode's interaction with other non-axisymmetric modes is small in the nonlinear transfer terms. For these reasons, unlike \citet{Herault_etal11,Riols_etal15,Riols_etal17}, we have preferred not to separate out the ${\bf k}_d$ dynamo mode, which carries the large-scale axisymmetric azimuthal field, and consider the dynamics of nearby non-axisymmetric modes against its background, but rather we treat it from a general standpoint of the dynamics and nonlinear
interactions of modes in Fourier space. In fact, the nonlinear
interaction of the ${\bf k}_d$ mode with other modes, whatever
effect it has, is already taken care of in the nonlinear terms
${\cal N}_i^{(b)}$, while its energy supply from the background shear flow is described by the Maxwell stress, which are all computed in Section \ref{sec:Spectraldynamics}. This is one of the advantages of our general analysis of the spectral dynamics of the turbulence that it gives a good overview of the active modes as well as the underlying linear and nonlinear processes in Fourier space.

\bibliographystyle{aasjournal}
\bibliography{biblio}

\end{document}